\documentclass[review]{elsarticle}

\usepackage{rotating}
\usepackage{lineno,hyperref}
\usepackage{amssymb}
\usepackage{amsmath}
\usepackage{latexsym}
\usepackage{graphicx}
\usepackage{bm}
\usepackage{color}
\usepackage{array,float}
\usepackage{booktabs}
\usepackage{pdflscape,color}
\modulolinenumbers[5]

\journal{Mathematical Biosciences}

\begin{document}

\begin{frontmatter}

\title{Inference on an heteroscedastic Gompertz tumor growth model}

\author[AlbanoAddress]{G. Albano\corref{mycorrespondingauthor}}
\ead{pialbano@unisa.it}
\author[GiornoAddress]{V. Giorno}
\ead{giorno@unisa.it}
\author[Granada,Granada2]{P. Rom{\'a}n-Rom{\'a}n}
\ead{proman@ugr.es}
\author[Roman2Address]{S. Rom{\'a}n-Rom{\'a}n}
\ead{sergio.Roman-Roman@curie.fr}
\author[Granada]{J.J. Serrano-P{\'e}rez}
\ead{jjserra@ugr.es}
\author[Granada,Granada2]{F. Torres-Ruiz}
\ead{fdeasis@ugr.es}
\address[AlbanoAddress]{Dipartimento di Studi Politici e Sociali, Universit\`a di Salerno, Italy}
\address[GiornoAddress]{Dipartimento di Informatica, Universit\`a di Salerno, Italy}
\address[Granada]{Departamento de Estad{\'\i}stica e Investigaci{\'o}n Operativa, Universidad de Granada, Spain}
\address[Granada2]{Instituto de Matem\'aticas (IEMath-GR), Universidad de Granada, Spain}
\address[Roman2Address]{D{\'e}p. de Recherche Translationnelle, Institut Curie, France}
\cortext[mycorrespondingauthor]{Corresponding author}


%

\begin{abstract}
We consider a non homogeneous Gompertz diffusion process whose parameters
are modified by generally time-dependent exogenous factors included in the infinitesimal moments.
The proposed model is able to describe tumor dynamics under the effect of anti-proliferative
and/or cell death-induced therapies. We assume that such therapies can modify also the
infinitesimal variance of the diffusion process. An estimation procedure, based on a control group
and two treated groups, is proposed to infer the model by estimating the constant parameters and
the time-dependent terms. Moreover, several concatenated hypothesis tests are considered in order
to confirm or reject the need to include time-dependent functions in the infinitesimal moments.
Simulations are provided to evaluate the efficiency of the suggested procedures and to
validate the testing hypothesis. Finally, an application to real data is considered.
\end{abstract}

\end{frontmatter}


\section{Introduction}
Diffusion processes are widely used in the literature to describe phenomena in a lot of fields,
ranging from economics \cite{Steele_2012, Shreve_2004} to biology \cite{Allen_2003, Biology2}.
Concerning the tumor growth modeling, many efforts have been devoted in the last years since
nowadays the cancer represents one of the main causes of death in our society. Further, the
availability of modern diagnostic and prognostic methodologies allows to build ever more faithful
models, giving useful insights into the dynamics of such disease
\cite{Ahn_2002,Ochab_2004,Talkington_2015}. On the other hand, the mathematical tractability of
the model must be taken into account because it allows to better handle explicit solutions of the
involved dynamics. In this context,  Gompertz growth model seems successfully overcome the \lq\lq
trade off\rq\rq\ between these two aspects. Indeed, it is widely accepted that such model is able
to capture dynamics of solid tumors and several models based on this growth have been proposed by
looking at deterministic and stochastic behaviors
\cite{Devladar_2004,Ferrante_2005,Lo_2007,Lo_2008,Tj_2017, Yang_2020}.

\par The Gompertz curve belongs to the
Richards family of  sigmoidal growth models, along with familiar models such as the negative
exponential, the logistic, and the Bertalanffy \cite{Tj_2010}. These curves, although born in
deterministic contexts, have been generalized to include stochastic effects aimed at bridging the
gaps that often exist between experimental data and theoretical results. Concerning the stochastic
version of the Gompertz growth, in the literature various contributions can be found concerning
both theoretical probabilistic properties and statistical characteristics
\cite{MB_2013,giorno_2019, Ascione_2020}.

\par Recently, the attention has been focused on non-homogeneous Gompertz
diffusion process describing the effect of some exogenous generally time-dependent factors. In
preclinical tumor growth studies it is useful to understand as experimental therapies can  modify
the natural cancer cells' growth rates.
A modified Gompertz equation is considered in Cabrales et al. \cite{Cabrales_2018} to describe
tumor responses to electrochemical treatments and  the  possible  decay  of  solutions is
investigated both from theoretical and numerical points of view.
In \cite{Shakeri_2018} a control approach to predict an optimal drug
dosage shrinking the cancer tumor-cell population was proposed. The predictive control problem is
hence based on the difference between the probability density function and the desired probability
density function calculated at each time instant. \par
In the mathematical framework, an untreated tumor volume can be modeled by an homogeneous Gompertz
stochastic process. In order to include the effect of an anti-angiogenic therapy in \cite{MB_2013, JTB_2006, CSDA_2012} the infinitesimal
moments of the homogeneous process were modified by introducing suitable continuous time-dependent functions modeling the
exogenous factors; in such a way, the therapy presence leads to a time non homogeneous stochastic
process. A statistical approach was also proposed in \cite{JTB_2011, JTB_2015, JTB_2016} to fit the modifications in the natural growth rates due to several therapies.

\subsection{Motivations and plan of the paper}
In the present paper we provide a natural generalization of the results provide in \cite{JTB_2015} and \cite{JTB_2016}. Specifically, we
assume that the two applied therapies (one of anti-proliferative and the other inducing the cancer
cells death) are generally able to modify both the drift and the infinitesimal variance of the process making it
time-dependent. We propose a statistical methodology to estimate the natural tumor rates and to
fit the exogenous term including also the infinitesimal variance of the resulting process. The
procedure uses a control group ${\cal G}$ described by a homogeneous Gompertz diffusion process and two treated groups
${\cal G}_1$ and ${\cal G}_2$ described by two time non-homogeneous processes. The group
${\cal G}_1$ is assumed to be treated with a therapy (anti-proliferative or cell death-induced)
while the group ${\cal G}_2$ is treated the same therapy of ${\cal G}_1$ together with an other
therapy of the other type. Further, we consider the case in which the two groups are characterized
by two generally different time-dependent infinitesimal variance. \par The procedure works as
follows. In a first step, the control group ${\cal G}$ is used to estimate the constant parameters
included in the homogeneous process, whereas in the subsequent steps the treated groups, ${\cal G}_1$ and
${\cal G}_2$, are used to fit the unknown time-dependent functions describing the effect of the
therapies. Precisely, via suitable mathematical relations, the group ${\cal G}_1$ is used to fit
the function related to the single therapy applied in it and the effect that this therapy has on
the infinitesimal variance. Then, the group ${\cal G}_2$ is used to fit the second therapy and its effect on the
infinitesimal variance.
\par Moreover, a bootstrap testing procedure able to evaluate the time
dependence of the exogenous factors is provided. Essentially, it is directed to establish if the
real effect of the various applied therapies has a known functional form. In particular the
proposed test is then used to establish if the therapy effect is null or constant.
\par We point out that both the estimation procedure and the testing hypothesis on the exogenous
factors related to the stochastic diffusion processes are of interest in various applicative and
theoretical contexts \cite{hook,Ignatieva,Mandal}.
 \par
The plan of the paper is  the following. In Section~2 the model is introduced and its probability
distribution  and some statistical  characteristics are derived. In Section~3 the procedure  to
estimate the parameters and to fit the unknown functions is proposed. Various simulated-based
examples are given to validate the fitting procedure. In Section~4 the hypothesis test procedure
is provided and several cases of particular interest in biological context are considered. In
particular, some concatenated tests are performed to evaluate the constant/null effect of the
therapy on the rates and on the infinitesimal variance. Finally, in Section~5 an application to
real data is provided to study the combined effect of Carboplatin and Taxol in ovarian cancer.

\section{The model}
In the mathematical framework, an untreated tumor volume can be modeled by an homogeneous Gompertz
stochastic process defined in $\mathbb{R}^+$ with infinitesimal moments
\begin{eqnarray}
&&A_1(x)=\alpha x-\beta x\log x,\nonumber\\
&&A_2(x)=\sigma^2 x^2,\label{control}
\end{eqnarray}
where $\alpha,\beta$ and $\sigma$ are positive constants. The parameters
$\alpha$ and  $\beta$  describe the cell's growth and death rates, respectively, $\sigma$ is
related to more or less intense environmental fluctuations introduced to justify discrepancies
between clinical data and theoretical predictions that quite often are detected.
\par Our approach for including the effect of an anti-angiogenic therapy consists to modify the infinitesimal
moments (\ref{control}) by introducing suitable continuous time-dependent functions modeling the
exogenous factors; in such a way, the therapy presence leads to a time non homogeneous stochastic
process.
Precisely, let $\{X(t): t\geq t_0\}$ with $t_0\geq 0$ be a stochastic process in $\mathbb{R}^+$ and satisfying
the following stochastic differential equation (SDE)

\begin{eqnarray}
&&dX(t) = \left\{(\alpha-C(t))-(\beta-D(t))\ln\,X(t)\right\}X(t)\,dt+\sigma\sqrt{V(t)}X(t)\,dW(t),\nonumber\\
&&X(t_0) = X_0.
\label{SDE}
\end{eqnarray}

Here, as in \eqref{control},  $\alpha,\beta$ and $\sigma$ are positive constants, while $C(t),D(t)$ and $V(t)$ are  functions in
$C_1[t_0,+\infty)$ with $V(t)>0$ for all $t\geq t_0$,  $X_0$ is a random variable describing the
initial state of the process, and $W(t)$ is a standard Wiener process independent from  $X_0$ for
$t\geq t_0$. In the model setting, $C(t)$ represents
tumor regression rate due to the therapy and has the same dimension as parameter $\alpha$, while
the function $D(t)$ modifies the death rate $\beta$ of the process \eqref{control} in $\beta-D(t)$. From a biological point of view, the function $C(t)$ describes the effect of an anti-proliferative therapy, that is, able to modify the natural birth rate of cancer cells, while $D(t)$ describes the effect of cell death-induced therapy (see \cite{MB_2013, JTB_2015}). Clearly $C(t)$ successfully applied when it assumes positive values, while $D(t)$ is effective when it is a negative function. Further it would be desirable to have small values of the function $V(t)$, describing fluctuations in the tumor volume. Anyway, in experimental studies the effectiveness of a therapy has to be tested, so we assume that the functions $C(t)$, $D(t)$ have real values and $V(t)>0.$
\par
The aim of this paper is to model the combined effect of two therapies, one anti-proliferative and
the other that induces the death of cancer cells. In this sense, model \eqref{SDE} can be viewed
as a modification of model \eqref{control} after transforming its infinitesimal moments by
introducing the functions $C(t)$, $D(t)$ and $V(t)$.

\vskip 0.2cm By considering
\begin{equation}
Z(t)=f(X(t),t)=k(t)\,\ln X(t),\label{transform}
\end{equation}
where
\begin{equation*}
k(t)=\exp\left(\int^t[\beta-D(s)]\,ds\right),
\end{equation*}
and by applying It{\^o}'s Lemma, we can transform process $X(t)$ into a non-homogeneous Wiener process
$Z(t)$ described by the following SDE:
\begin{equation}
dZ(t)=a(t)\,dt+b(t)\,dW(t), \qquad Z(t_0)=Z_0,\label{SDE_trasformate}
\end{equation}
with
\begin{equation*}
a(t)= k(t)\,\left[\alpha-C(t)-\frac{\sigma^2 V(t)}{2}\right],\qquad
b(t)=\sigma\sqrt{V(t)}\,k(t),
\end{equation*}
whose solution is
\begin{equation*}
Z(t)=Z_0+\int_{t_0}^ta(s)ds+\int_{t_0}^tb(s)\,dW(s). 
\end{equation*}
Finally, undoing the change (\ref{transform}), we obtain
\begin{equation*}
X(t)=\exp\Biggr\{\frac{1}{k(t)}\Biggl[k(t_0)\ln X_0+\int_{t_0}^ta(s)ds +
\int_{t_0}^tb(s)\,dW(s)\Biggr]\Biggr\}. 
\end{equation*}

\subsection{Distribution of the process}

From (\ref{SDE_trasformate}), and if $Z_0$ is a degenerate random variable, i.e. $P(Z_0=z_0)=1$, with $z_0 \in \mathbb R$
or normally distributed, i.e. $Z_0\sim N_1[\mu_0,\sigma_0^2]$, then $Z(t)$ is a Gaussian process,
so, $\forall \, n \in \mathbb{N}$ and $t_1<\cdots<t_n$, vector $(Z(t_1),\ldots,Z(t_n))^T$ has a
$n$-dimensional normal distribution $N_n[\bm\varepsilon,\bm\Sigma]$, where the components of
vector $\bm\varepsilon$ and matrix $\bm\Sigma$ are
$$
\varepsilon_i=E[Z_0]+\int_{t_0}^{t_i}k(s)\left[\alpha-C(s)-\frac{\sigma^2 V(s)}{2}\right]\,ds, \ \
i= 1, \ldots, n
$$
and
$$
\sigma_{ij}=Var[Z_0]+\sigma^2\int_{t_0}^{\min(t_i,t_j)}k^2(s)V(s)\,ds, \ \ i,j=1,\ldots,n,
$$
respectively.

Therefore, by \eqref{transform} all the finite-dimensional distributions of the process $X(t)$ are
lognormal; specifically, $\forall \, n \in \mathbb{N}$,
\begin{equation}
\label{DistrX1_Xn} (X(t_1), \ldots, X(t_n))^T
\sim \Lambda_n[\bm\xi,\bm\Delta],
\end{equation}
where $\xi_i=\dfrac{\varepsilon_i}{k(t_i)}$ and $\delta_{ij}=\dfrac{\sigma_{ij}}{k(t_i)k(t_j)}$,
$i,j=1\ldots,n$, are the components of $\bm\xi$ and $\bm\Delta$, respectively.

\vskip 8pt In particular, by considering $X_0\sim\Lambda_1[\mu_0;\sigma_0^2]$, we have

\begin{equation*}
\label{DistX(t)} X(t)\sim \Lambda_1\left[M^*(t|\mu_0,t_0);V^*(t|\sigma_0^2,t_0)\right],
\end{equation*}
where, for $\tau < t$,
\[
M^*(t|u, \tau)= u \bar{k}(t|\tau)+\displaystyle\int_{\tau}^{t}
\displaystyle\left(\alpha-C(s)-\frac{\sigma^2V(s)}{2}\right)\bar{k}(t|s) \, ds,
\]
and
\[
{V}^*(t|u, \tau)=u  \bar{k}^2(t|\tau)+\displaystyle\int_{\tau}^{t} \sigma^2 V(s)\bar{k}^2(t|s) \,
ds,
\]
with $\bar{k}(t|\tau) =k(\tau)/k(t)$.

\vskip 8pt In the following we will assume that $X_0$ is a degenerate random variable in $x_0$.
This assumption is quite common in the context of tumor growth since the variable of interest is
usually the relative volume of the tumor, and $x_0=1$ is the relative volume at the detection of
the tumor. So, we will assume $P[X_0=x_0]=1$. In this case
\begin{equation*}
\label{DistrX_X0d} X(t)\sim \Lambda_1\left[m_1(t);u(t)\right],
\end{equation*}
with
\begin{equation}
\label{ElogX} m_1(t) = M^*(t|\ln x_0,t_0)
\end{equation}
and
\begin{equation}
\label{VarlogX} u(t) = V^*(t|0,t_0).
\end{equation}

\noindent So, the mean and the variance functions of $X(t)$ are:
\begin{align}
  E[X(t)] & = \exp\left(m_1(t)+\frac{1}{2} u(t) \right) , \nonumber\\
& \label{momento}\\
  Var[X(t)] & = \exp\left(2 m_1(t) + u(t) \right) \times \left[\exp\left(u(t) \right)-1\right] , \nonumber
\end{align}
respectively.

\section{Estimation of the model}

In this section we propose a procedure to estimate the parameters $\alpha,\beta,\sigma$, and to
approximate the functions $C(t)$, $D(t)$ and $V(t)$ in $[t_0,T]$. To this end, in practice it is
necessary to have data from three experimental groups of individuals. Concretely:
\begin{itemize}
  \item an untreated (control) group, say ${\cal G}$,
  \item a first group, ${\cal G}_1$, treated with a single therapy that affects only one of the two rates that model the untreated tumor
  volume,
  \item a second group, ${\cal G}_2$, treated with two therapies. One of them must be the same therapy applied in ${\cal G}_1$, whereas the other one affects the rate not modified in ${\cal G}_1$.
\end{itemize}

The control group is associated to the stochastic process $X(t)$ described by the SDE
\begin{equation}
dX(t)= [\alpha-\beta\ln X(t)] X(t)dt+ \sigma X(t)dW(t)\qquad X(t_0)=x_0.\label{gruppo_G}
\end{equation}
Moreover, group ${\cal G}_1$ is modeled by a stochastic process $X_1(t)$ for which two cases can
be considered:

\begin{itemize}
\item ${\cal G}_1$ is treated with an anti-proliferative therapy, i.e. mainly affecting cell growth. In this case, $X_1(t)$ follows the SDE
\begin{equation}
d X_1(t)=\left\{[\alpha-C(t)]-\beta\ln X_1(t)\right\} X_1(t) dt+\sigma \sqrt{V_1(t)}\,
X_1(t)dW(t),\,\, X_1(t_0)=x_0. \label{case_a_X1}
\end{equation}
\item ${\cal G}_1$ is treated with a therapy that induces, or mainly induces, the death of cancer cells. Now the SDE followed by $X_1(t)$ is
\begin{equation}
d X_1(t)=\left\{\alpha-[\beta-D(t)]\ln X_1(t)\right\} X_1(t)dt + \sigma \sqrt{V_1(t)}\,
X_1(t)dW(t),\,\, X_1(t_0)=x_0. \label{case_b_X1}
\end{equation}
\end{itemize}
Finally, group ${\cal G}_2$ is described by a stochastic process $X_2(t)$ solution of
\begin{equation}
d X_2(t)=\left\{[\alpha-C(t)]-[\beta-D(t)]\ln X_2(t)\right\} X_2(t)dt + \sigma \sqrt{V_2(t)}\,
X_2(t)dW(t),\,\, X_2(t_0)=x_0. \label{case_a_b_X2}
\end{equation}

The basic idea is to use data from the control group to estimate the parameters $\alpha,\beta$ and
$\sigma^2$, whereas the treated groups are used to fit the functions $C(t)$, $D(t)$, $V_1(t)$ and
$V_2(t)$.

\subsection{Some basic expressions}
In this subsection we introduce some expressions that are the basis of the estimation procedure
developed in the next one.

From \eqref{momento}, we define
\begin{equation*}
m_2(t)=\ln E[X(t)]=m_1(t)+\frac{1}{2} u(t), \label{m2}
\end{equation*}
and by considering \eqref{ElogX} and \eqref{VarlogX}, after some algebra, the following
relationships are obtained:
\begin{align}
\label{C} C(t)&=
\alpha-(\beta -D(t))(m_1(t)+u(t)) - m_1^{\prime}(t) - \frac{1}{2} u^{\prime}(t)\nonumber \\
&=\alpha-(\beta -D(t))(2m_2(t)-m_1(t))- m_2^\prime (t)
\end{align}

\begin{align}
\label{D} D(t)&=
\beta +\displaystyle\dfrac{m_1^{\prime}(t) + \displaystyle\frac{1}{2} u^\prime(t) - \alpha + C(t)}{m_1(t) + u(t)}=\beta +\displaystyle\dfrac{m_2^\prime (t) - \alpha + C(t)}{2 m_2(t) - m_1(t)}
\end{align}

\begin{align}
\label{V} V(t)&=
\displaystyle\frac{1}{\sigma^2} \displaystyle(  u^\prime (t) +2 (\beta -D(t)) u(t) \displaystyle)\nonumber\\
&=\displaystyle\frac{2}{\sigma^2} \displaystyle[  (m_2^\prime (t) - m_1^\prime (t)) +2
(\beta-D(t))(m_2(t)-m_1(t)) \displaystyle].
\end{align}

We point out that the two expressions obtained for $C(t), D(t)$ and $V(t)$ in \eqref{C}, \eqref{D} and \eqref{V} respectively, can be alternatively used to fit the functions depending on the behavior of the sampling versions of these functions in real applications.

\subsection{The estimation procedure}

Let us consider $d$ sample-paths from the control group, observed at the same time instants $t_j$,
$j=0,\ldots,n-1$, in the interval $[t_0,T]$. Let $\{x_{ij}, \, i=1,\ldots,d;\,j=0,\ldots,n-1\}$ be
the observed values of the sample paths. Moreover, let $\{x^{(k)}_{ij}, \,
i=1,\ldots,d_k;\,j=0,\ldots,n-1\}$ be the values of $d_k$ sample paths from the treated group
${\cal G}_k$, $k=1,2$, observed at the same previous time instants.
\medskip

Making use of Equations (\ref{C})-(\ref{V}), and denoting $m_1^{(k)}(t)=E[\ln\,X_k(t)]$,
$m_2^{(k)}(t)=\ln E[X_k(t)]$ and $u^{(k)}(t)=Var[\ln\,X_k(t)]$, $k=1,2$, we can estimate the three
models from data provided by the control and the two treated groups. To this end we provide the
following stepwise procedure:
\begin{itemize}
\item Obtain the maximum likelihood (ML) estimates of $\alpha$, $\beta$ and $\sigma^2$ by solving the likelihood equation system (\ref{vero-alpha})-(\ref{vero-sigma2}) in Appendix for $C(t)=D(t)=0$ and $V(t)=1$, from the data of group ${\cal G}$.
Denote by $\widehat\alpha$, $\widehat\beta$ and $\widehat\sigma^2$ such estimates. \vspace{5pt}
\item Calculate at each time instant $t_j$, $j=0, \ldots, n-1$, the values
\[
\widehat{m}_1^{(k)}(t_j)=\bar{y}_j^{k}, \,\, \widehat{m}_2^{(k)}(t_j)=\ln(\bar{x}_j^{(k)}), \,\,
\widehat{u}^{(k)}(t_j)= {s_j^{2}}^{(k)}, \, \, k=1, 2
\]
where
\begin{itemize}
\item[-] $\bar{y}_j^{k}$ is the sample mean of the logarithms of the values of the sample paths of the group ${\cal G}_k$ ($k=1,2$) at $t_j$,
\item[-] $\bar{x}_j^{(k)}$ is the sample mean of the values of the sample paths of  ${\cal G}_k$ ($k=1,2$) at $t_j$,
\item[-] $ {s_j^{2}}^{(k)}$ is the unbiased sample variance of the logarithms of the values of the sample paths of ${\cal G}_k$ ($k=1,2$) at $t_j$.
\end{itemize}
\item For $k=1,2$, approximate the derivatives of  $m_1^{(k)}(t)$, $m_2^{(k)}(t)$ and $u^{(k)}(t)$, at $t_j$ from the values obtained in the previous step.
Denote by $\widehat{m}_1^{(k)^{\prime}}(t_j)$, $\widehat{m}_2^{(k)^{\prime}}(t_j)$ and
$\widehat{u}^{(k)^{\prime}}(t_j)$ the obtained values. \vspace{5pt}

\item Estimating $C(t)$, $D(t)$, $V_1(t)$ and $V_2(t)$ as follows:

\begin{itemize}
\item If ${\cal G}_1$ is modeled by \eqref{case_a_X1}, i.e. it is treated with an anti-proliferative therapy, obtain an initial estimate of
$C(t_j)$ and $V_1(t_j)$ by applying the observed data of this group to expressions (\ref{C}) and
(\ref{V}), with $D(t)=0$. This leads to
\begin{align*}
\widehat{C}_j &= \widehat{\alpha}- \widehat{\beta} \left( \widehat{m}_1^{(1)}(t_j) + \widehat{u}^{(1)}(t_j)\right) - \widehat{m}_1^{(1)^{\prime}}(t_j) - \frac{1}{2} \widehat{u}^{(1)^{\prime}}(t_j) \\
&=\widehat{\alpha}- \widehat{\beta} \left(2 \widehat{m}_2^{(1)}(t_j) -
\widehat{m}_1^{(1)}(t_j)\right)- \widehat{m}_2^{(1)^{\prime}}(t_j)
\end{align*}
and
\begin{align*}
\widehat{V}_{1,j} &=
\displaystyle\frac{1}{\widehat{\sigma}^2} \left( \widehat{u}^{(1)^{\prime}}(t_j)+ 2 \widehat{\beta} \widehat{u}^{(1)}(t_j) \right)  \\
&=\displaystyle\frac{2}{\widehat{\sigma}^2} \left(\widehat{m}_2^{(1)^{\prime}}(t_j) -
\widehat{m}_1^{(1)^{\prime}}(t_j) + 2 \widehat{\beta} \left(
\widehat{m}_2^{(1)}(t_j)-\widehat{m}_1^{(1)}(t_j) \right) \right).
\end{align*}

Next, for each $t_j$, calculate  initial estimates of $D(t_j)$ and $V_2(t_j)$ for process
$X_2(t)$, by considering (\ref{D}) and (\ref{V}) for the data of group ${\cal G}_2$ and the
previous $\widehat{C}_j$ values. In this way the following values are obtained:

\begin{align*}
\widehat{D}_j &=
\widehat{\beta}+ \displaystyle\frac{\widehat{m}_1^{(2)^{\prime}}(t_j) + \frac{1}{2} \widehat{u}^{(2)^{\prime}}(t_j) - \widehat{\alpha} + \widehat{C}_j}{\widehat{m}_1^{(2)}(t_j) + \widehat{u}^{(2)}(t_j)}  \\
&=\widehat{\beta} + \displaystyle\frac{\widehat{m}_2^{(2)^{\prime}}(t_j) - \widehat{\alpha} +
\widehat{C}_j}{2 \widehat{m}_2^{(2)}(t_j) - \widehat{m}_1^{(2)}(t_j)}
\end{align*}
and
\begin{align*}
\widehat{V}_{2,j} &=
\displaystyle\frac{1}{\widehat{\sigma}^2} \left( \widehat{u}^{(2)^{\prime}}(t_j)+ 2 (\widehat{\beta}-\widehat{D}_j) \widehat{u}^{(2)}(t_j) \right)  \\
&=
\displaystyle\frac{2}{\widehat{\sigma}^2} \left(\widehat{m}_2^{(2)^{\prime}}(t_j)-
\widehat{m}_1^{(2)^{\prime}}(t_j) + 2 (\widehat{\beta}-\widehat{D}_j) \left(
\widehat{m}_2^{(2)}(t_j)-\widehat{m}_1^{(2)}(t_j) \right) \right).
\end{align*}

\item If ${\cal G}_1$ is treated with a therapy that induces the death of cancer cells, i.e. the model \eqref{case_b_X1} is now considered,
determine values $\widehat{D}_j$ and $\widehat{V}_{1,j}$ (initial estimates of $D(t_j)$ and
$V_1(t_j)$, $j=0,\ldots,n-1$) from (\ref{D}) and (\ref{V}) by considering $C(t)=0$ and the data of
${\cal G}_1$, thus obtaining
\begin{align*}
\widehat{D}_j &=
\widehat{\beta}+ \displaystyle\frac{\widehat{m}_1^{(1)^{\prime}}(t_j) + \frac{1}{2} \widehat{u}^{(1)^{\prime}}(t_j) - \widehat{\alpha}}{\widehat{m}_1^{(1)}(t_j) + \widehat{u}^{(1)}(t_j)}  \\
&=\widehat{\beta}+ \displaystyle\frac{\widehat{m}_2^{(1)^{\prime}}(t_j)- \widehat{\alpha}}{2
\widehat{m}_2^{(1)}(t_j) - \widehat{m}_1^{(1)}(t_j)}
\end{align*}
and
\begin{align*}
\widehat{V}_{1,j}&=
\displaystyle\frac{1}{\widehat{\sigma}^2} \left( \widehat{u}^{(1)^{\prime}}(t_j)+ 2 (\widehat{\beta}-\widehat{D}_j) \widehat{u}^{(1)}(t_j)  \right)  \\
&=\displaystyle\frac{2}{\widehat{\sigma}^2} \left(\widehat{m}_2^{(1)^{\prime}}(t_j)-
\widehat{m}_1^{(1)^{\prime}}(t_j) + 2 (\widehat{\beta}-\widehat{D}_j) \left(
\widehat{m}_2^{(1)}(t_j)-\widehat{m}_1^{(1)}(t_j) \right) \right).
\end{align*}

Then, for process $X_2(t)$, compute initial estimates of $C(t_j)$ and $V_2(t_j)$,  $t_j$, $j=0,
\ldots,n-1$, from (\ref{C}) and (\ref{V}) by taking the data of group ${\cal G}_2$ and the values
$\widehat{D}_j$ previously estimated. This leads to

\begin{align*}
\widehat{C}_j &=
\widehat{\alpha}- (\widehat{\beta}-\widehat{D}_j) \left( \widehat{m}_1^{(2)}(t_j) + \widehat{u}^{(2)}(t_j)\right) - \widehat{m}_1^{(2)^{\prime}}(t_j) - \frac{1}{2} \widehat{u}^{(2)^{\prime}}(t_j) \\
&=\widehat{\alpha}- (\widehat{\beta}-\widehat{D}_j) \left(2 \widehat{m}_2^{(2)}(t_j) -
\widehat{m}_1^{(2)}(t_j)\right)- \widehat{m}_2^{(2)^{\prime}}(t_j)
\end{align*}
and
\begin{align*}
\widehat{V}_{2,j}&=
\displaystyle\frac{1}{\widehat{\sigma}^2} \left( \widehat{u}^{(2)^{\prime}}(t_j)+ 2 (\widehat{\beta}-\widehat{D}_j) \widehat{u}^{(2)}(t_j)  \right)  \\
&=\displaystyle\frac{2}{\widehat{\sigma}^2} \left(\widehat{m}_2^{(2)^{\prime}}(t_j)-
\widehat{m}_1^{(2)^{\prime}}(t_j) + 2 (\widehat{\beta}-\widehat{D}_j) \left(
\widehat{m}_2^{(2)}(t_j)-\widehat{m}_1^{(2)}(t_j) \right) \right).
\end{align*}

\end{itemize}

\item Obtain $\widehat{C}(t)$, $\widehat{D}(t)$, $\widehat{V}_1(t)$ and $\widehat{V}_2(t)$ as follows:
\begin{itemize}
  \item Calculate the final estimated values $\widehat{C}(t_j)$, $\widehat{D}(t_j)$, $\widehat{V}_1(t_j)$ and $\widehat{V}_2(t_j)$  by using
  local regression of $\widehat{C}_j$, $\widehat{D}_j$, $\widehat{V}_{1,j}$ and $\widehat{V}_{2,j}$ on $t_j$,
  respectively. \vspace{5pt}
  \item Interpolate, by means of  spline functions\footnote{Since the functions $C(t)$, $D(t)$ and $V(t)$ are sufficiently smooth (they are $C^1$-class), we use the natural cubic spline interpolation.}, the data points $(t_j, \widehat{C}(t_j))$, $(t_j, \widehat{D}(t_j))$, $(t_j, \widehat{V}_1(t_j))$ and $(t_j,\widehat{V}_2(t_j))$, respectively.
\end{itemize}

\end{itemize}

\subsection{Simulation-based applications} 

In order to validate the proposed estimation procedure, we have developed two applications based
on simulated data:

\begin{itemize}
  \item In the former, we consider an untreated group (${\cal G}$), a first group (${\cal G}_1$) treated with an anti-proliferative therapy (so, it is modeled by (\ref{case_a_X1})), and a second group (${\cal G}_2$) that is treated with the same therapy as the first group together with another one inducing the death of cancer cells. This group is modeled by (\ref{case_a_b_X2}).
  \item In the second case, in addition to the control group (${\cal G}$), we consider a group ${\cal G}_1$ treated with a therapy that induces the death of cancer cells (modeled by (\ref{case_b_X1})),
  whereas ${\cal G}_2$ is treated with the same therapy as the first group together with an anti-proliferative therapy, so this group is modeled by (\ref{case_a_b_X2}).
\end{itemize}

In the two applications the untreated group is modeled by (\ref{gruppo_G}) with the same
parameters. Table \ref{Ap} summarizes the parameters and functions considered. The choice of $\alpha$, $\beta$ and $\sigma$ values has been made so that the simulated paths present values similar to real situations. On the other hand, the therapeutical functions in our simulation experiment are in line with \cite{MB_2013, JTB_2015}. In Application 1 we consider the case in which the group ${\cal G}_1$ is treated with an anti-proliferative linear therapy, while the group ${\cal G}_2$ is treated with a cell death-induced therapy having a \lq\lq bump effect\rq\rq\  when it is applied and asymptotically reduces of $12\%$ the natural death rate of the tumor. In Application 2 the two therapies are reversed. The infinitesimal variances $V_1(t)$ and $V_2(t)$ involve two lognormal probability density functions since we expect that the variability of the process is greatly influenced by the therapies when they are applied, then they restore to natural values. This assumption is close to what is observed in real situations like the one presented in Section 5.
\medskip

\begin{table}[h]
  \caption{Parameters and functions considered in each example (being $\Lambda_1(t, \mu, \sigma^2)$ the density function of a lognormal
  distribution $\Lambda_1(\mu,\sigma^2)$).} \label{Ap} \smallskip
  \centering
  \tabcolsep 0pt
  \begin{tabular}{ccc}
     \toprule
     \textbf{Group} & \textbf{Application 1} & \textbf{Application 2} \\
     \midrule
     $\mathbf{\cal G}$ & \multicolumn{2}{c}{\fboxrule 0pt \fbox{$\alpha=0.5$, \ $\beta=0.2$, \ $\sigma=0.01$}} \\ \midrule
     $\mathbf{\cal G}_1$ & \fboxrule 0pt \fbox{$\setlength\extrarowheight{2pt} \begin{array}{c} C(t) = 0.005 \thinspace t \\
     V_1(t) = (0.7 \negthinspace + \negthinspace 10 \thinspace \Lambda_1(t, 3, 0.5))^2
     \end{array}$} & $\setlength\extrarowheight{2pt} \begin{array}{c} D(t) = -0.12 \thinspace t^2/(50 + t(t-10)) \\
     V_1(t) = (0.7 \negthinspace + \negthinspace 10 \thinspace \Lambda_1(t, 3, 0.5))^2
     \end{array}$ \\ \midrule
     $\mathbf{\cal G}_2$ & \fboxrule 0pt \fbox{$\setlength\extrarowheight{2pt} \begin{array}{c} D(t) = -0.12 \thinspace t^2/(50 + t(t-10)) \\
     V_2(t) = (0.7 \negthinspace + \negthinspace 15 \thinspace \Lambda_1(t, 3, 0.5))^2
     \end{array}$} & $\setlength\extrarowheight{2pt} \begin{array}{c} C(t) = 0.005 \thinspace t \\
     V_2(t) = (0.7 \negthinspace + \negthinspace 10 \thinspace \Lambda_1(t, 3, 0.5))^2
     \end{array}$ \\
     \bottomrule
   \end{tabular}
\end{table}

The estimation procedure has been replicated 100 times in both the examples. In each replication,
25 sample paths  have been simulated by considering 51 time instants equally spaced in the
interval $[0,50]$. The sample paths have been simulated using the \textit{snssde1d} function from
the R package \textit{Sim.DiffProc} \cite{simdiffproc}. This function allows to simulate the
solution of a stochastic differential equation from its discretization by using differents
numerical schema as Euler-Maruyama and Milstein among others (see Iacus \cite{Iacus} for details).
Further, a degenerate initial distribution at $x_0=1$ has been considered. This choice has been
made because in real studies on the evolution of tumors (such as the one shown in section 5) the
data provided are relative volumes of the tumors. \smallskip

\begin{landscape}
\begin{figure}[H]
\hspace{-0.35cm} \tabcolsep 0pt
\begin{tabular}{cccc}
  \includegraphics[height=0.35\textheight]{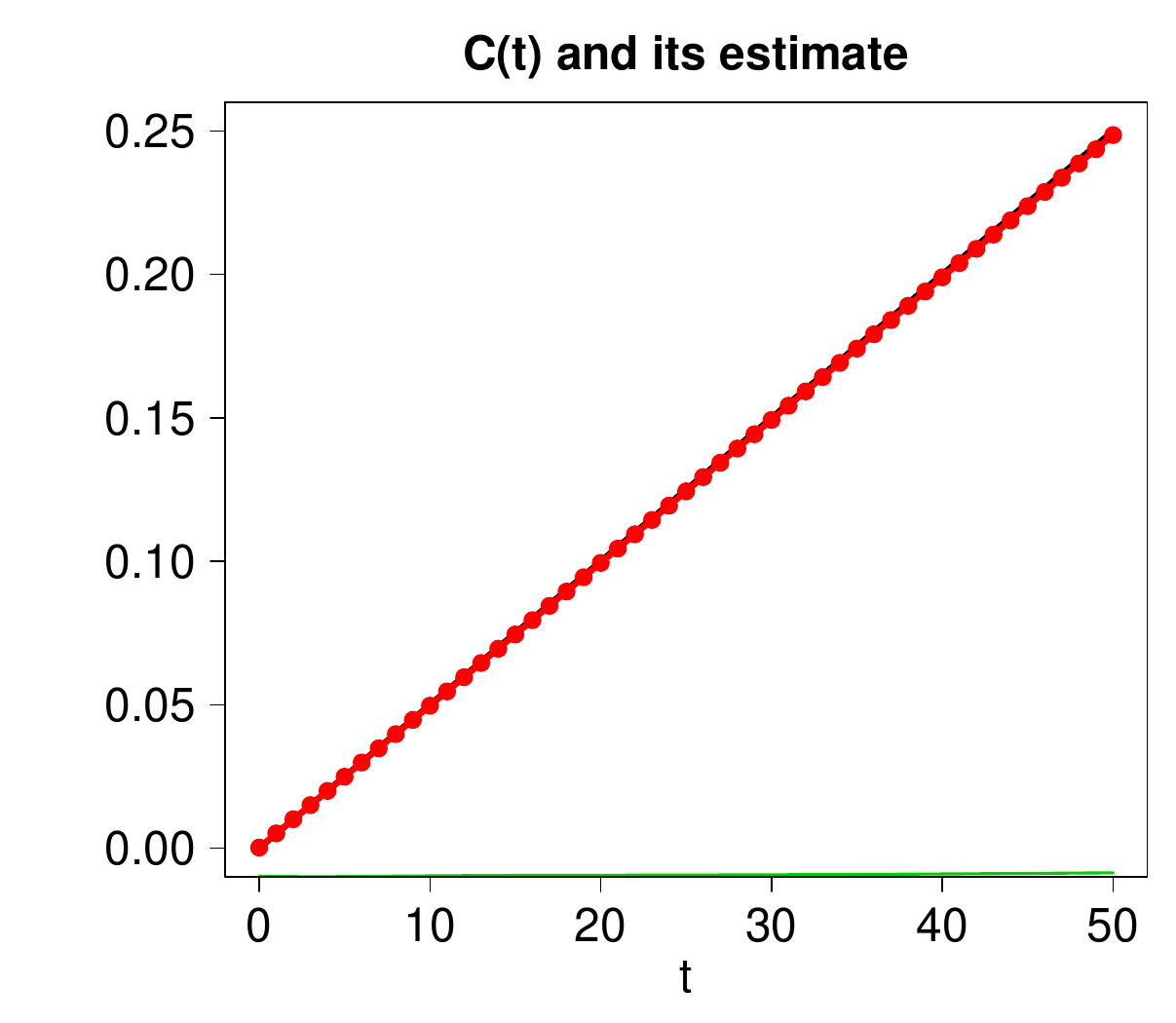} & \includegraphics[height=0.35\textheight]{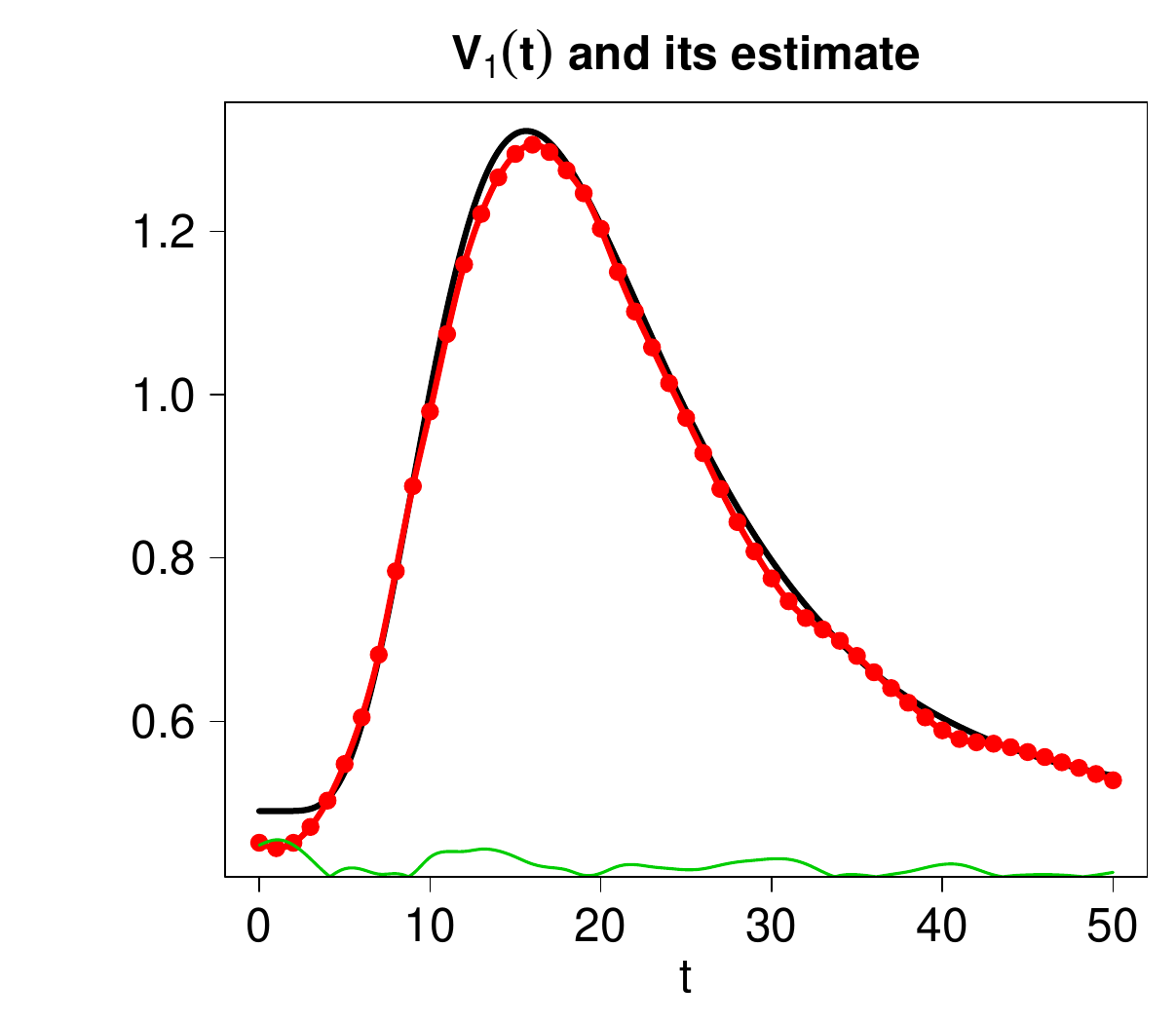} &
  \includegraphics[height=0.35\textheight]{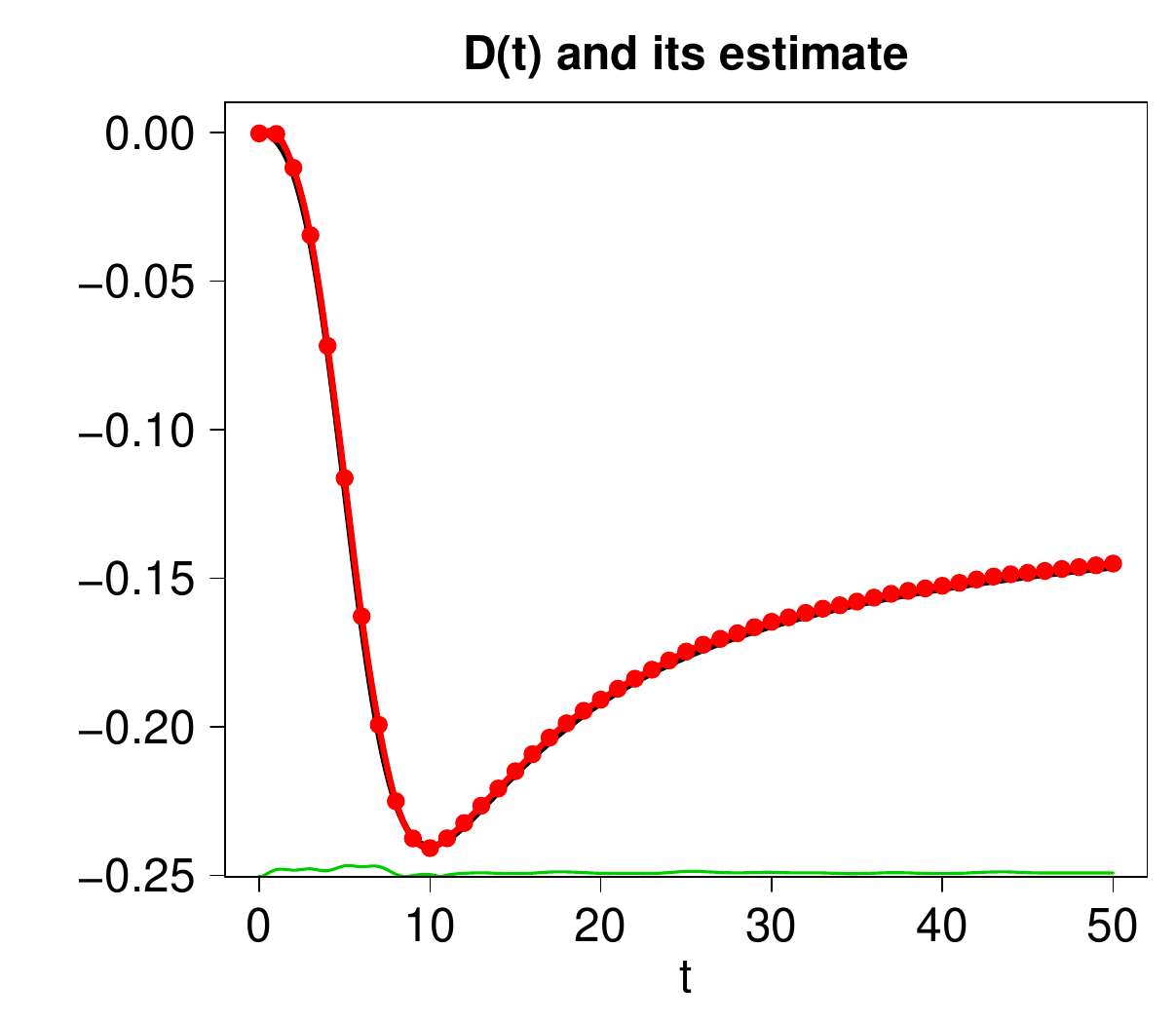} & \includegraphics[height=0.35\textheight]{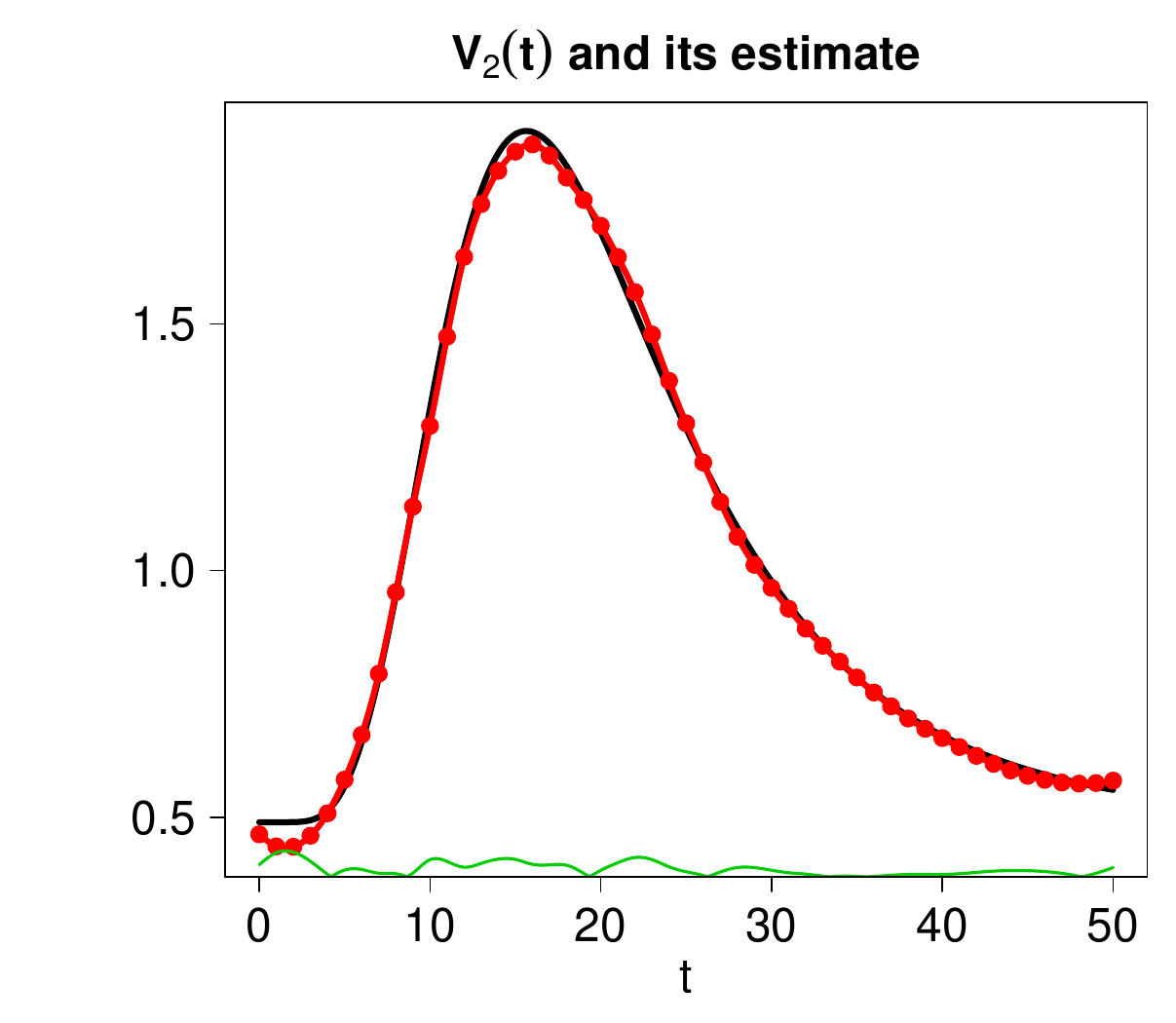} \\
  \includegraphics[height=0.35\textheight]{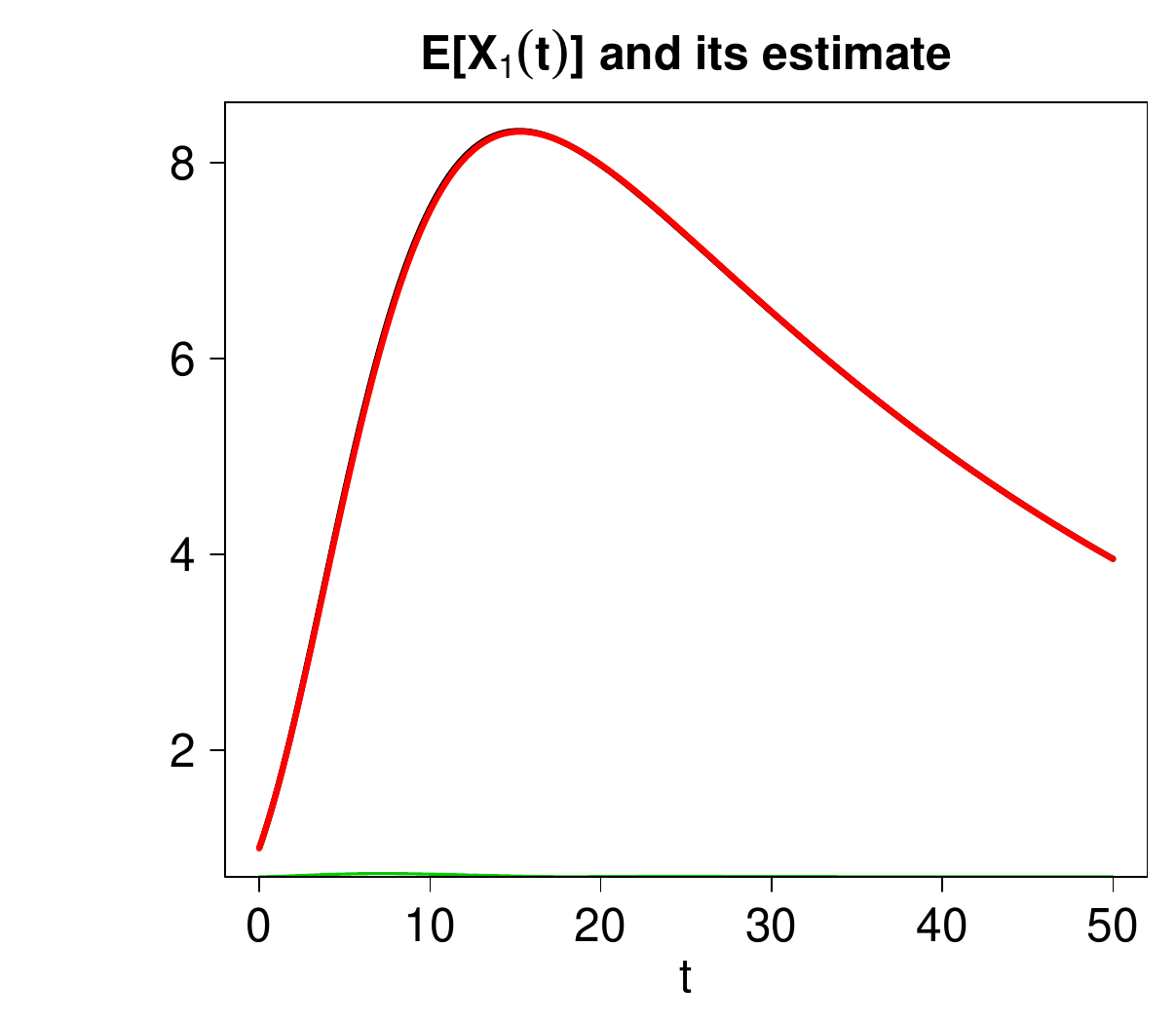} & \includegraphics[height=0.35\textheight]{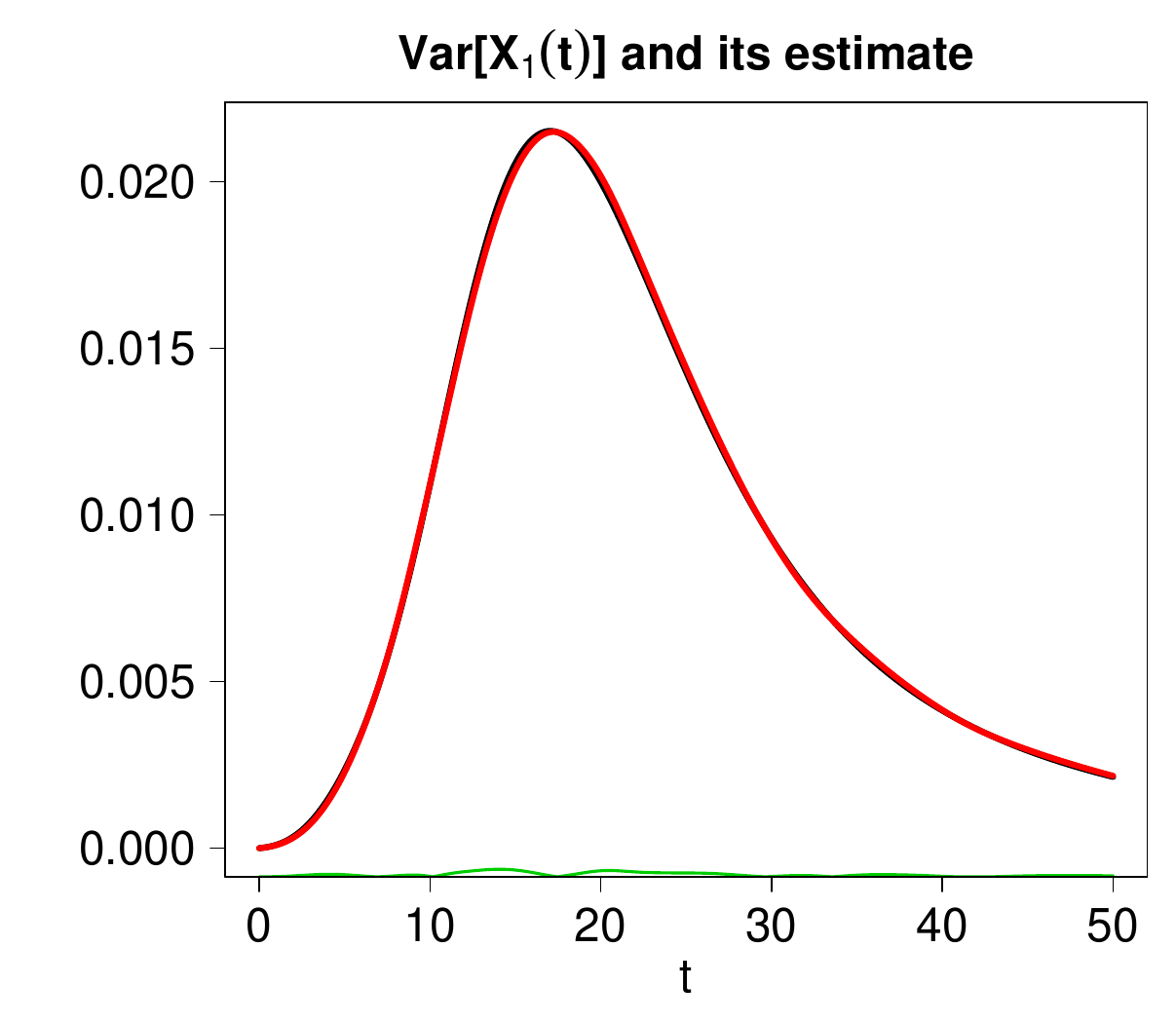} &
  \includegraphics[height=0.35\textheight]{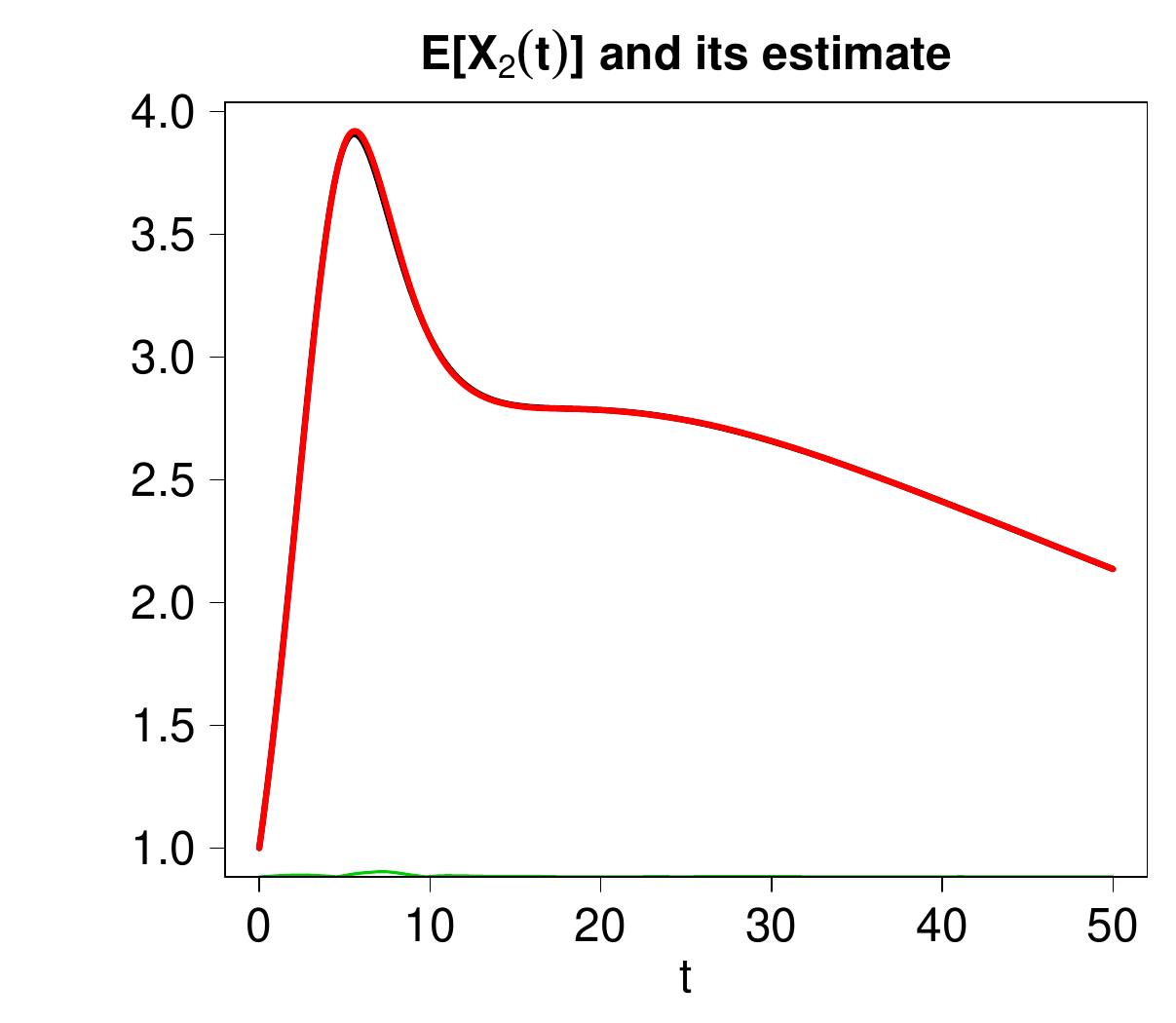} & \includegraphics[height=0.35\textheight]{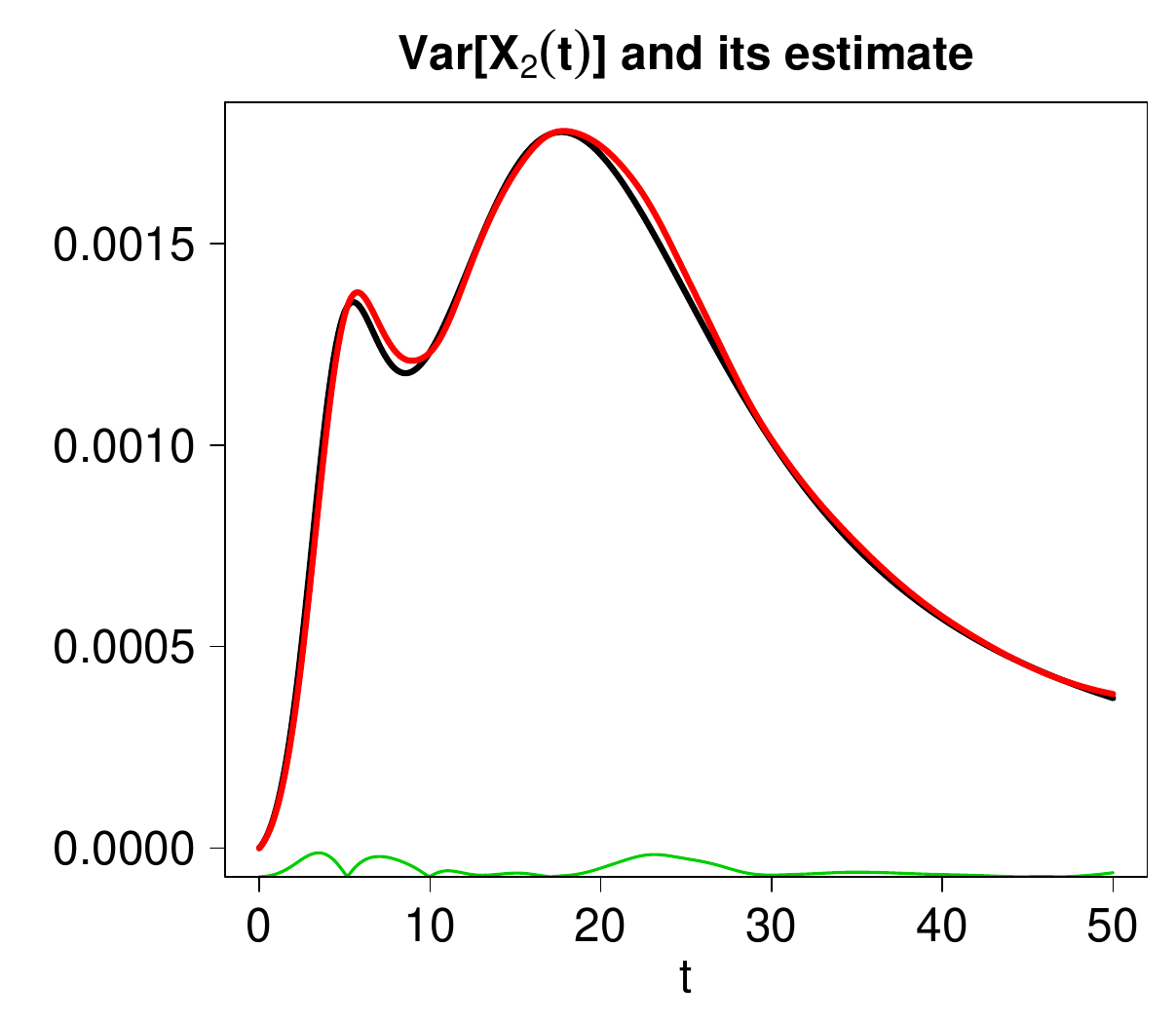} \\
\end{tabular}
\vspace{-5pt}
  \caption{Fit of the functions $C(t), V_1(t), D(t)$ and $V_2(t)$ in the models (\ref{case_a_X1}) and (\ref{case_a_b_X2}) in
  Application 1 are plotted on the top. The mean and variances of the processes $X_1(t)$ and $X_2(t)$ along with their fitted  versions (in red) are also shown on the bottom. The absolute difference functions between the simulated and fitted function are represented in green.}
  \label{Rep_FitCV_Cncte_V1ncte}
\end{figure}
\end{landscape}

\begin{landscape}
\begin{figure}[H]
\hspace{-0.35cm} \tabcolsep 0pt
\begin{tabular}{cccc}
  \includegraphics[height=0.35\textheight]{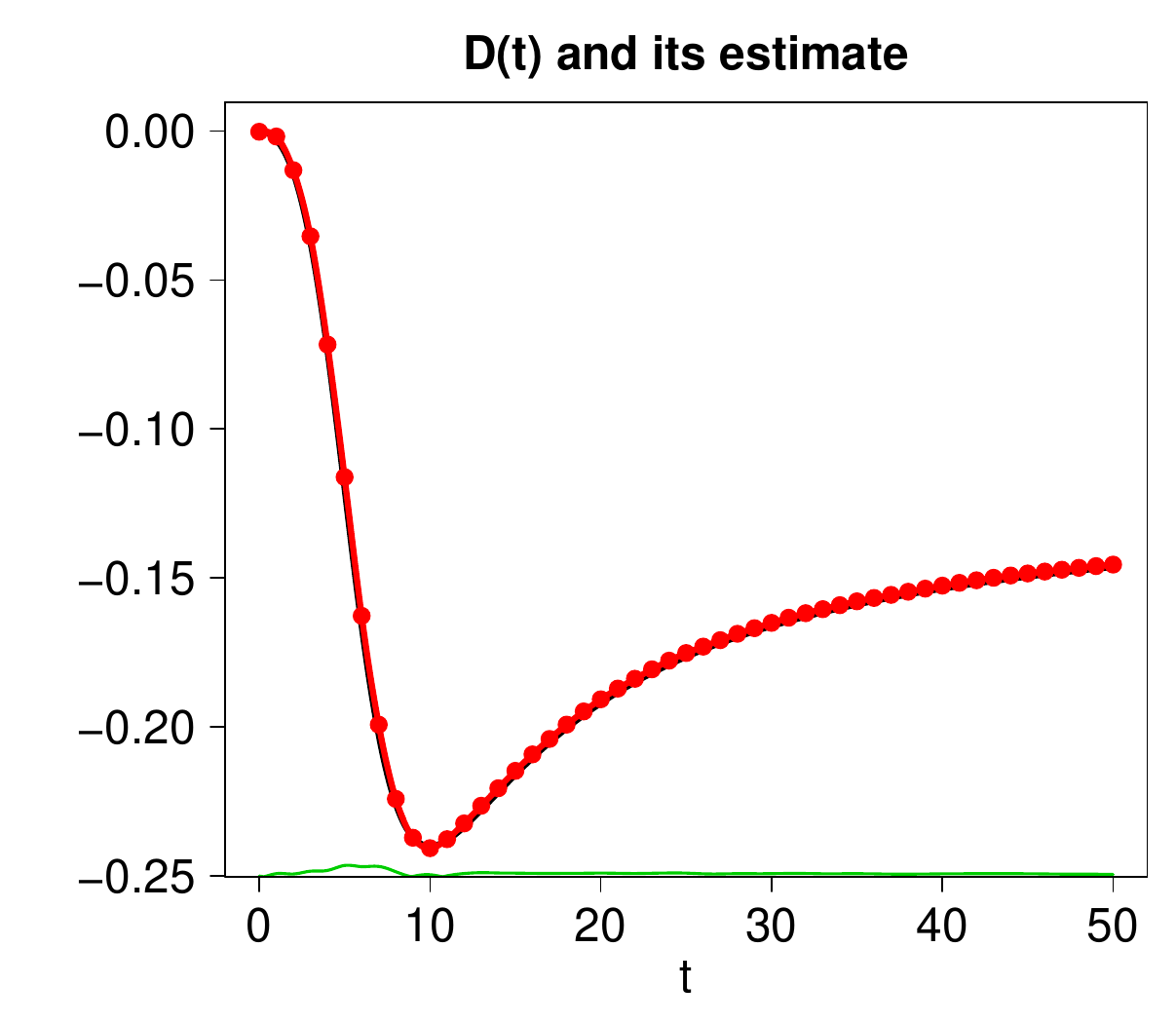} & \includegraphics[height=0.35\textheight]{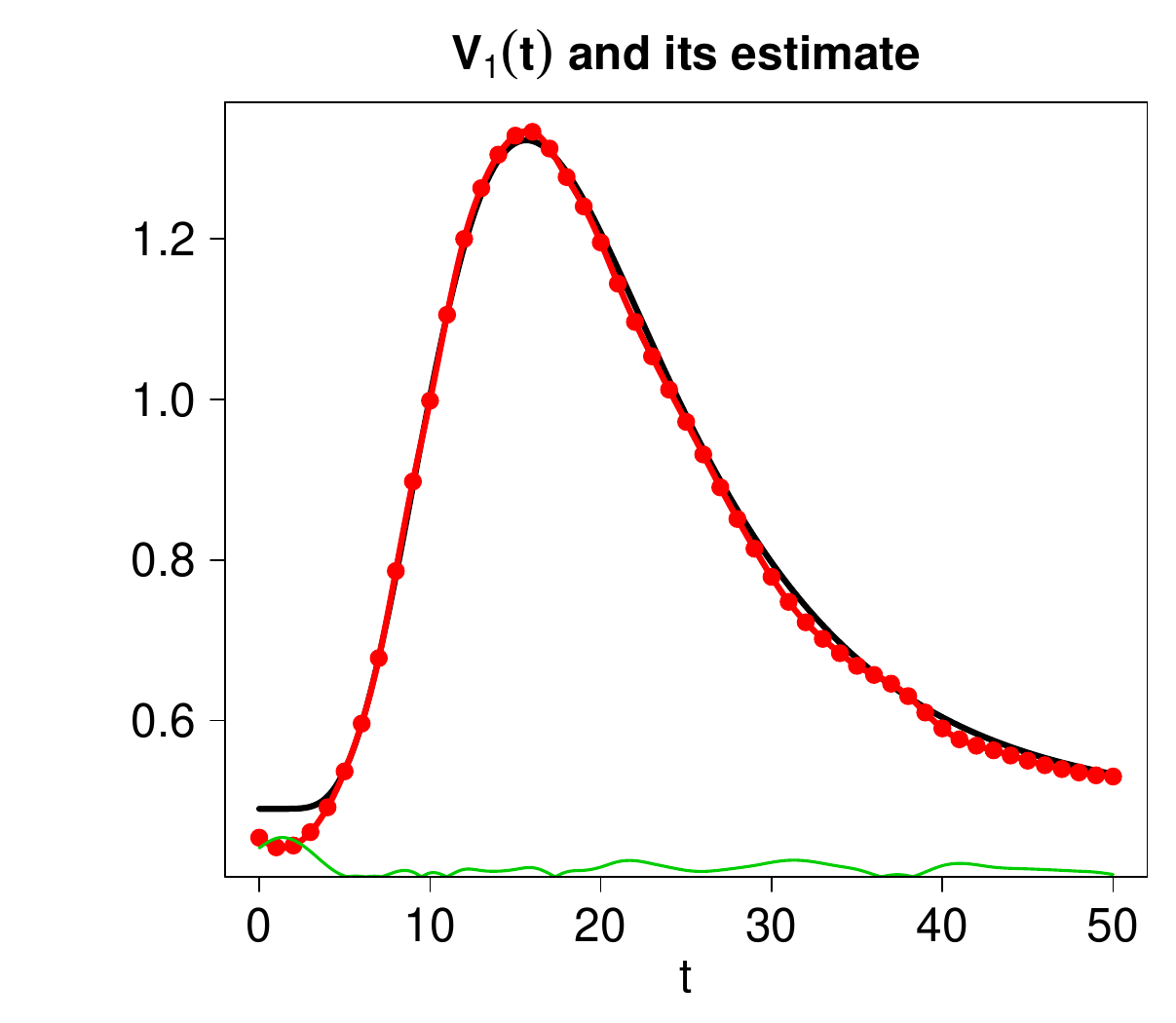} &
  \includegraphics[height=0.35\textheight]{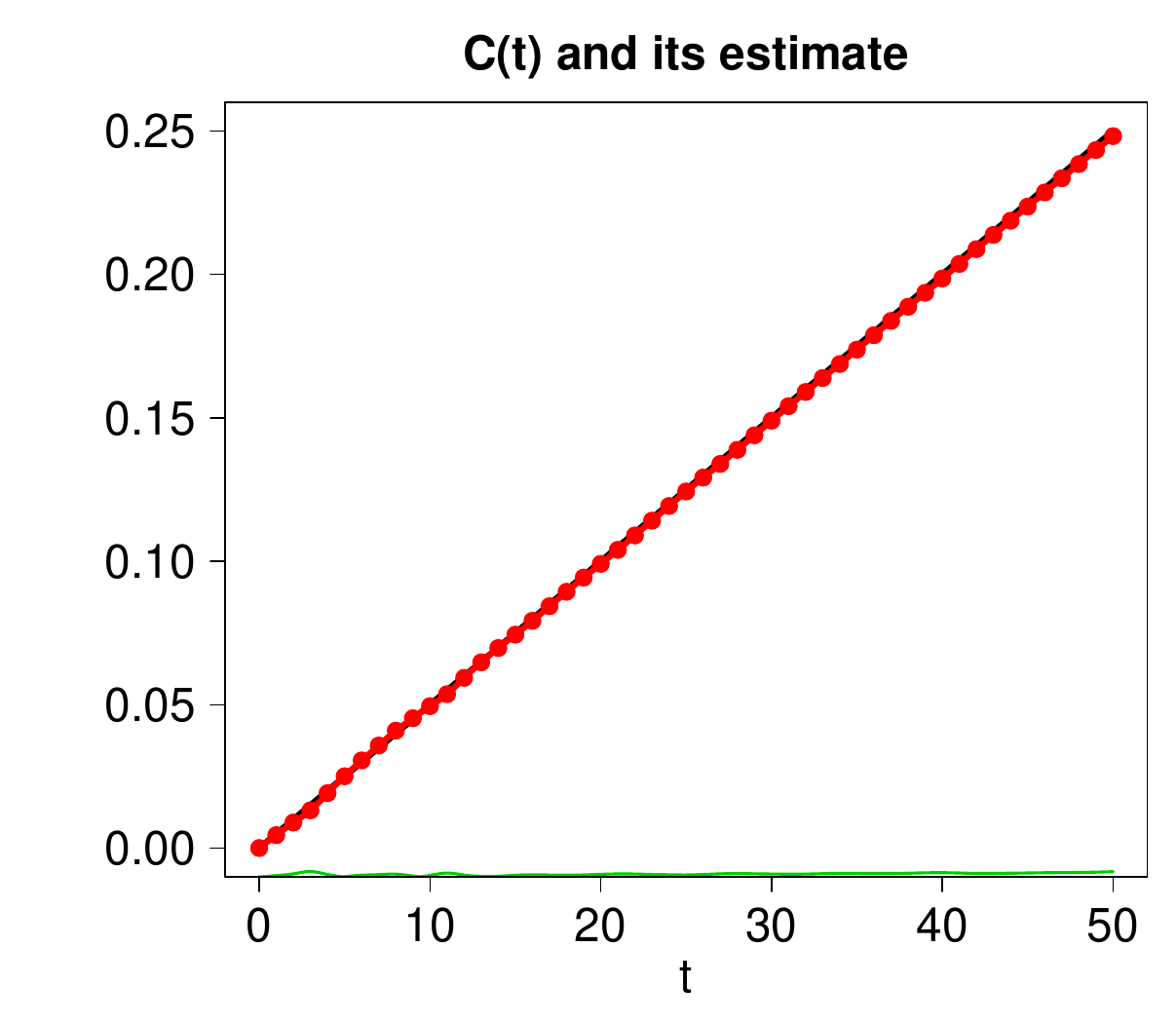} & \includegraphics[height=0.35\textheight]{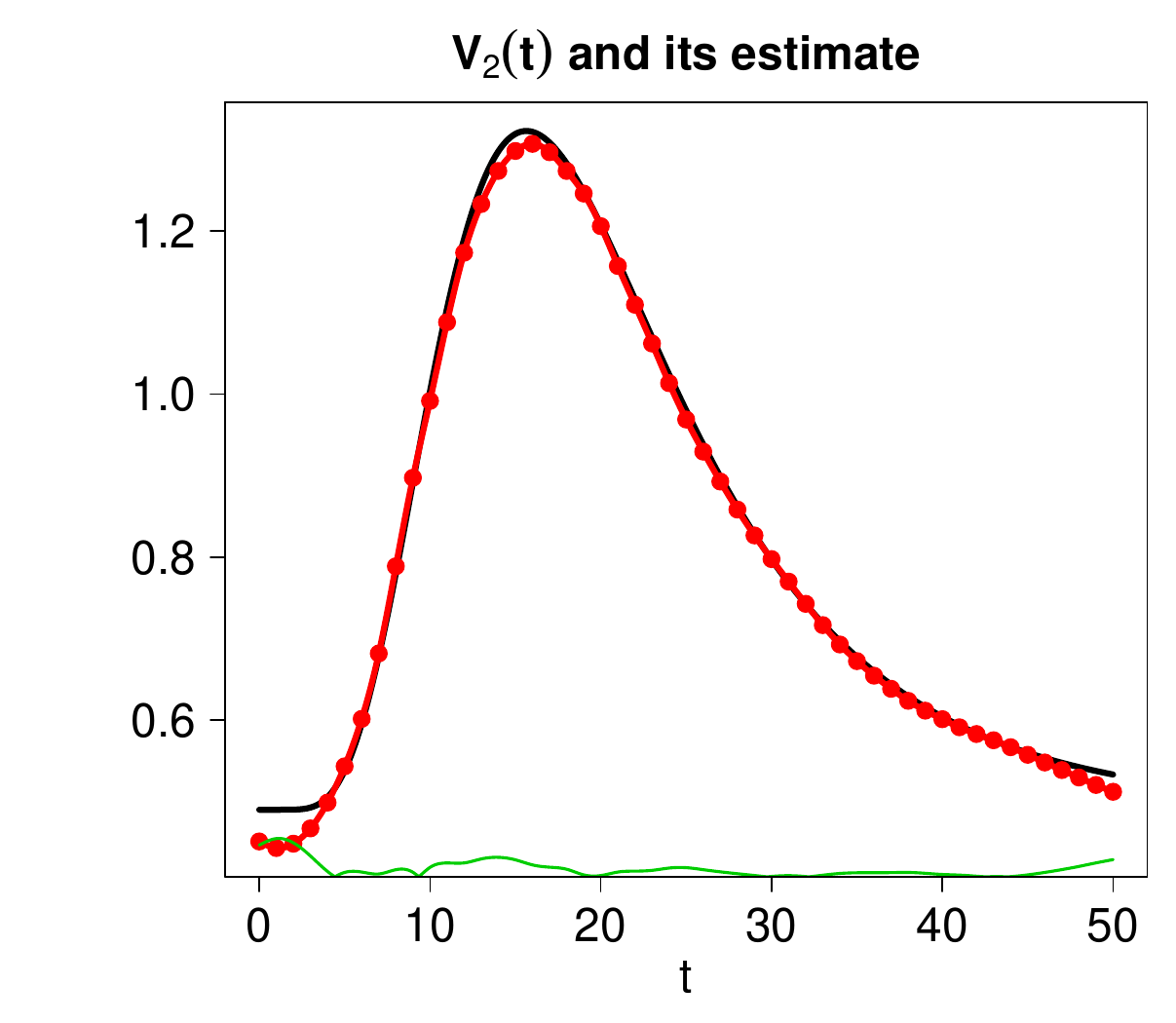} \\
  \includegraphics[height=0.35\textheight]{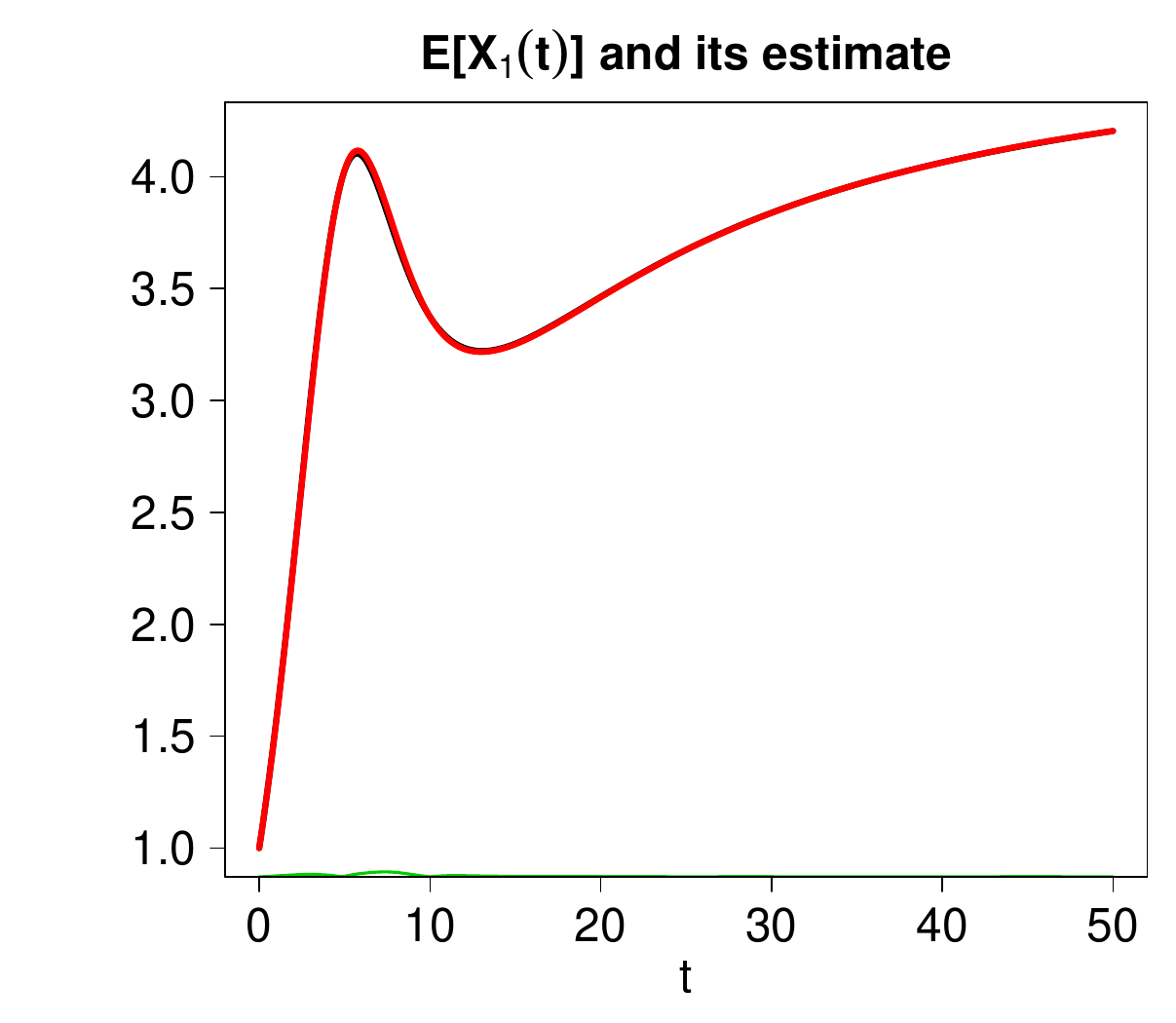} & \includegraphics[height=0.35\textheight]{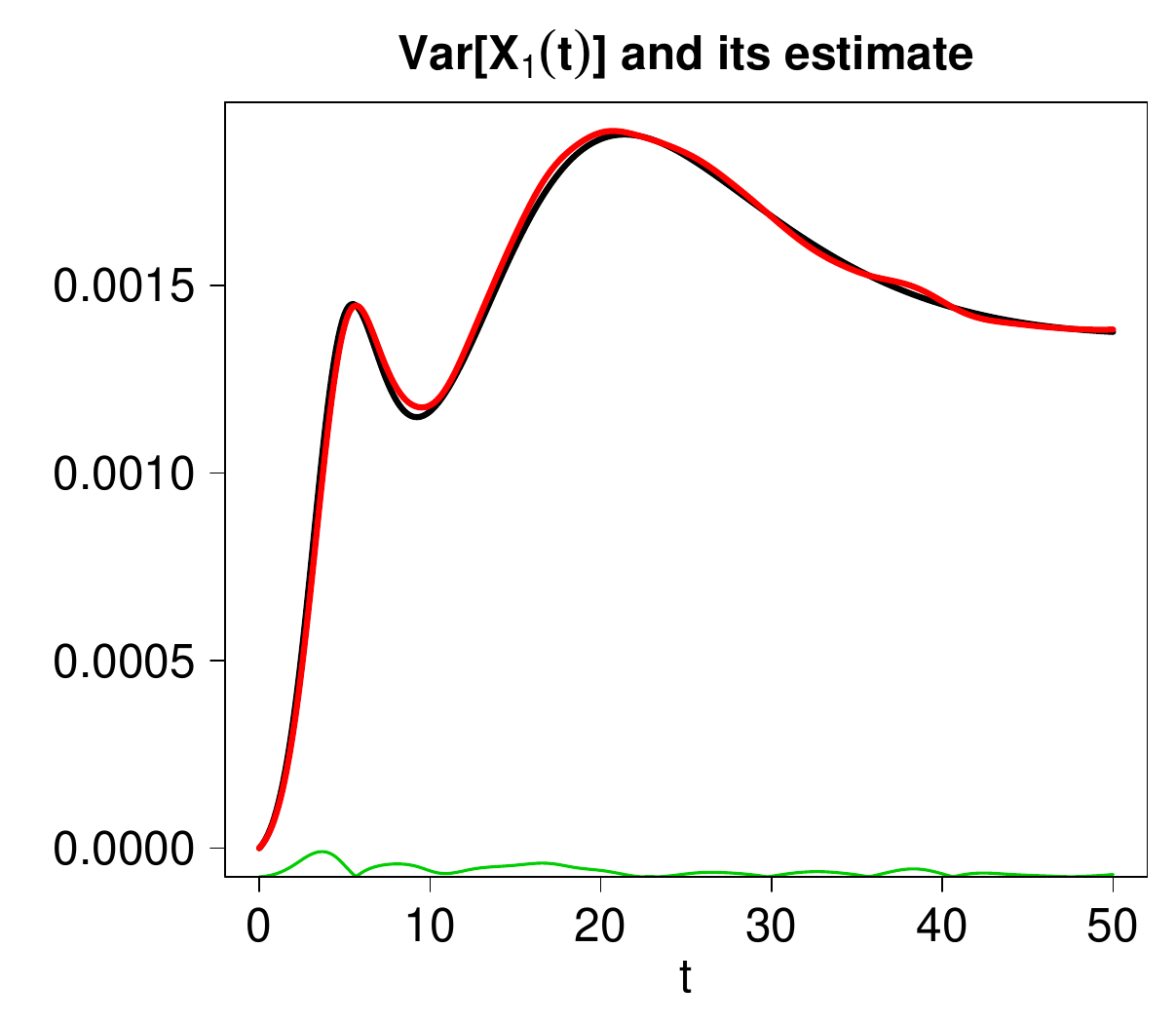} &
  \includegraphics[height=0.35\textheight]{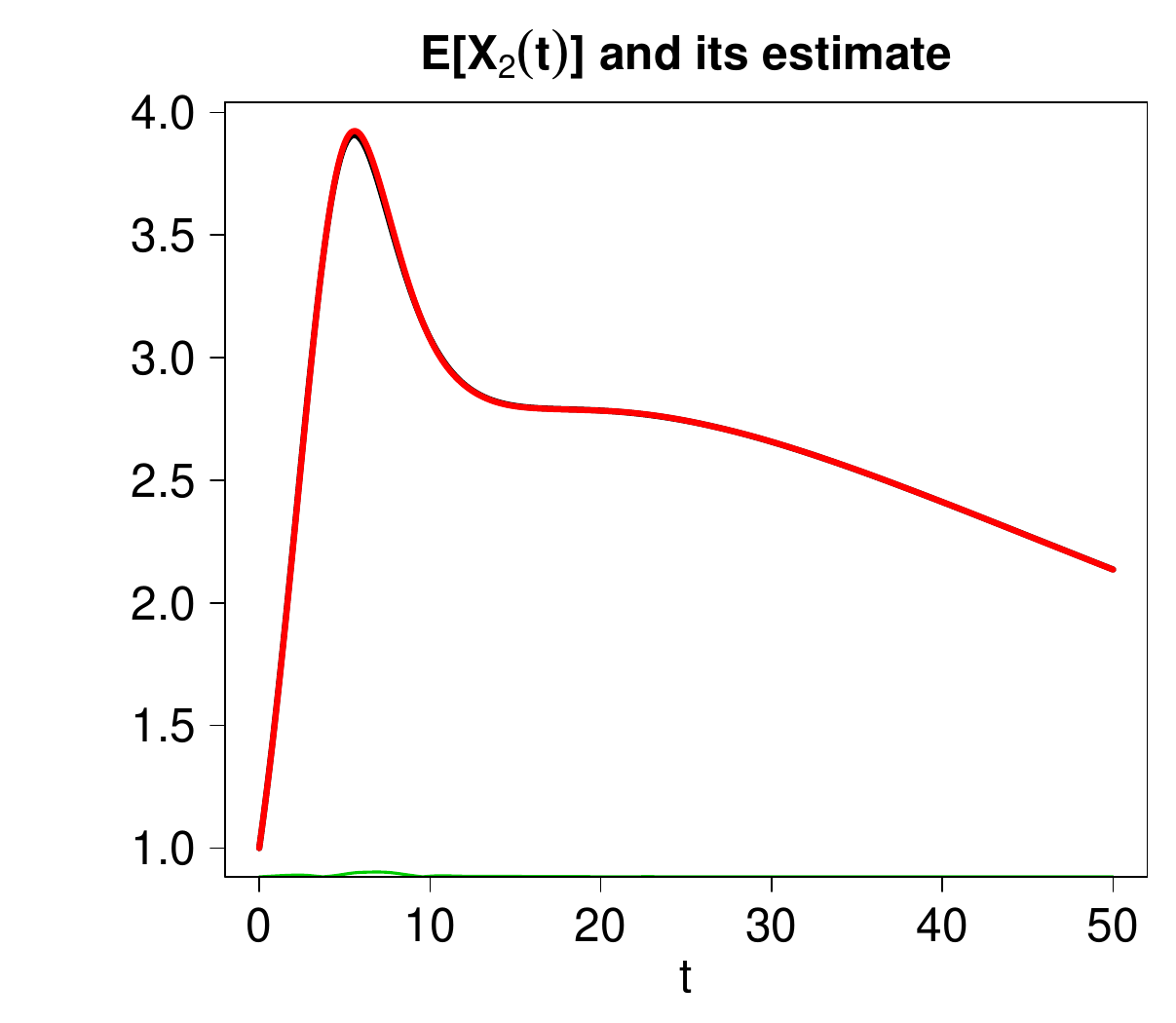} & \includegraphics[height=0.35\textheight]{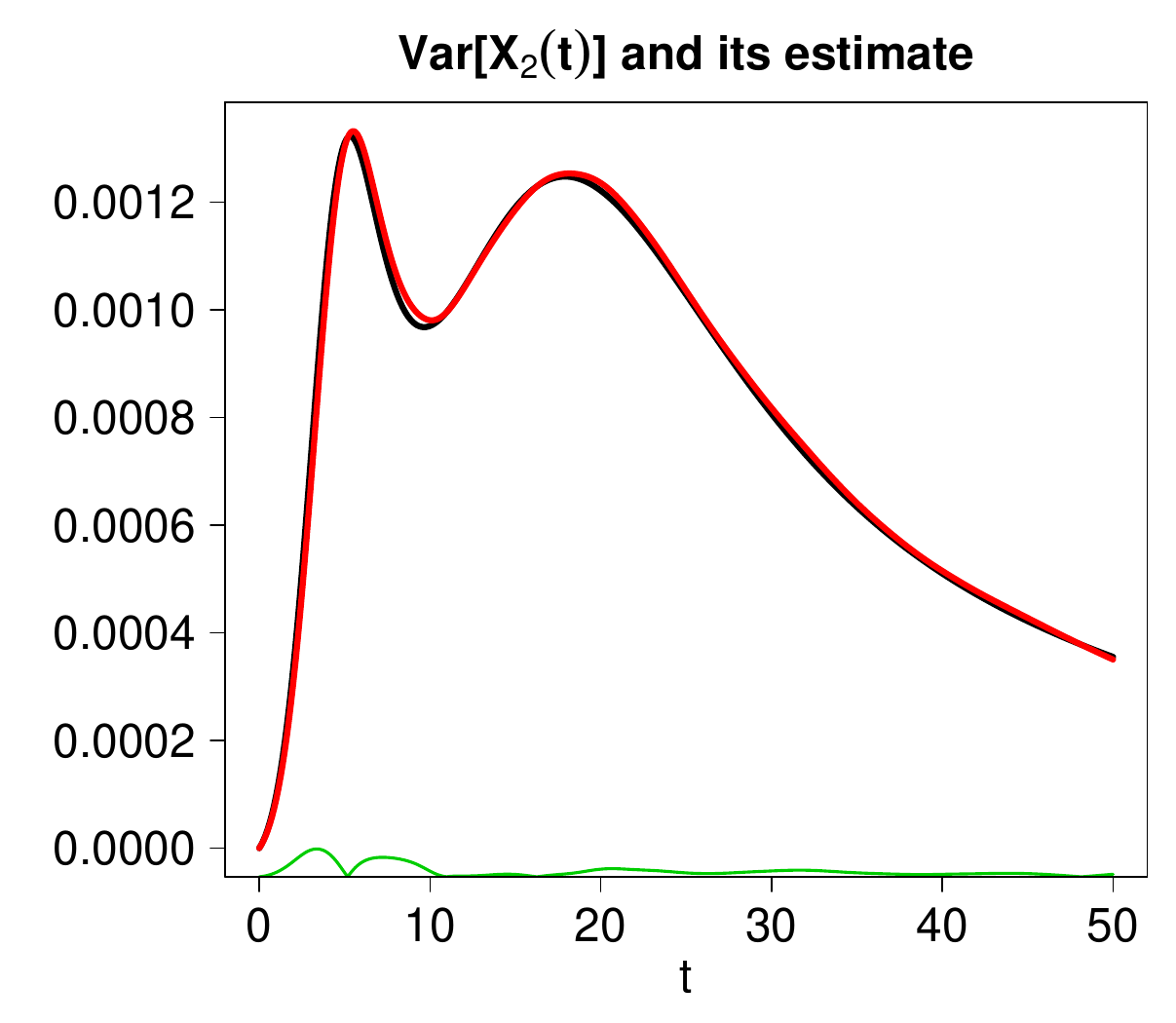} \\
\end{tabular}
\vspace{-5pt}
  \caption{Fit of the functions $D(t), V_1(t), C(t)$ and $V_2(t)$ in the models (\ref{case_b_X1}) and (\ref{case_a_b_X2}) in Application 2 are plotted on the top. The mean and variances of the processes $X_1(t)$ and $X_2(t)$ along with their
  fitted versions (in red) are also shown on the bottom. The absolute difference functions between the simulated and fitted function are represented in green.}
  \label{Rep_FitDV_Dncte_V1ncte}
\end{figure}
\end{landscape}

The estimates in the control group were $\widehat{\alpha}=0.496477$, $\widehat{\beta}=0.198469$
and $\widehat{\sigma}=0.010043$.
In Figure
\ref{Rep_FitCV_Cncte_V1ncte} the fit of the functions $C(t), V_1(t), D(t)$ and $V_2(t)$ in the models (\ref{case_a_X1}) and (\ref{case_a_b_X2}) in Application 1 are plotted on the top.
The mean and variances of the processes $X_1(t)$ and $X_2(t)$ along with their fitted versions are also shown on the bottom. The absolute difference functions between the simulated and fitted function are also represented in green.
Results related to Application 2 are shown in Figure \ref{Rep_FitDV_Dncte_V1ncte}. In both the applications, the procedure provides estimated functions (red lines) very close to the theoretical ones (black lines).
%

Moreover, in order to have a measure of goodness of fit of the obtained estimates, we have considered the
following mean squared error

$$
MSE(H)=\frac{1}{n}\sum_{j=0}^{n-1}\left(\widehat{H}(t_j)-H(t_j)\right)^2,
$$
where $H$ represents one of the $C, D, V_1, V_2$ functions. Table \ref{MSEs} includes the values
obtained for these errors, confirming the closeness between theoretical and estimated functions.

\begin{table}[h]
  \caption{Mean squared errors of estimated functions in groups ${\cal G}_1$ and ${\cal G}_2$ for Applications 1 and 2}
  \label{MSEs} \smallskip \centering
  \begin{tabular}{cccc}
  \toprule
  \multicolumn{4}{c}{Application 1} \\ \midrule
  \multicolumn{2}{c}{Group ${\cal G}_1$} & \multicolumn{2}{c}{Group ${\cal G}_2$}
  \\ \midrule
  Function & MSE & Function & MSE \\ \midrule
  $C(t)$ & $5.779772e-07$ & $D(t)$ & $2.637550e-06$  \\
  $V_1(t)$ & $3.006884e-04$ & $V_2(t)$ & $4.192476e-04$  \\
  $E(X_1(t))$ & $1.684190e-04$ & $E(X_2(t))$ & $2.448863e-05$  \\
  $Var(X_1(t))$ & $7.953060e-09$ & $Var(X_2(t))$ & $5.742807e-10$ \\
  \bottomrule \toprule
  \multicolumn{4}{c}{Application 2} \\ \midrule
  \multicolumn{2}{c}{Group ${\cal G}_1$} & \multicolumn{2}{c}{Group ${\cal G}_2$}
  \\ \midrule
  Function & MSE & Function & MSE \\ \midrule
  $D(t)$  & $1.942332e-06$ & $C(t)$ & $1.111295e-06$  \\
  $V_1(t)$ & $2.428458e-04$ & $V_2(t)$ & $2.102767e-04$  \\
  $E(X_1(t))$ & $3.193397e-05$ & $E(X_2(t))$ & $2.437206e-05$  \\
  $Var(X_1(t))$ &  $4.368352e-10$ & $Var(X_2(t))$ &  $2.315756e-10$ \\
  \bottomrule
  \end{tabular}
\end{table}

We point out that the fit functions for Applications 1 and 2 are obtained by considering as data-generating process the model $X_1(t)$ and $X_2(t)$ in (\ref{case_a_X1}) and (\ref{case_a_b_X2}) for Application 1 and Eqs (\ref{case_b_X1}) and (\ref{case_a_b_X2})  for Application 2.

In the next section we consider the problem of testing if the influence over time of a therapy follows a given functional scheme and, in particular, if its application results in a constant modification of the natural parameters of the process.

\vspace{10pt}
\section{Testing hypothesis about functions $C(t)$, $D(t)$ and $V(t)$}

In the tumor context outlined in this paper, in order to evaluate the effectiveness of
experimental therapies, the following questions are of special interest:

\begin{itemize}
  \item Is the real effect on the growth rate of an anti-proliferative therapy null?
  \item Is the real effect on the death rate of a therapy that induces the death of cancer cells null?
  \item Does the therapy, or combination of therapies, affect the infinitesimal variance?
  \item Do the effects of a therapy or combination of therapies depend on time?
  \item Do the functions that model the effect of the therapy or combination of therapies have a specific form?
\end{itemize}

Since the functions included in model (\ref{SDE}) represent different effects of a therapy, or
combination of therapies on tumor growth, to answer these questions we propose to perform
hypothesis testing about the functions $C(t)$, $D(t)$ and $V(t)$ in model (\ref{SDE}) (note that
models (\ref{case_a_X1}), (\ref{case_b_X1}) and (\ref{case_a_b_X2}) are particular cases of this).

The null hypothesis can be formulated in a unified way as
$$
H_0: H(t) = h(t),
$$
with $H(t)$ any of the functions in model (\ref{SDE}) and $h(t)$ a given function. \medskip

To test the null hypothesis we propose to use a bootstrap test (b-Test) based
on the statistic
$
D = \displaystyle\sum_{j=0}^{n-1} |\widehat{H}(t_j) - h(t_j)|.
$
Calculation of values of this statistic is based on a bootstrap procedure following the line proposed in Rom{\'a}n-Rom{\'a}n et al. \cite{JTB_2016,SII_2017}. Concretely, the schema is the following:
\begin{itemize}
\item Generate $m$ bootstrap samples of the considered model. Each bootstrap sample consists of $d_k$ sample paths (depending of the treated group ${\cal G}_k$ being considered) simulated in the same way as the one previously exposed by taking function $h(t)$ in $H_0$ and the estimates of the parameters and the rest of functions via the procedure proposed in the previous section.
\item Estimate $H(t)$ from the sample paths of each bootstrap sample, and calculate
a value $D_l$, $l=1,\ldots,m$, of the statistic $D$.
\item Calculate the $p$-value as the proportion of values $D_l$ greater than or equal to $D$.
\end{itemize}

The case $h(t)=h$ is of special interest because it means that the effect of therapy represented
by $H(t)$ does not depend on time. Even more, $C(t)=0$ means that the therapy has no
anti-proliferative effect; $D(t)=0$ signifies that the therapy does not induce the death of cancer
cells, whereas $V(t)=1$ leads to the non-influence of the therapy on the infinitesimal variance of
the process. \medskip

In such  case, the constant $h$ to be included in $H_0$ has to be chosen.
If it is not known a priori, as usual in applications, we propose to choose $h $ as the value obtained from the ML estimation of $h(t)=h$ in model (\ref{SDE}). Concretely,
\begin{itemize}
  \item Testing $C(t)$ constant. In this case,  $H_0:C(t)={c}$.
      The value of ${c}$ is obtained from the ML estimate of the growth rate in model \eqref{SDE} by solving (\ref{vero-alpha}) in Appendix taking $C(t)=0$, $D(t) = \widehat{D}(t)$ (if $D(t) \neq 0$), $V(t) = \widehat{V}(t)$ (if $V(t) \neq 1$),
      and considering $\beta = \widehat{\beta}$ and $\sigma = \widehat{\sigma}$, the ML estimates of $\beta$ and $\sigma$ for group ${\cal G}$. In this way we obtain $\widehat{\alpha-c}$, from which $c=\widehat{\alpha}-\widehat{\alpha-c}$.

  \item Testing $D(t)$ constant.
   Now, $H_0: D(t) = d$,
%
 where the value ${d}$ is obtained as in the previous case, by changing $C(t)$ with $D(t)$ and $\alpha$ with $\beta$. Note that in this case the equation (\ref{vero-beta}) in Appendix must be solved.

  \item Testing $V(t)$ constant. This case is performed considering $H_0: V(t) = {v}$,
where the constant $v$ is obtained from ${v}=\widehat{\sigma^2\,v}/\widehat{\sigma^2}$, were $\widehat{\sigma^2\,v}$ matches the ML estimate of $\sigma^2$ in the model (\ref{SDE})  by solving (\ref{vero-sigma2}) in Appendix taking $V(t) = 1$, $\alpha = \widehat{\alpha}$, $\beta = \widehat{\beta}$, $C(t) = \widehat{C}(t)$ (if $C(t) \neq 0$) and $D(t) = \widehat{D}(t)$ (if $D(t) \neq 0$).
\end{itemize}


\subsection{Simulation study}

In order to show the behavior of the proposed bootstrap tests we have performed a simulation study
considering a control group ${\cal G}$ and two treated groups
 ${\cal G}_1$ and ${\cal G}_2$, modeled by (\ref{gruppo_G}), (\ref{case_a_X1}) and (\ref{case_a_b_X2}), respectively. The present study is limited to the case in which ${\cal G}_1$ is treated with an anti-proliferative therapy. The study in the case in which ${\cal G}_1$ is treated with a therapy that induces the death of cancer cells would be carried out in a similar way. \medskip

Once the models are estimated following the corresponding procedure in Section 3.1, we test
hypotheses about all the functions included in (\ref{case_a_X1}) and (\ref{case_a_b_X2}) as
follows.

\begin{enumerate}
  \item For group ${\cal{G}}_1$, $H_0: V_1(t)=v_1$ is tested, where $v_1$ is proposed following the comments mentioned above. Then, we test $H_0: C(t) = c$,  taking into account that:
\begin{itemize}
  \item If  $H_0: V_1(t) = v_1$ is not rejected, the value of $c$ is determined considering $V_1(t) = v_1$. \vspace{0pt}

  \item If $H_0: V_1(t) = v_1$ is rejected, $c$ is determined considering $V_1(t) =
  {\widehat V}_1(t)$. \vspace{5pt}
\end{itemize}

  \item For group ${\cal{G}}_2$, $H_0: V_2(t)= v_2$ is tested. To this end, $v_2$ is determined making use of $C(t)=c$, if $H_0: C(t) = c$ \thinspace is not rejected,
  or $C(t) = \widehat{C}(t)$ otherwise, and $D(t) = \widehat{D}(t)$. Then we test $H_0: D(t) = d$, noting in this case that:
\begin{itemize}
  \item If $H_0: V_2(t) = v_2$ is not rejected,  $d$ is determined making use of $C(t) = \widehat{C}(t)$, if $H_0: C(t) = c$ is rejected, or $C(t) = c$ on
  the contrary, and $V_2(t) = v_2$. \vspace{10pt}

  \item If  $H_0: V_2(t) = v_2$ is rejected, the value of $d$ is determined considering $C(t) = \widehat{C}(t)$ , if $H_0: C(t) = c$ is rejected, or $C(t)=c$
  otherwise, and $V_2(t)=\widehat{V}_2(t)$.
\end{itemize}

\end{enumerate}

\vskip 10pt In the simulation study we have considered models  (\ref{gruppo_G}), (\ref{case_a_X1})
and (\ref{case_a_b_X2}) with $\alpha=0.5$, $\beta=0.2$ and $\sigma=0.01$ for all possible
combinations of the functions in Table \ref{fun}. Precisely, we consider 3 choices for functions
$C(t)$, $D(t)$ and $V_1(t)$ and four cases for $V_2(t)$, obtaining 9 cases for group ${\cal G}_1$
and 108 for group ${\cal G}_2$. These functions have been selected in order to simulate the tumor
growth in the groups treated with therapies of diverse effects, ranging from therapies that do not
produce any improvement, until therapies that produce a significant reduction both in the mean
relative volume of the tumor and in its variability.
\medskip

For each model, $25$ sample paths have been simulated over $51$ equally time instants in $[0,50]$.
The estimates obtained for model (\ref{gruppo_G}), from the data of the control group ${\cal G}$,
were $\widehat{\alpha} = 0.4972273$, $\widehat{\beta} = 0.1987757$ and $\widehat{\sigma} =
0.0100692$. Further, in each case for groups ${\cal G}_1$ and ${\cal G}_2$, the number of
bootstrap samples used to perform each b-Test was $m=1500$. \smallskip

The complete simulation study is presented in schematic form in Supplementary Material.
In each case, for groups ${\cal G}_1$ and ${\cal G}_2$, we show: 
\begin{itemize}
  \item the sample paths simulated for each model,
  \item the estimates of the functions included in each model,
  \item the results of the hypothesis tests listed above (concatenated b-Tests)
  \item the sample mean and variance functions,
  \item the theoretical mean and variance functions, together with their estimated versions, before and after the concatenated b-Tests are performed.
\end{itemize}

\begin{table}[h]
  \caption{Different values of functions $C(t)$, $V_1(t)$,
  $D(t)$ and $V_2(t)$ for the simulation study, being $\Lambda_1(t, \mu, \sigma^2)$ the density
  function of a lognormal distribution $\Lambda_1(\mu, \sigma^2).$} \label{fun} \smallskip
  \centering
  \begin{tabular}{cccc}
    \toprule $\mathbf{C(t)}$ & $\mathbf{V_1(t)}$ & $\mathbf{D(t)}$ & $\mathbf{V_2(t)}$ \\ \midrule
    $0$ & $1$ & $0$ & $1$ \smallskip \\
    $0.025$ & $0.49$ & $-0.05$ & $0.49$ \smallskip \\
    $0.005 \thinspace t$ & $(0.7 \negthinspace + \negthinspace 10 \thinspace \Lambda_1(t, 3, 0.5))^2$ & $\dfrac{-0.12 \thinspace t^2}{50 +
    t(t-10)}$ & $(0.7 \negthinspace + \negthinspace 10 \thinspace \Lambda_1(t, 3, 0.5))^2$ \smallskip \\ & & & $(0.7 \negthinspace + \negthinspace 15 \thinspace
    \Lambda_1(t, 3, 0.5))^2$ \\ \bottomrule
  \end{tabular}

\end{table}

In this section we focus on two cases that have a specific meaning. These two cases show how the
proposed tests allow obtaining a better estimation of the mean functions and variances of the
simulated processes.

\begin{itemize}
  \item \textbf{Case 1}. $C(t)=0$, $V_1(t)=1$, $D(t)=0$ and $V_2(t)=1$.

  In this case, in the group ${\cal G}_1$, the anti-proliferative effect of the first therapy is null, and this therapy also does not affect the infinitesimal variance of process $X_1(t)$. Moreover, in group ${\cal G}_2$, the effect of therapy that induces the death of cancer cells is null, and the combined effect of the two therapies  does not affect the infinitesimal variance of the process $X_2(t)$. \smallskip

  Figure \ref{Paths_X1_V1eq1_Ceq0_X2_Ceq0_V2eq1_Deq0} shows the simulated sample paths of models (\ref{gruppo_G}), (\ref{case_a_X1}) and (\ref{case_a_b_X2}).

  \begin{figure}[h]
    \centering
    \includegraphics[height=0.2\textheight]{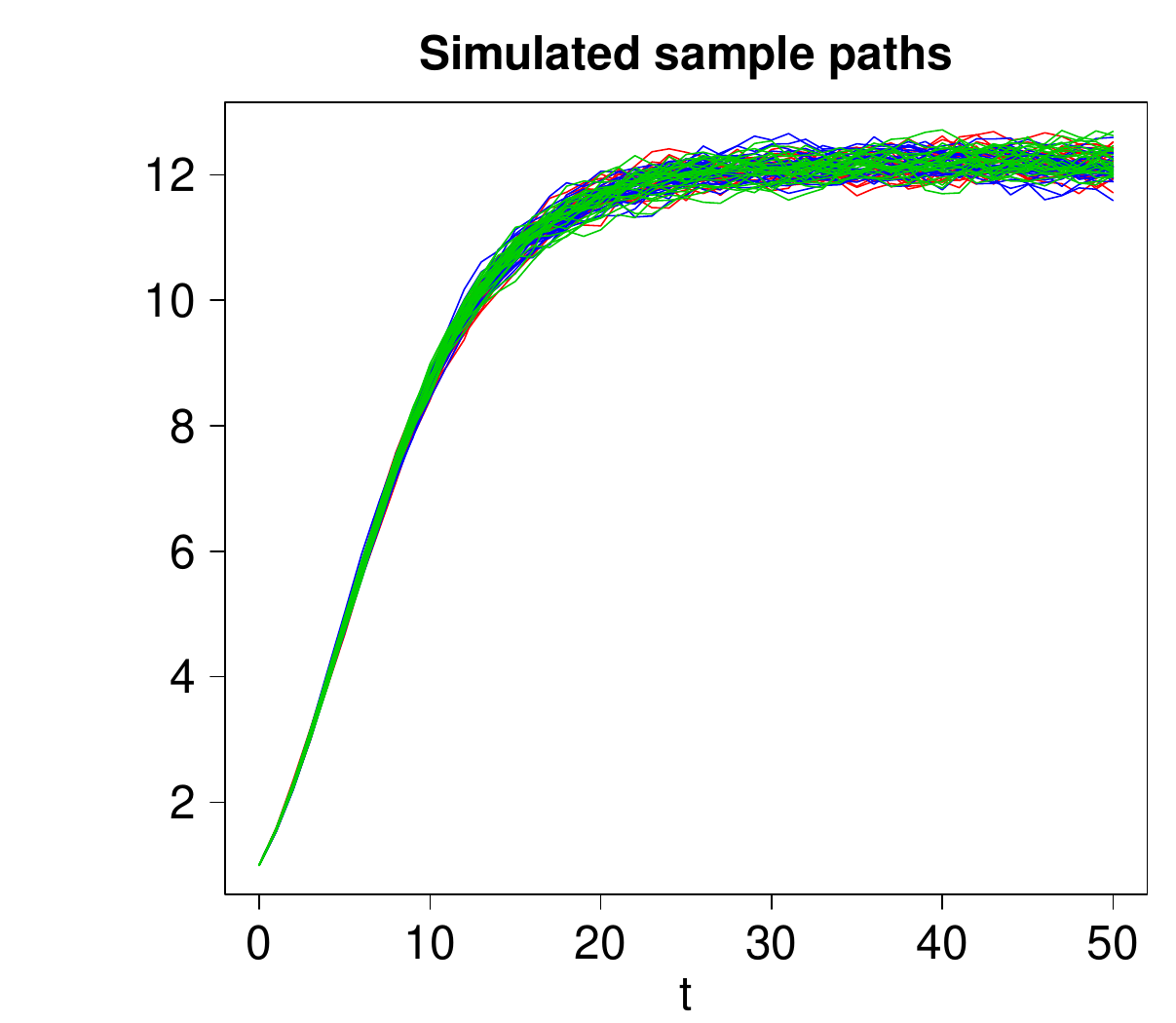}
    \caption{Simulated sample paths of models (\ref{gruppo_G}), (\ref{case_a_X1}) and (\ref{case_a_b_X2}), in red, blue and
    green, respectively, with $\alpha=0.5$, $\beta=0.2$, $\sigma=0.01$, $C(t)=0$, $V_1(t)=1$, $D(t)=0$ and $V_2(t)=1$.}
    \label{Paths_X1_V1eq1_Ceq0_X2_Ceq0_V2eq1_Deq0}
  \end{figure}

   Model (\ref{case_a_X1}) has been adjusted from the data of treated group  ${\cal G}_1$
   with $\alpha = \widehat{\alpha}$, $\beta = \widehat{\beta}$ and $\sigma = \widehat{\sigma}$, obtaining the estimates $\widehat{C}(t)$ and $\widehat{V}_1(t)$. These estimates are shown in Figure \ref{Fit_V1eq1_Ceq0} as well as the estimated mean and variance
   functions of the process $X_1(t)$. Specifically, in Figures \ref{Fit_V1eq1_Ceq0}(a) and \ref{Fit_V1eq1_Ceq0}(b), the red points correspond to the estimated values $\widehat{C}_i$ and $\widehat{V}_{1,i}$, the red solid lines represent the estimated functions $\widehat{C}(t)$ and $\widehat{V}_1(t)$, while the black solid lines indicate the theoretical functions. In Figures \ref{Fit_V1eq1_Ceq0}(c) and \ref{Fit_V1eq1_Ceq0}(d), the red points correspond to the sample mean and variance functions, respectively;
   the red solid lines represent the estimated mean and variance functions of the process $X_1(t)$ whereas the black solid lines indicate the
   theoretical mean and variance functions of process $X_1(t)$. In all the cases, the absolute difference functions between the simulated and fitted function are also represented in green.

  \begin{figure}[H]
    \centering
    \begin{tabular}{cc}
    \includegraphics[height=0.2\textheight]{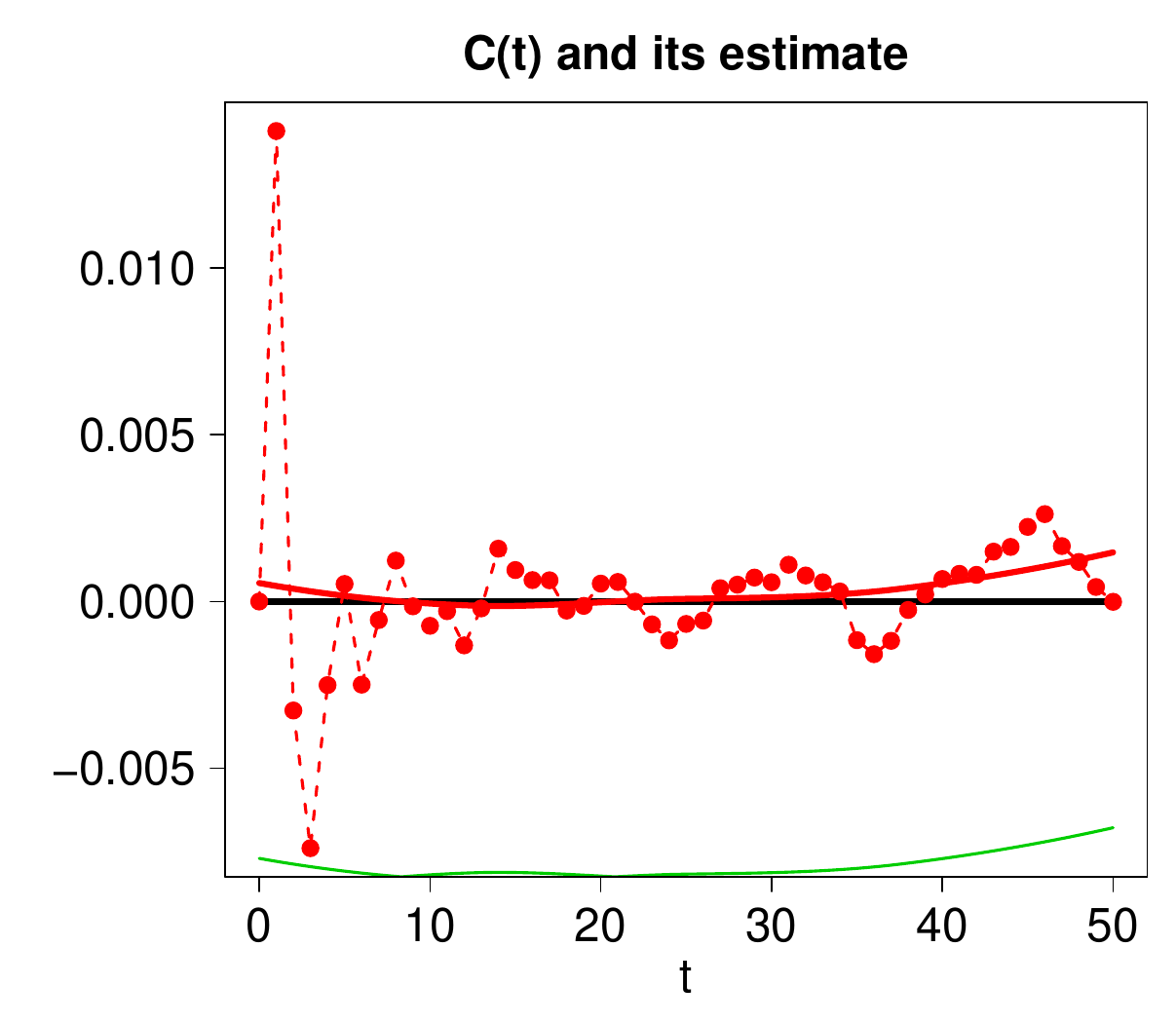} &
    \includegraphics[height=0.2\textheight]{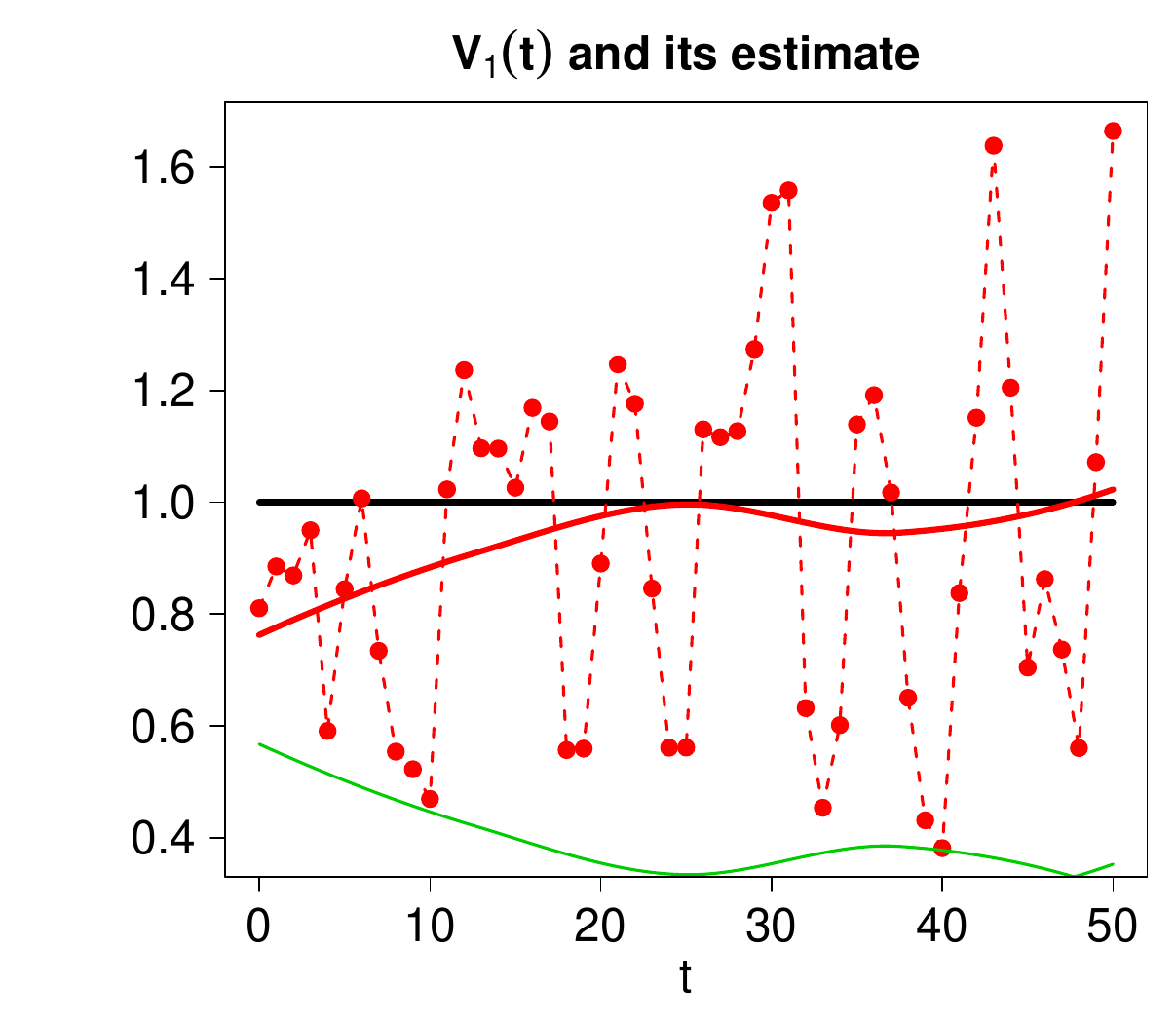} \\ \qquad \ {\scriptsize (a)} & \qquad \ {\scriptsize (b)} \\
    \includegraphics[height=0.2\textheight]{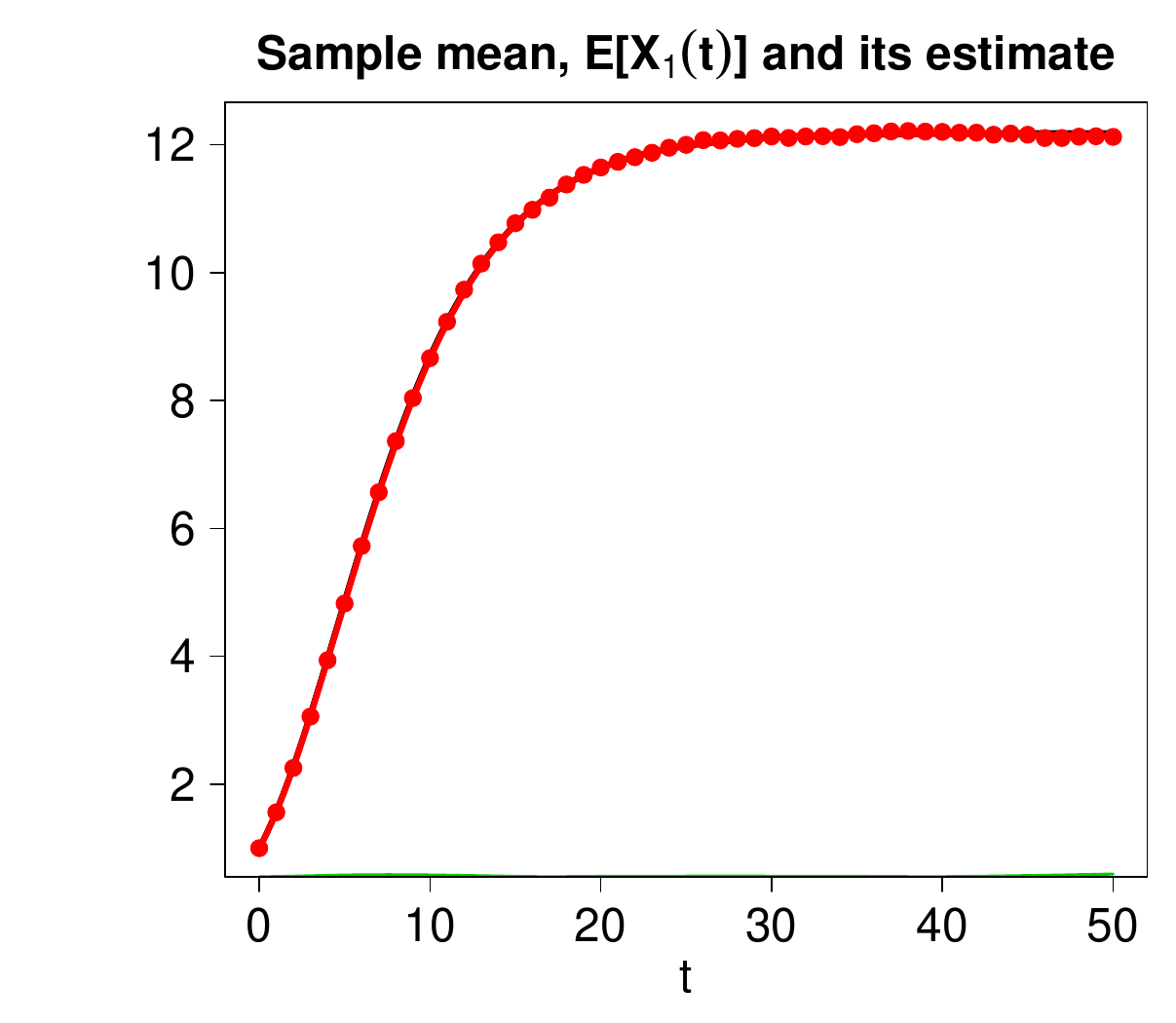} &
    \includegraphics[height=0.2\textheight]{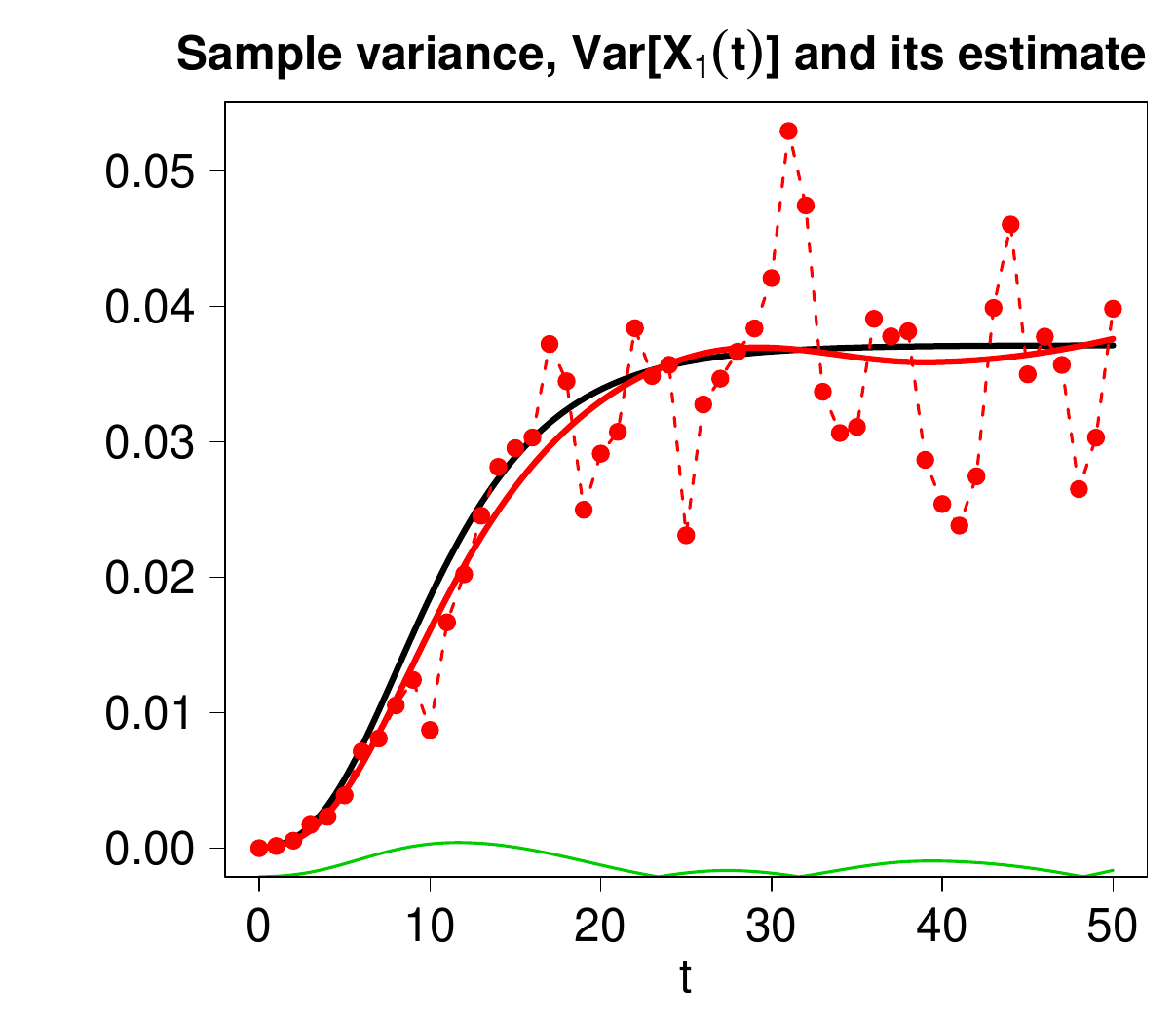} \\ \qquad \ {\scriptsize (c)} & \qquad \ {\scriptsize (d)}
    \end{tabular}
    \caption{Fit of simulated data of model (\ref{case_a_X1}) with $\alpha=0.5$, $\beta=0.2$, $\sigma=0.01$, $C(t)=0$ and $V_1(t)=1$. The absolute difference functions between the simulated and fitted function are represented in green.}
    \label{Fit_V1eq1_Ceq0}
  \end{figure}

  Figure \ref{Fit_V1eq1_Ceq0} seems to indicate that the estimated values $\widehat{C}_i$ and $\widehat{V}_{1,i}$ vary around values close to $0$ and $1$, respectively, and it makes sense to test if the functions $C(t)$ and $V_1(t)$ are constant. \smallskip

 First, we test if $V_1(t)$ is constant. The constant is chosen via the ML estimation as previously indicated. In this case,  $v_1 = 0.9710093$. The value of the $D$-statistics is $D = 2.5203844$ and the associated $p$-value is $0.974$, so there is no evidence to reject that the effect of the anti-proliferative therapy on the infinitesimal variance of the process that models tumor growth does not depend on time.
  Moreover, $v_1 \approx 1 $ suggests that the anti-proliferative therapy hardly affects such infinitesimal variance. \smallskip

Then, under the assumption $V_1(t) = 0.9710093$, we test if $C(t)$ is constant, i.e. $H_0: C(t) = c$ where $c = 0.0002401$ has been determined as described before. The value of the $D$-statistics is $D = 0.0158729$ and the associated $p$-value is $0.755$, so there is no evidence to reject that the effect of the anti-proliferative therapy on the rate of growth does not
  depend on time. In fact, since $c \approx 0$, we can conclude that the supposed anti-proliferative effect of the therapy on the growth rate has been almost null. \smallskip

  Figure \ref{Fit_V1eq1_Ceq0_PostTest}, similarly to Figures \ref{Fit_V1eq1_Ceq0}(c) and \ref{Fit_V1eq1_Ceq0}(d), shows the estimated mean and variance functions
  in the group ${\cal G}_1$ with $\widehat{\alpha}$, $\widehat{\beta}$, $\widehat{\sigma}$,
  $\widehat{V}_1(t) = 0.9710093$ and $\widehat{C}(t) = 0.0002401$, together with the sample and theoretical mean and variance functions of process $X_1(t)$.
  Observe how now the estimated mean and variance functions better reproduce the theoretical ones.

  \begin{figure}[h]
    \centering
    \begin{tabular}{cc}
    \includegraphics[height=0.2\textheight]{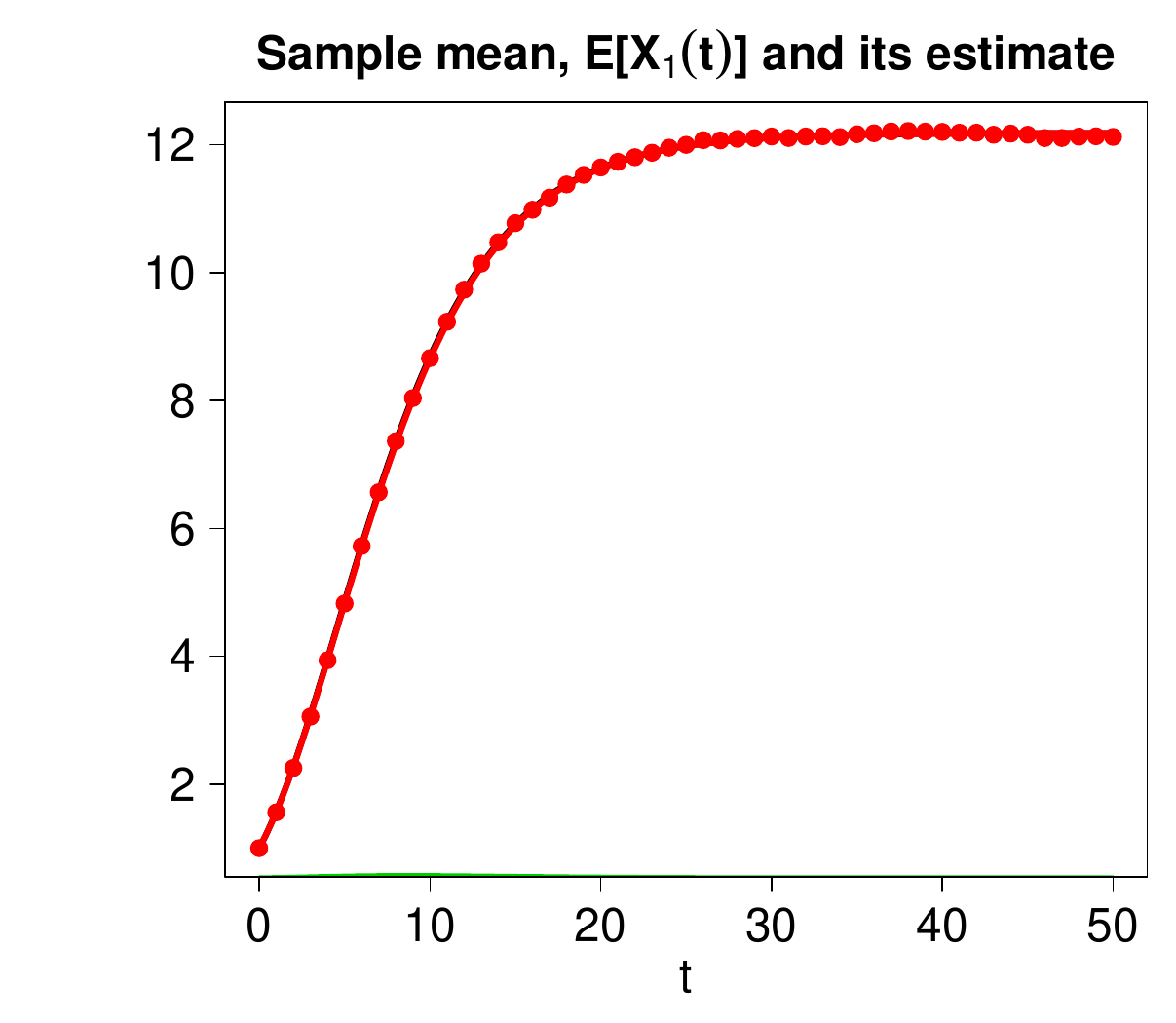} &
    \includegraphics[height=0.2\textheight]{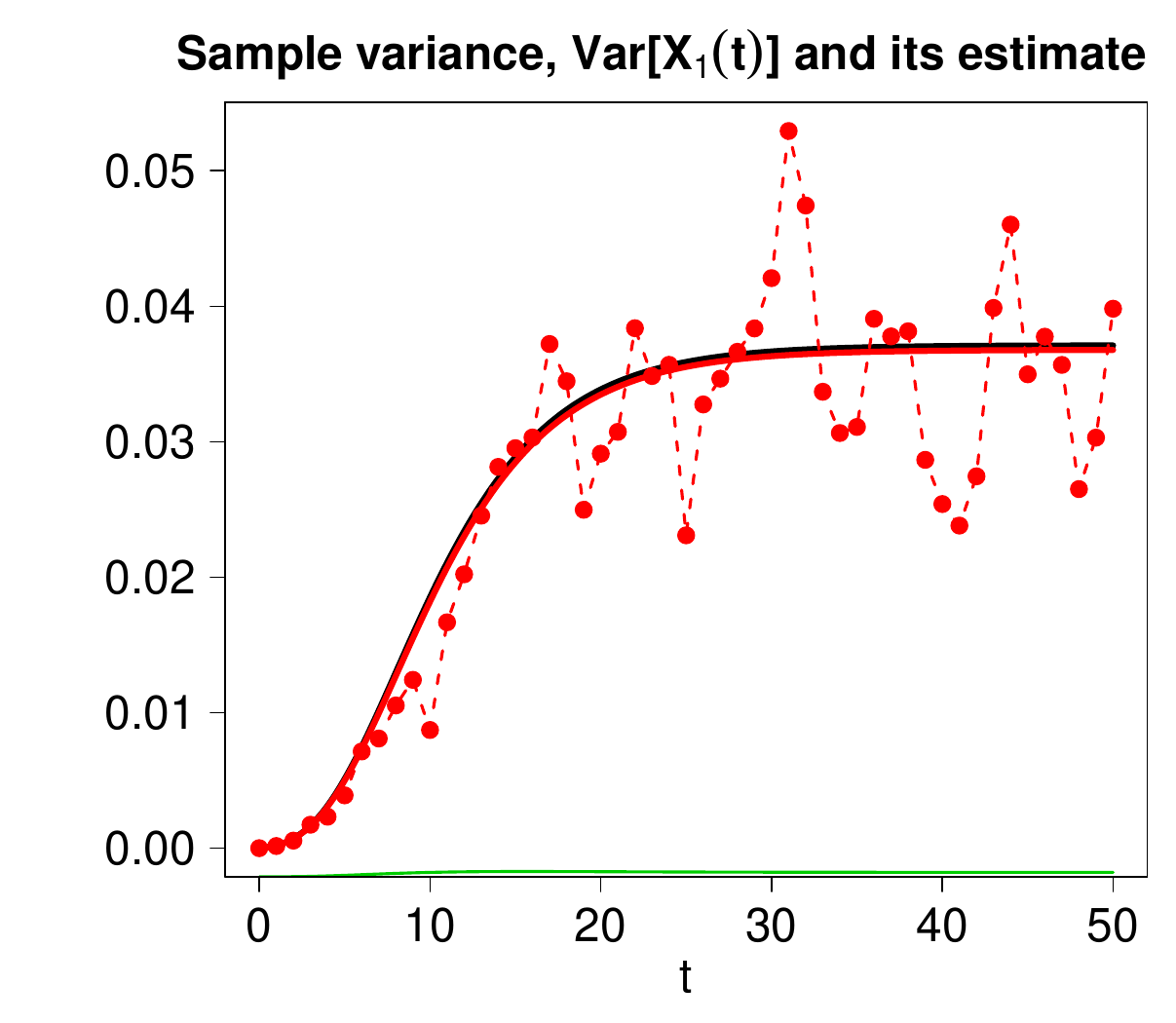}
    \end{tabular}
    \caption{Fit of simulated data of model (\ref{case_a_X1}) with $\alpha=0.5$, $\beta=0.2$, $\sigma=0.01$, $C(t)=0$ and $V_1(t)=1$,
    assuming that it is accepted first that $V_1(t)$ is constant, and then, that $C(t)$ is constant. The absolute difference functions
    between the simulated and fitted function are represented in green.} \label{Fit_V1eq1_Ceq0_PostTest}
  \end{figure}

  Next, from data of the treated group ${\cal G} _2$, model (\ref{case_a_b_X2}) have been adjusted by using $\widehat{\alpha}$,
  $\widehat{\beta}$, $\widehat {\sigma}$ and $\widehat{C}(t) = 0.0002401$. As in Figure \ref{Fit_V1eq1_Ceq0}, the estimated functions $\widehat{D}(t)$
  and $\widehat{V}_2(t)$ in Figure \ref{Fit_V1eq1_Ceq0_V2eq1_Deq0} are plotted as well as the estimated mean and
  variance functions of the process $X_2(t)$, showing how the values $\widehat{D}_i$ and $\widehat{V}_{2,i}$ are close to $0$ and $1$,
  respectively, which leads us to test whether the functions $D(t)$ and $V_2(t)$ are constant.

\begin{figure}[H]
    \centering
    \begin{tabular}{cc}
    \includegraphics[height=0.2\textheight]{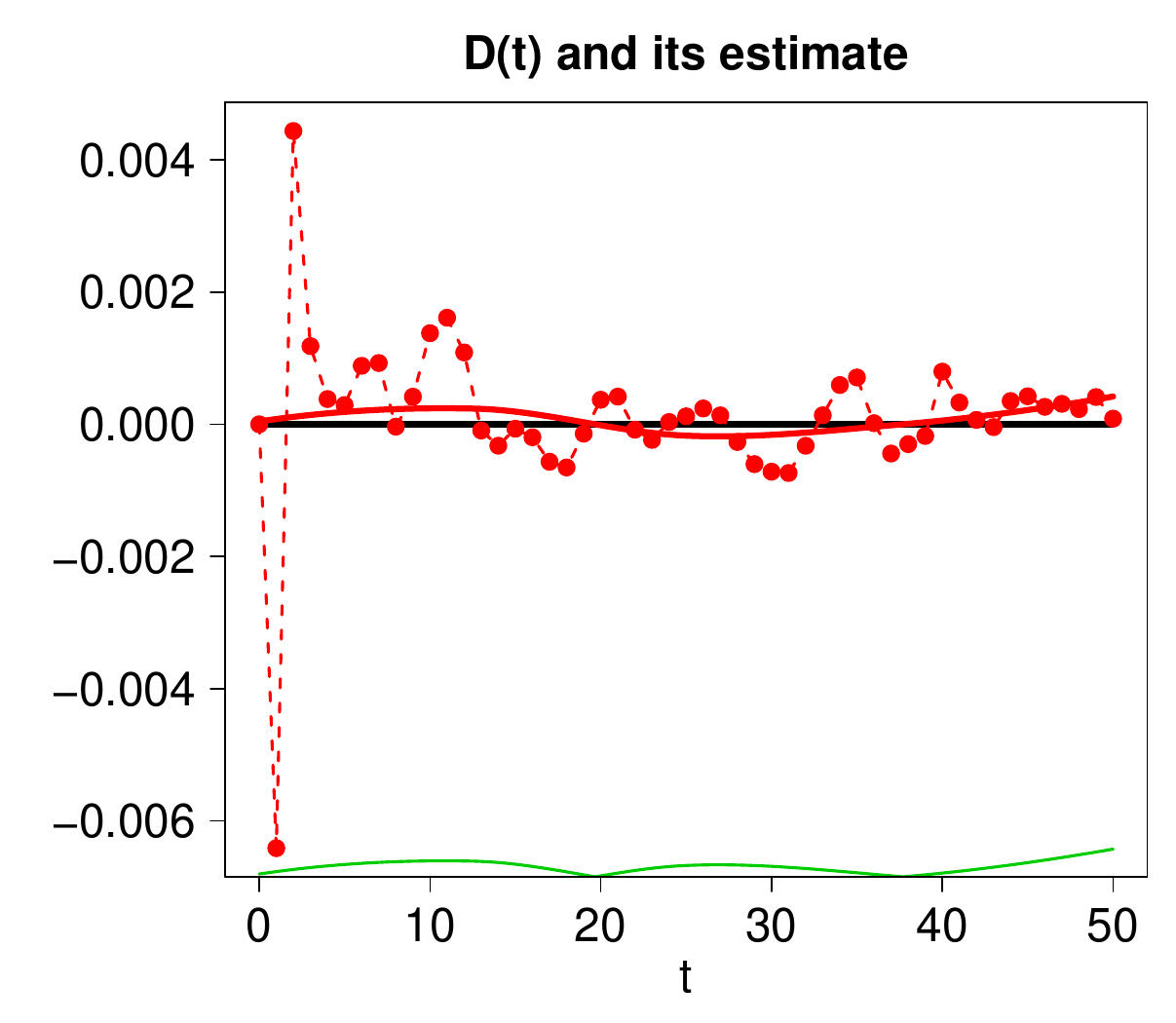} &
    \includegraphics[height=0.2\textheight]{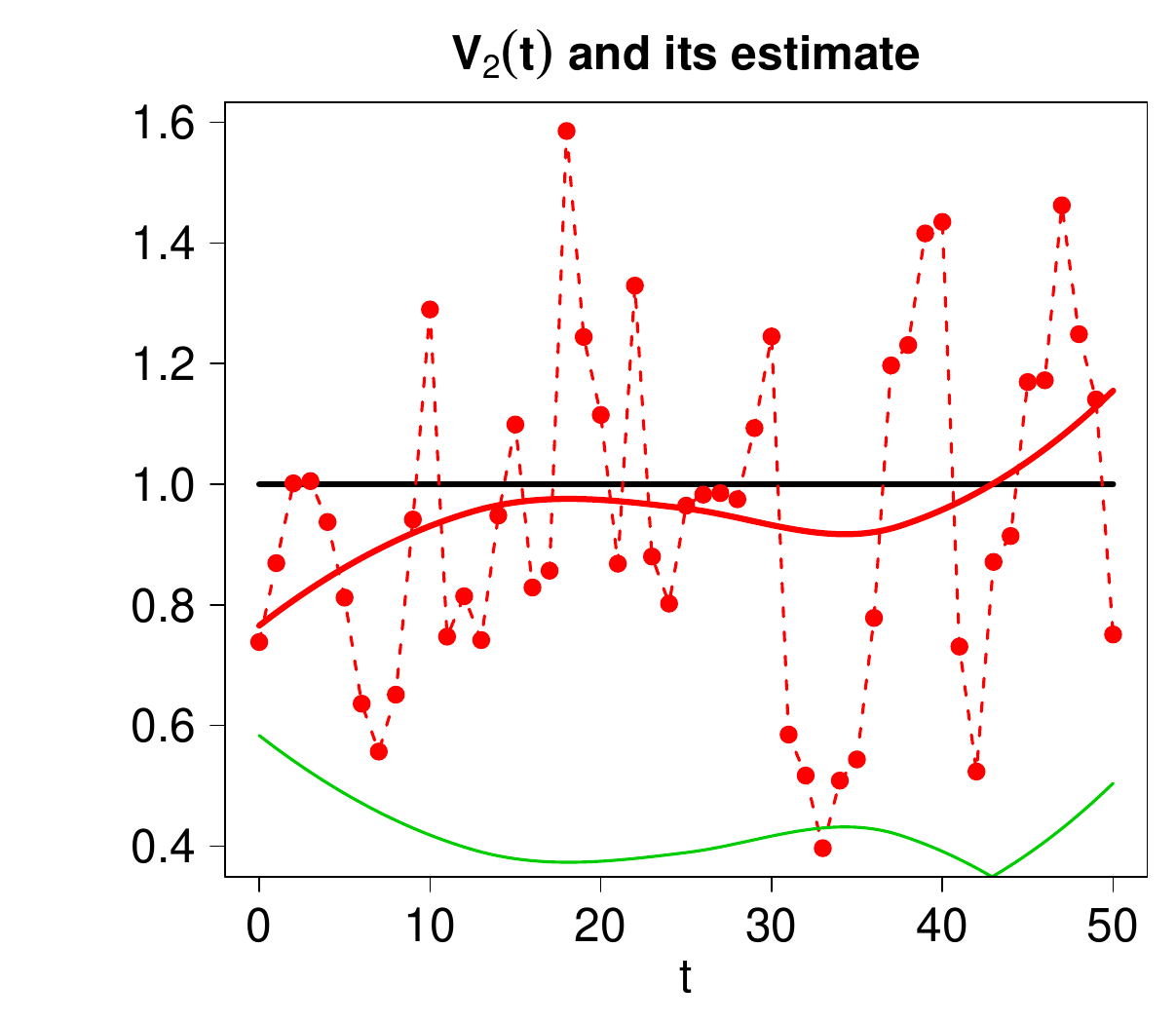} \\ \qquad \ {\scriptsize (a)} & \qquad \ {\scriptsize (b)} \\
    \includegraphics[height=0.2\textheight]{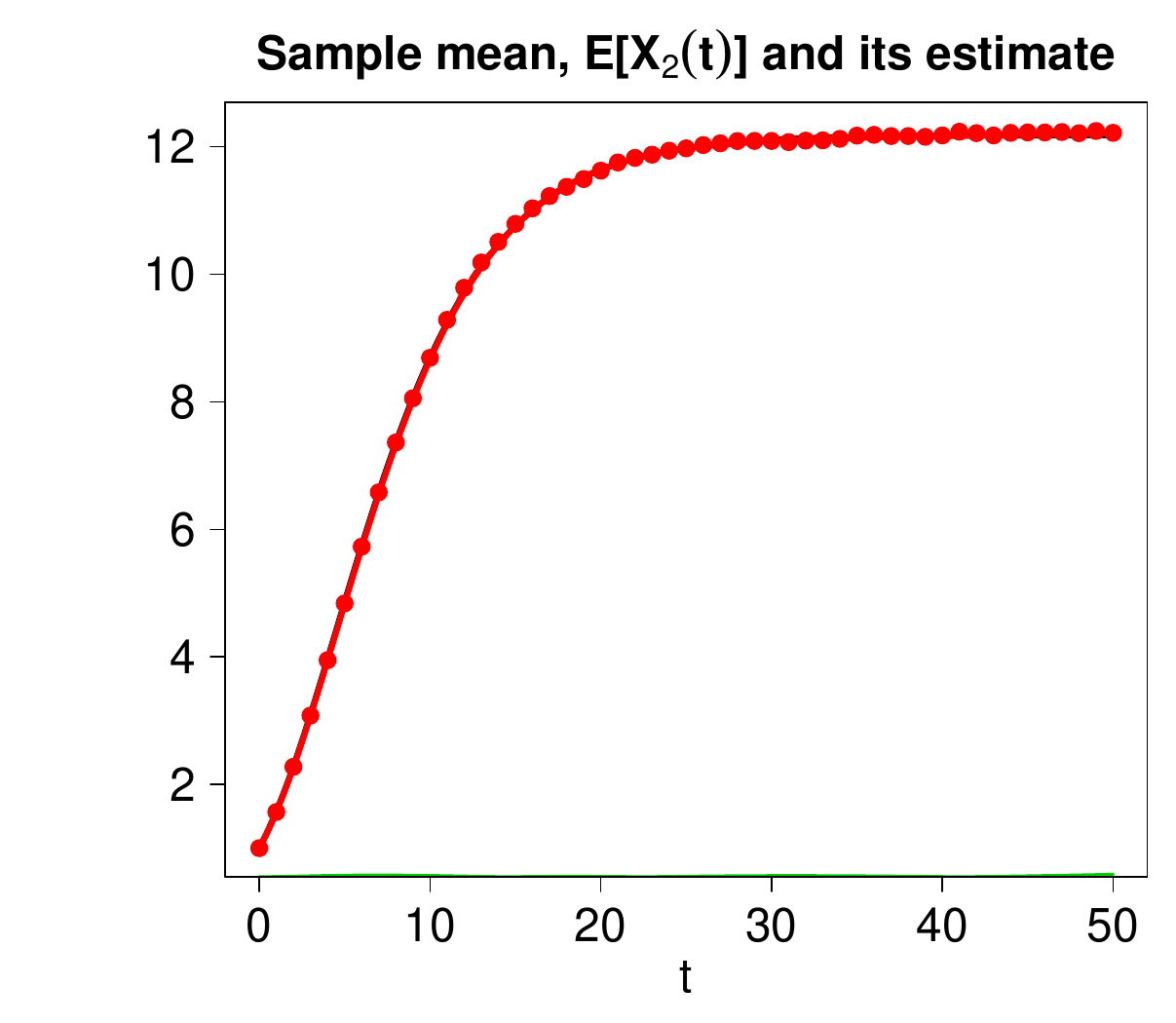} &
    \includegraphics[height=0.2\textheight]{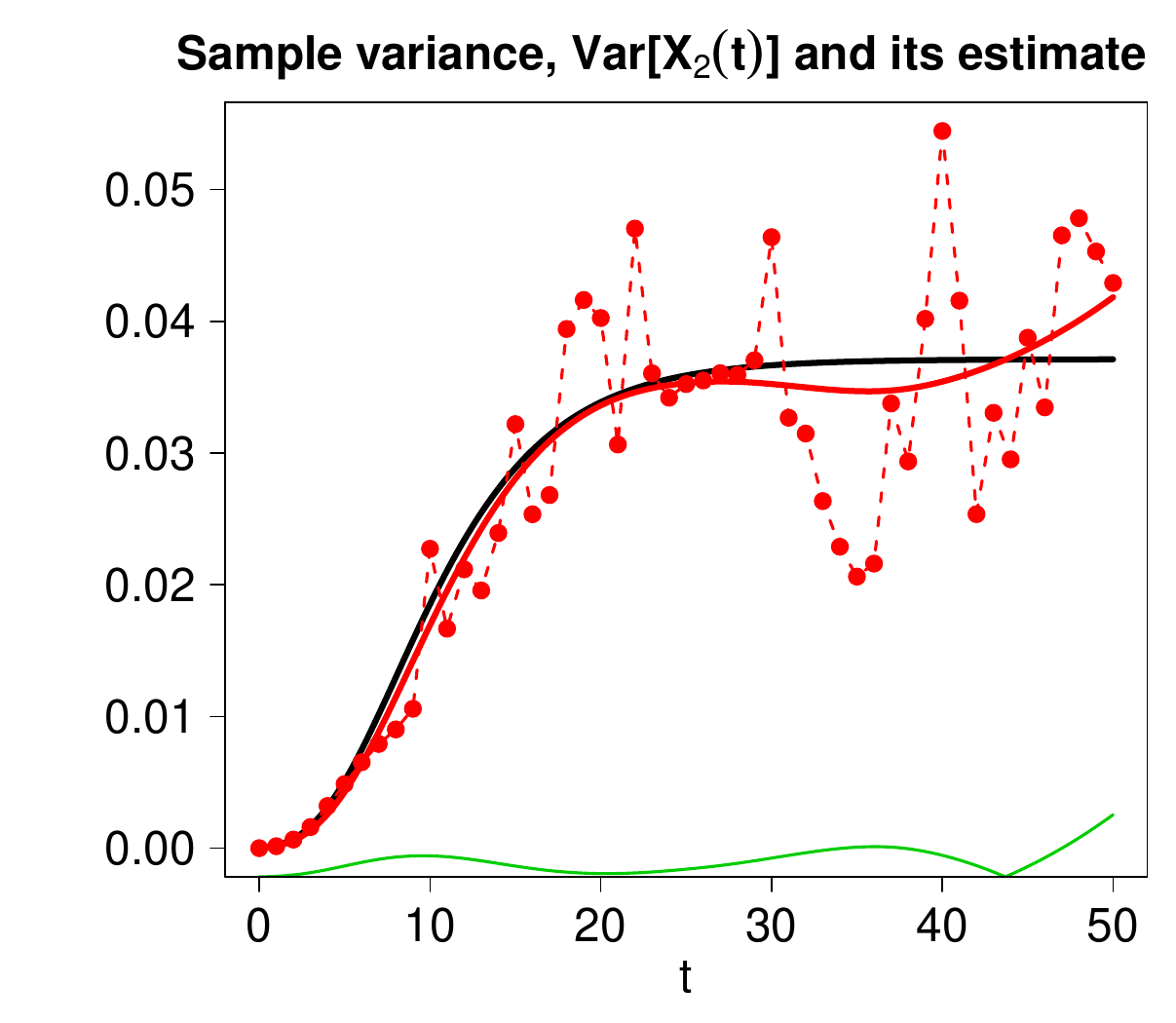} \\ \qquad \ {\scriptsize (c)} & \qquad \ {\scriptsize (d)}
    \end{tabular}
    \caption{Fit of simulated data of model (\ref{case_a_b_X2}) with $\alpha=0.5$, $\beta=0.2$, $\sigma=0.01$, $C(t)=0$, $D(t)=0$ and $V_2(t)=1$. The absolute difference functions between the simulated and fitted function are represented in green.}
    \label{Fit_V1eq1_Ceq0_V2eq1_Deq0}
  \end{figure}

  First, we test the hypothesis $H_0: V_2(t) = v_2$, where now $v_2 = 0.9778234$.
  The bootstrap test results in $D=2.9449791$ and the associated $p$-value$=0.931$ and, therefore, there is no evidence to reject that the effect of the combination of therapies on the infinitesimal variance of the process $X_2(t)$ does not depend on the time. In addition,
  as $v_2 \approx 1 $, the combination of therapies has a negligible effect on
  such infinitesimal variance. \smallskip

  Thus, assuming $V_2 (t) = 0.9778234$, we test $H_0: D(t) = d$, being now
  $d = 0.0002078$. The value of the $D$-statistics is $0.0084735$ and the $p$-value is $0.897$, so that there is no evidence to reject that the effect of the therapy inducing
  the death of cancer cells does not depend on time. Furthermore, since $d\approx 0 $, we can conclude
  that the therapy has hardly induced the death of cancer cells. \smallskip

Figure \ref{Fit_V1eq1_Ceq0_V2eq1_Deq0_PostTest} shows the estimated mean and variance functions
  in the group ${\cal G}_2$ for $\widehat{\alpha} = 0.4972273$, $\widehat{\beta} = 0.1987757$, $\widehat{\sigma} = 0.0100692$,
  $\widehat{C}(t)=0.0002401$, $\widehat{V}_2(t)=0.9778234$ and $\widehat{D}(t) = 0.0002078$, together with the sample and the theoretical mean and
  variance functions of process $X_2(t)$. Comparing this figure with Figures \ref{Fit_V1eq1_Ceq0_V2eq1_Deq0}(c) and \ref{Fit_V1eq1_Ceq0_V2eq1_Deq0}(d),
  we can see that the estimated mean and variance functions better reproduce the theoretical ones.

  \begin{figure}[H]
    \centering
    \begin{tabular}{cc}
    \includegraphics[height=0.2\textheight]{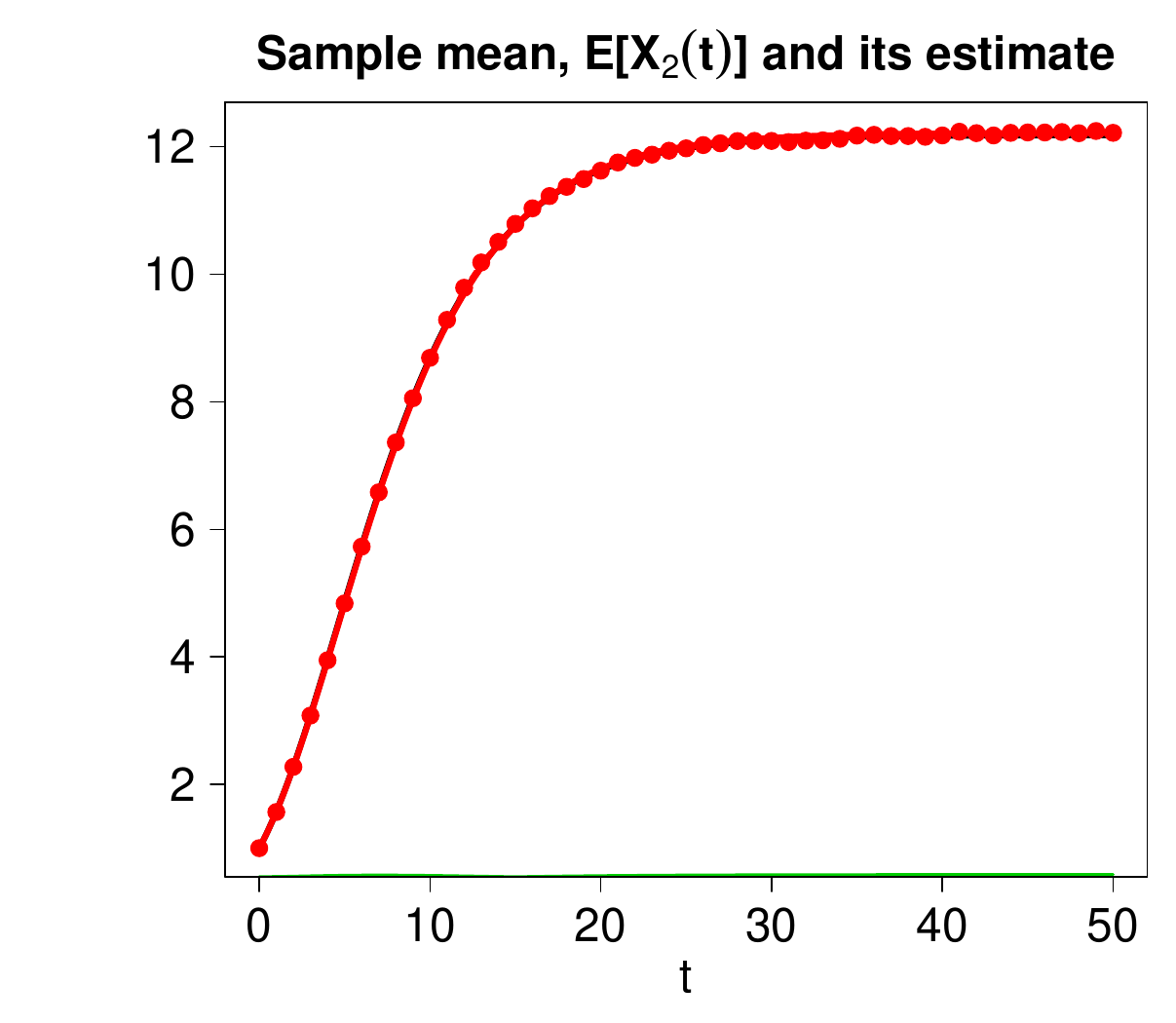} &
    \includegraphics[height=0.2\textheight]{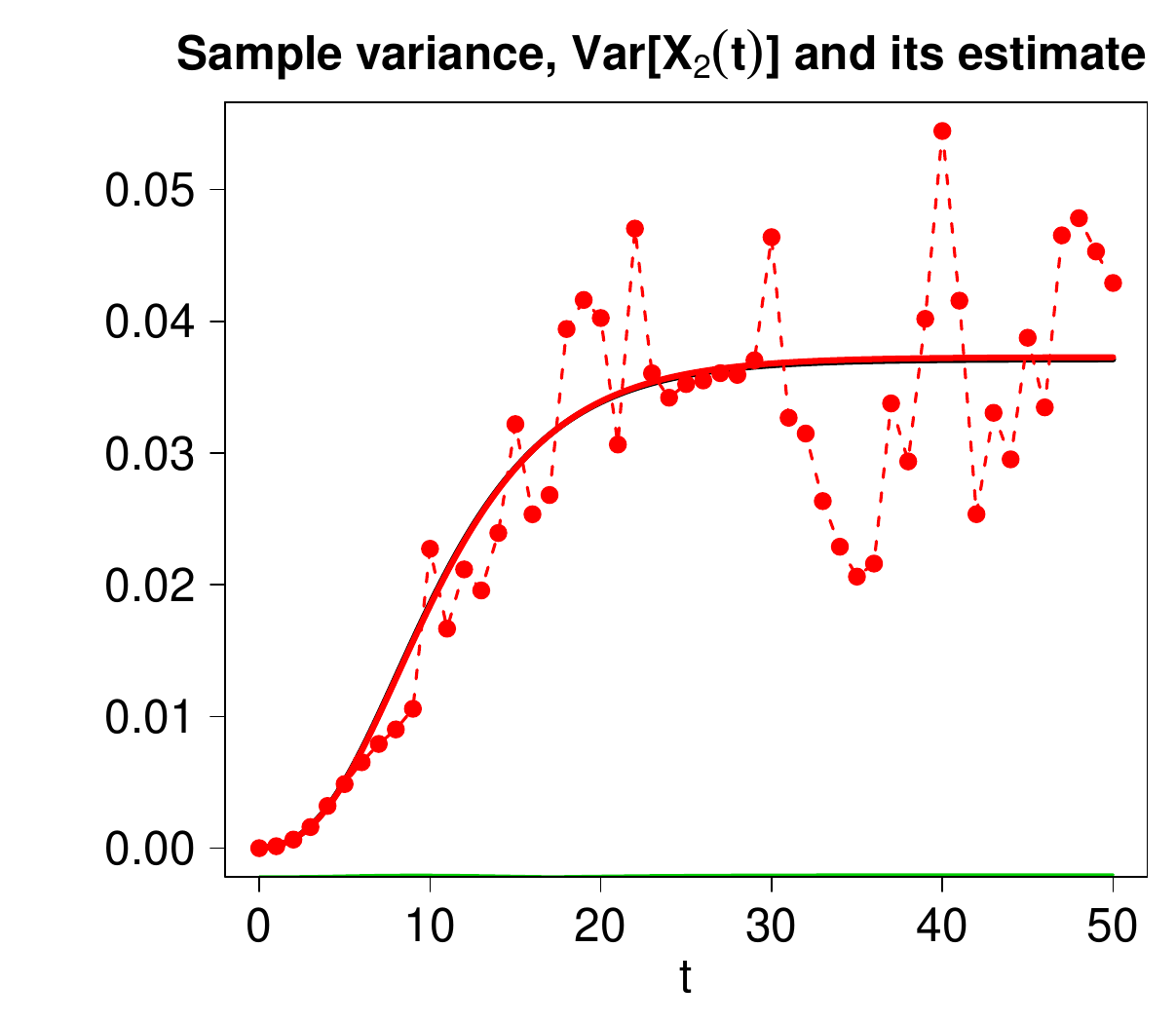}
    \end{tabular}
    \caption{Fit of simulated data of model (\ref{case_a_b_X2}) with $\alpha=0.5$, $\beta=0.2$, $\sigma=0.01$, $C(t)=0$, $D(t)=0$ and $V_2(t)=1$,
    knowing that $C(t)$ is constant and assuming that it is accepted at first that $V_2(t)$ is constant, and then, that $D(t)$ is constant}
    \label{Fit_V1eq1_Ceq0_V2eq1_Deq0_PostTest}
  \end{figure}

In Figure \ref{kernelDensityCase1} the Gaussian kernel density estimations of the $D$-statistics
for the tests just discussed are plotted based on $m=1500$ runs. The bandwidth is chosen by using
pilot estimation of derivatives as indicated in \cite{Royal}. To summarize the results of the
tests the values of the D-statistics (green points) and the critical region with significance
$0.05$ (red) are also shown.

 \begin{figure}[h]
    \centering
    \begin{tabular}{cc}
    \includegraphics[height=0.2\textheight]{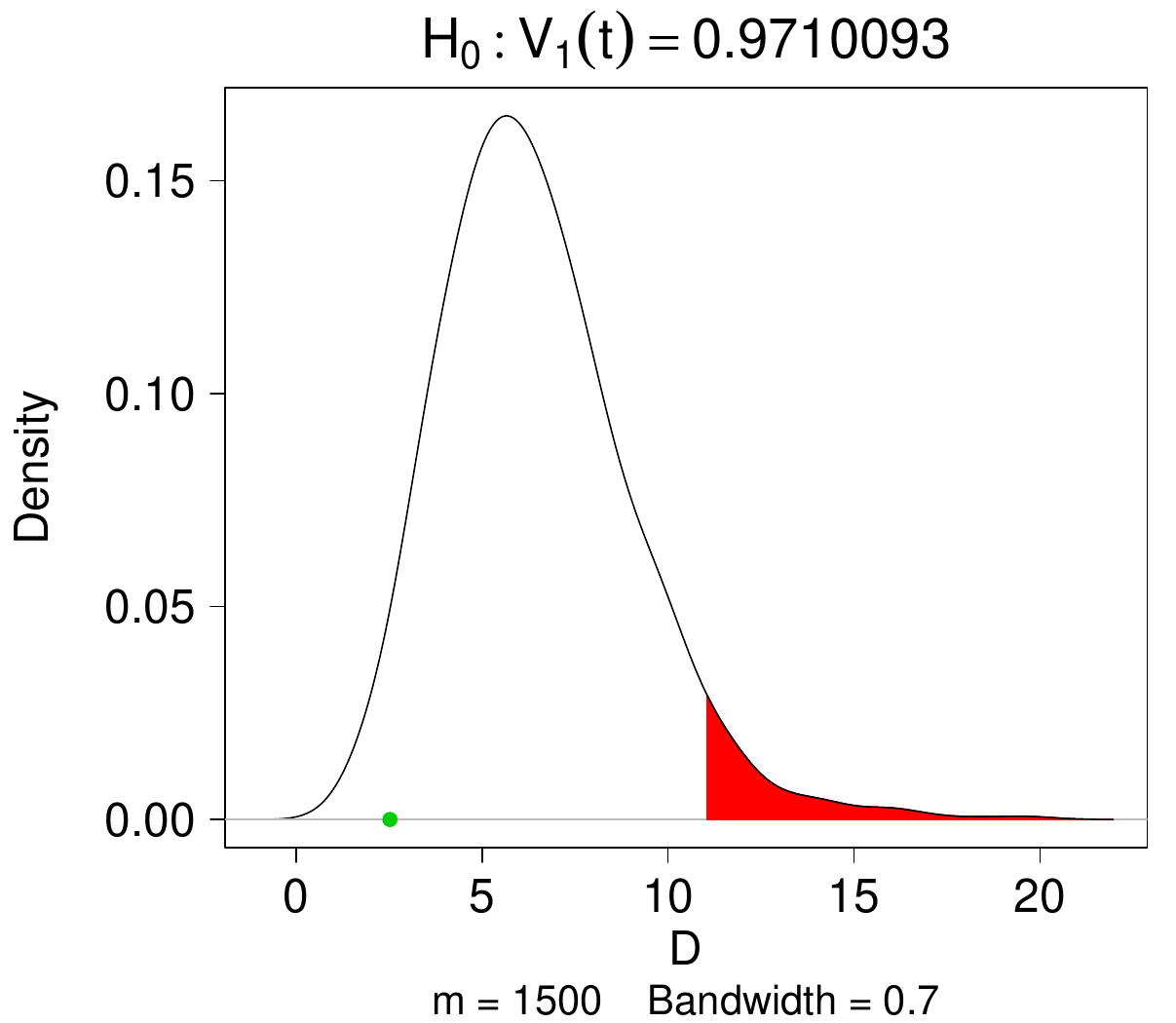} &
    \includegraphics[height=0.2\textheight]{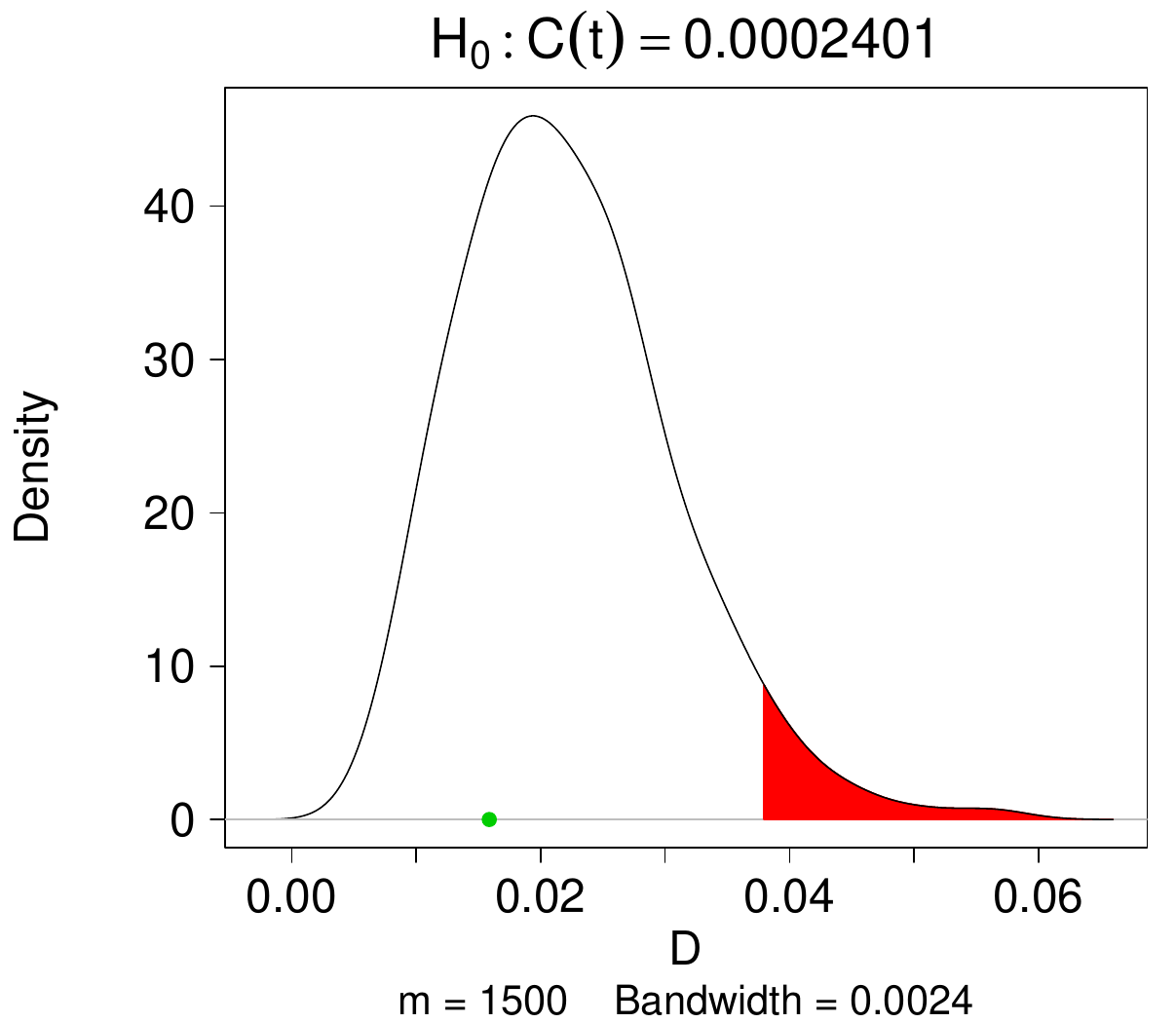}\\
\includegraphics[height=0.2\textheight]{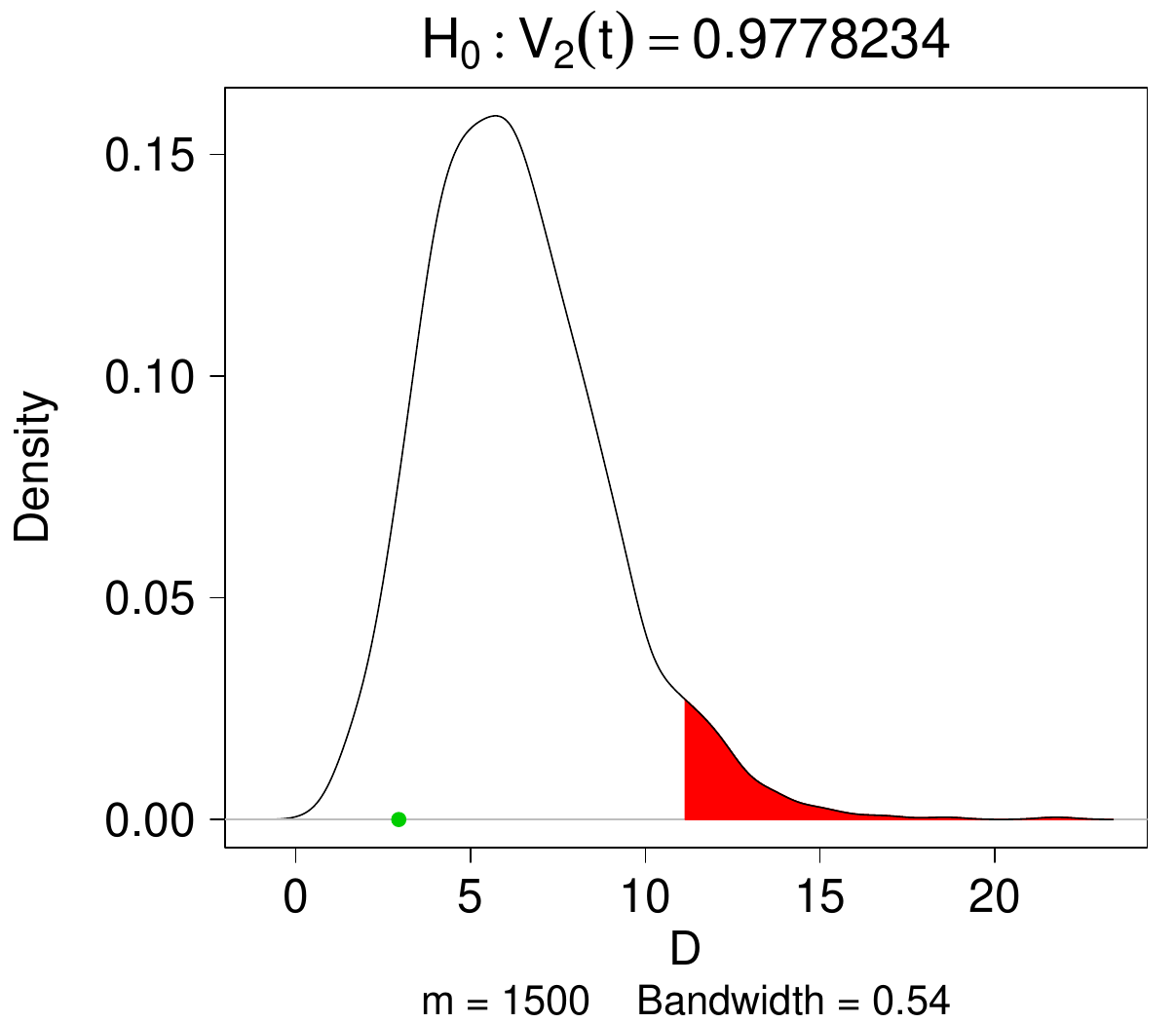} &
    \includegraphics[height=0.2\textheight]{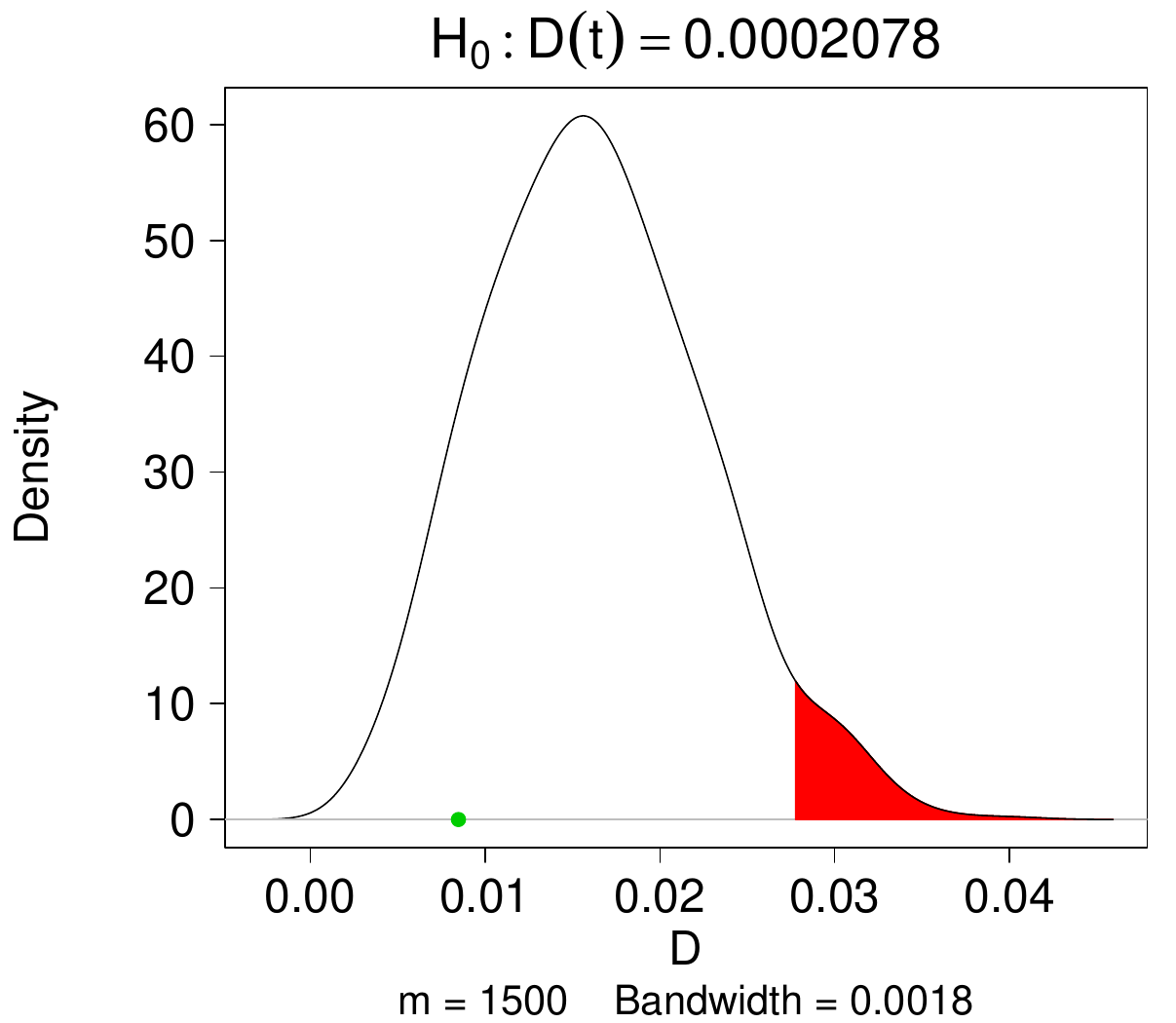}
    \end{tabular}
    \caption{Gaussian kernel density estimation of $D$-statistics for the tests associated to Case 1. Green points are the values of the D-statistics in our simulation experiment. In red the critical region with significance $0.05$ is shown.}
\label{kernelDensityCase1}
  \end{figure}

  \medskip

\item \textbf{Case 2.} $C(t)=0.025$, $V_1(t)=0.49$, $D(t) = -0.05$ and $V_2(t)=0.49$

In this case, in group ${\cal G}_1$, an anti-proliferative therapy that affects the infinitesimal
variance of the process $X_1(t)$ is considered, although its effects do not depend on time. In
group ${\cal G}_2$, the effect of the therapy that induces the death of cancer cells does not
depend on time and such therapy does not affect the infinitesimal variance previously modified by
the anti-proliferative therapy. \smallskip

Figure \ref{Paths_X1_V1cte_Ccte_X2_Ccte_V2cte_Dcte} shows the simulated sample paths of models
(\ref{gruppo_G}), (\ref{case_a_X1}) and (\ref{case_a_b_X2}) in red, blue and green, respectively.

Figure \ref{Fit_V1cte_Ccte} shows the estimated functions $\widehat{C}(t)$ and $\widehat{V}_1(t)$
in model (\ref{case_a_X1}) as well as the estimated mean and variance functions of process
$X_1(t)$. We can see that the estimated values $\widehat{C}_i$ and $\widehat{V}_{1,i}$ vary around
values close to $0.025$ and $0.49$, respectively, which suggests that the functions $C(t)$ and
$V_1(t)$ could be constant.

\begin{figure}[H]
    \centering
    \includegraphics[height=0.2\textheight]{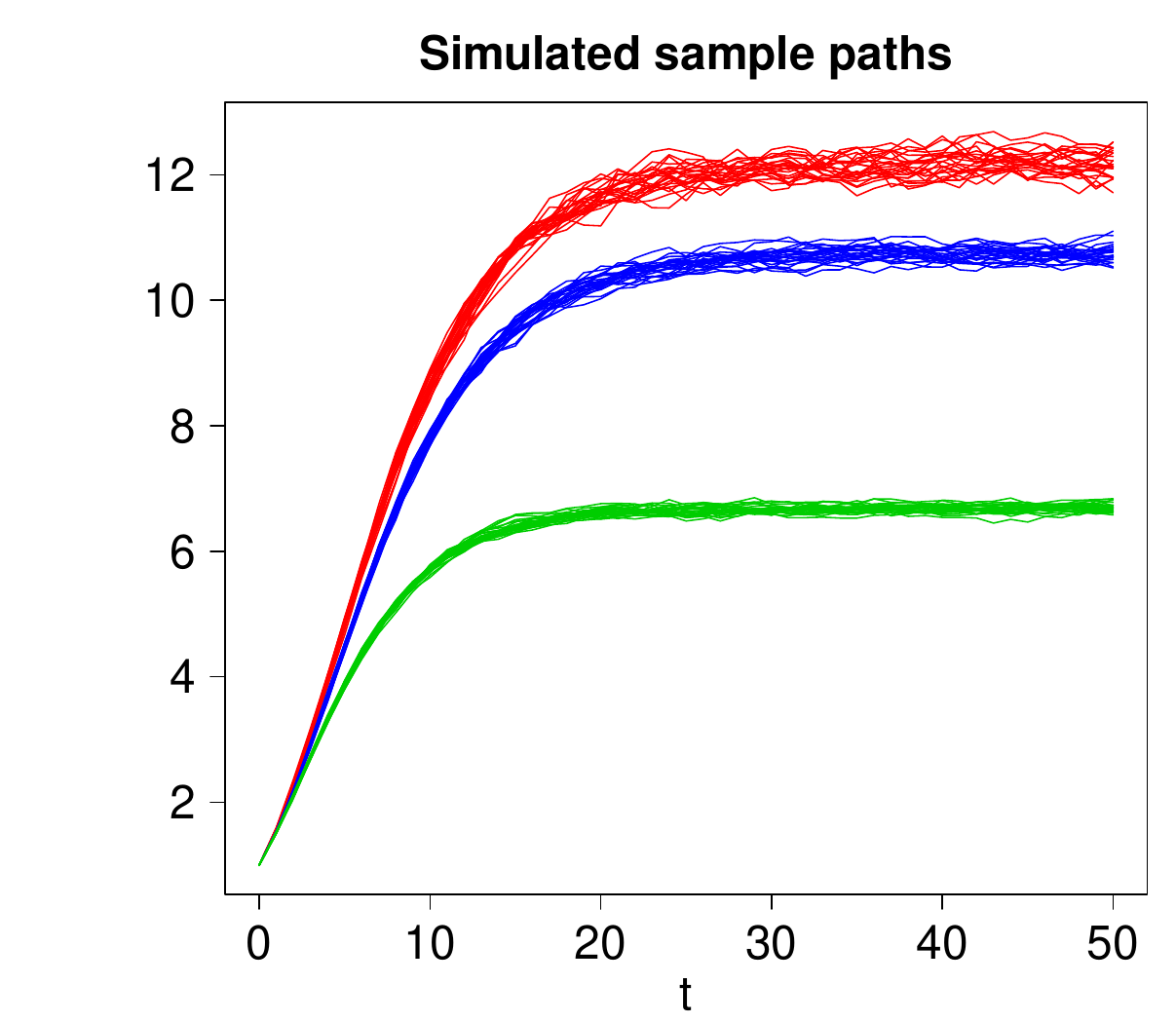}
    \caption{Simulated sample paths of models (\ref{gruppo_G}), (\ref{case_a_X1}) and (\ref{case_a_b_X2}),
    in red, blue and green, respectively, with $\alpha=0.5$, $\beta=0.2$, $\sigma=0.01$, $C(t)=0.025$, $V_1(t)=0.49$,
    $D(t) = -0.05$ and $V_2(t)=0.49$.} \label{Paths_X1_V1cte_Ccte_X2_Ccte_V2cte_Dcte}
\end{figure}

\begin{figure}[h]
  \centering
  \begin{tabular}{cc}
  \includegraphics[height=0.2\textheight]{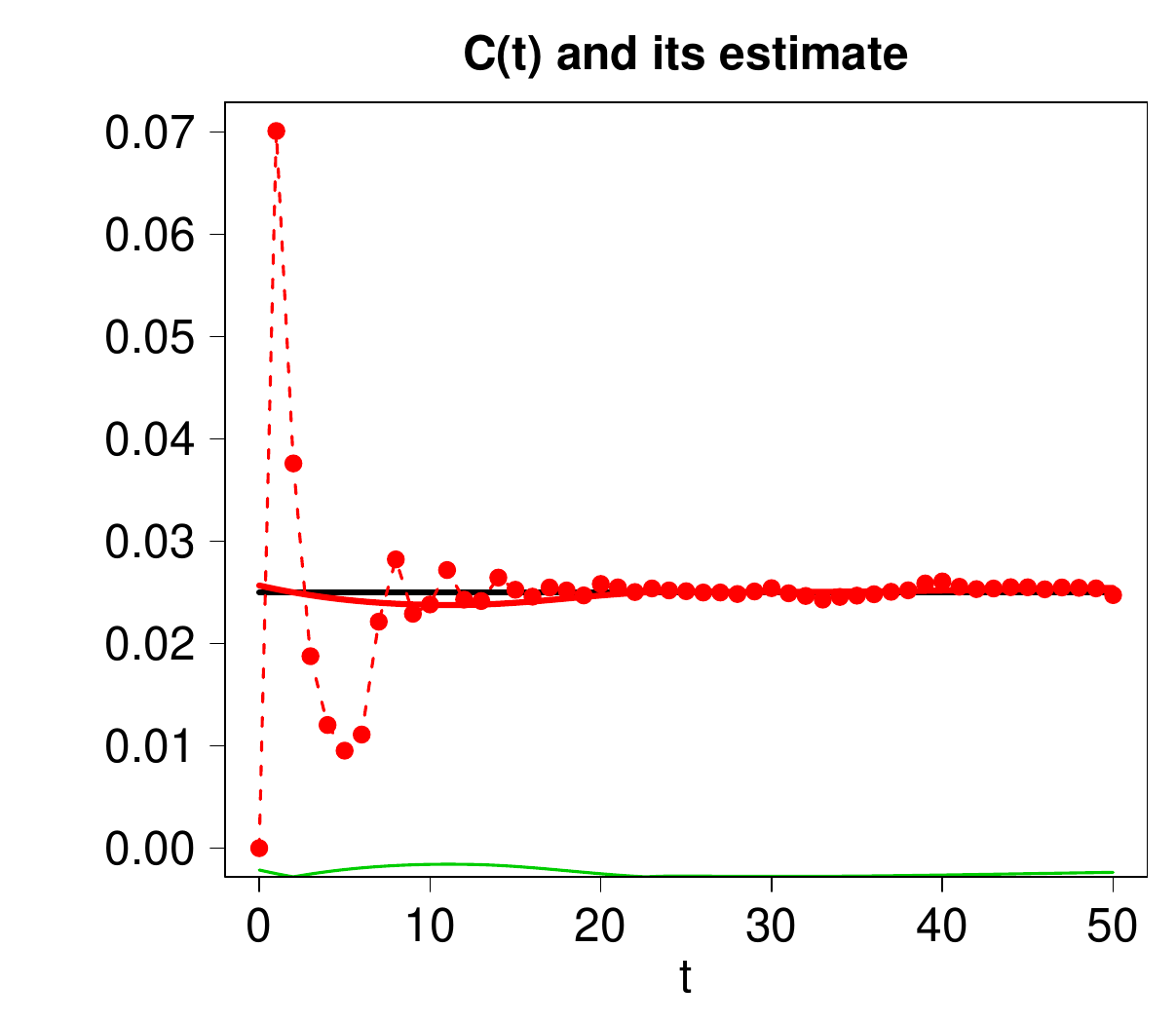} &
  \includegraphics[height=0.2\textheight]{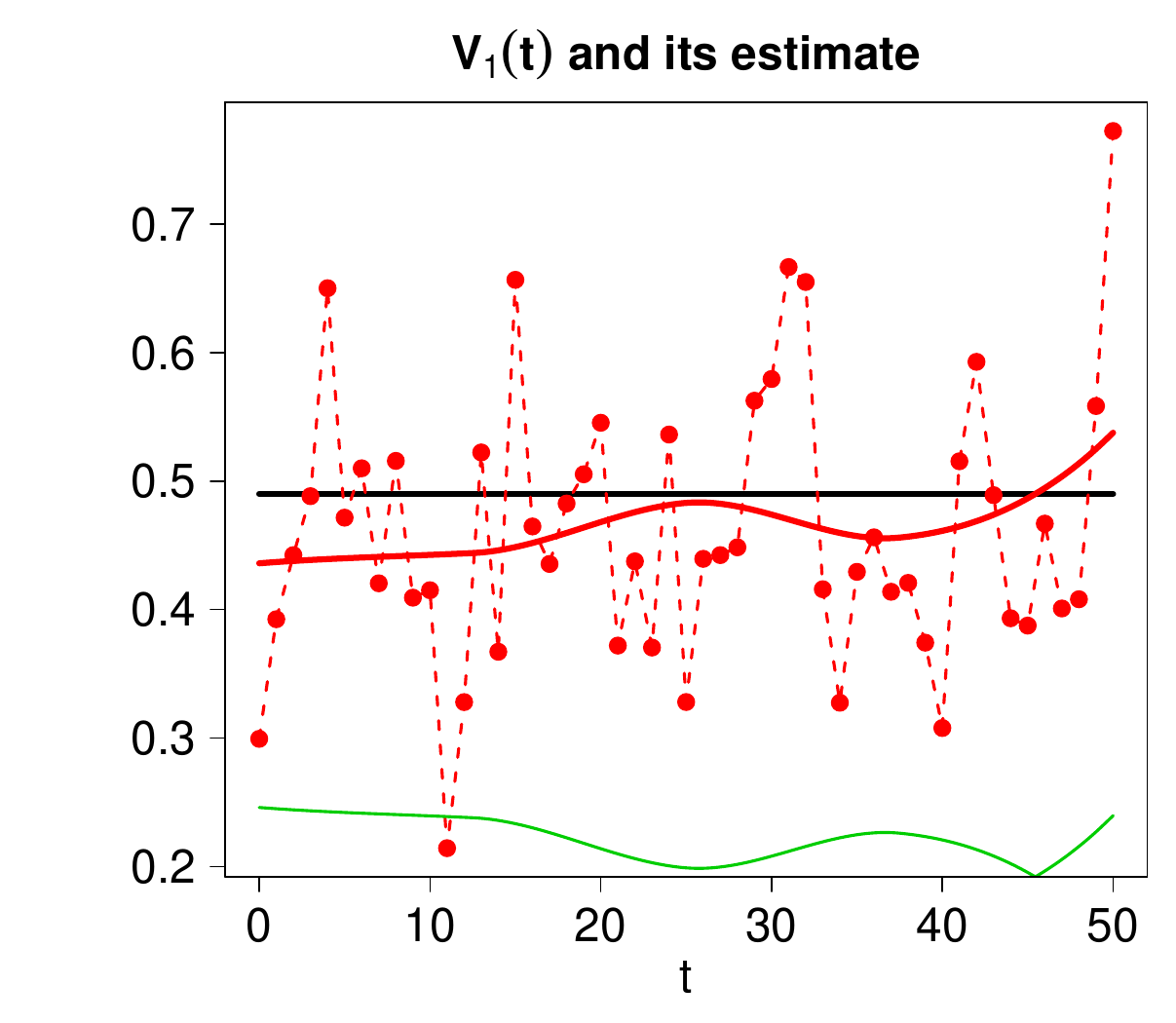} \\
  \includegraphics[height=0.2\textheight]{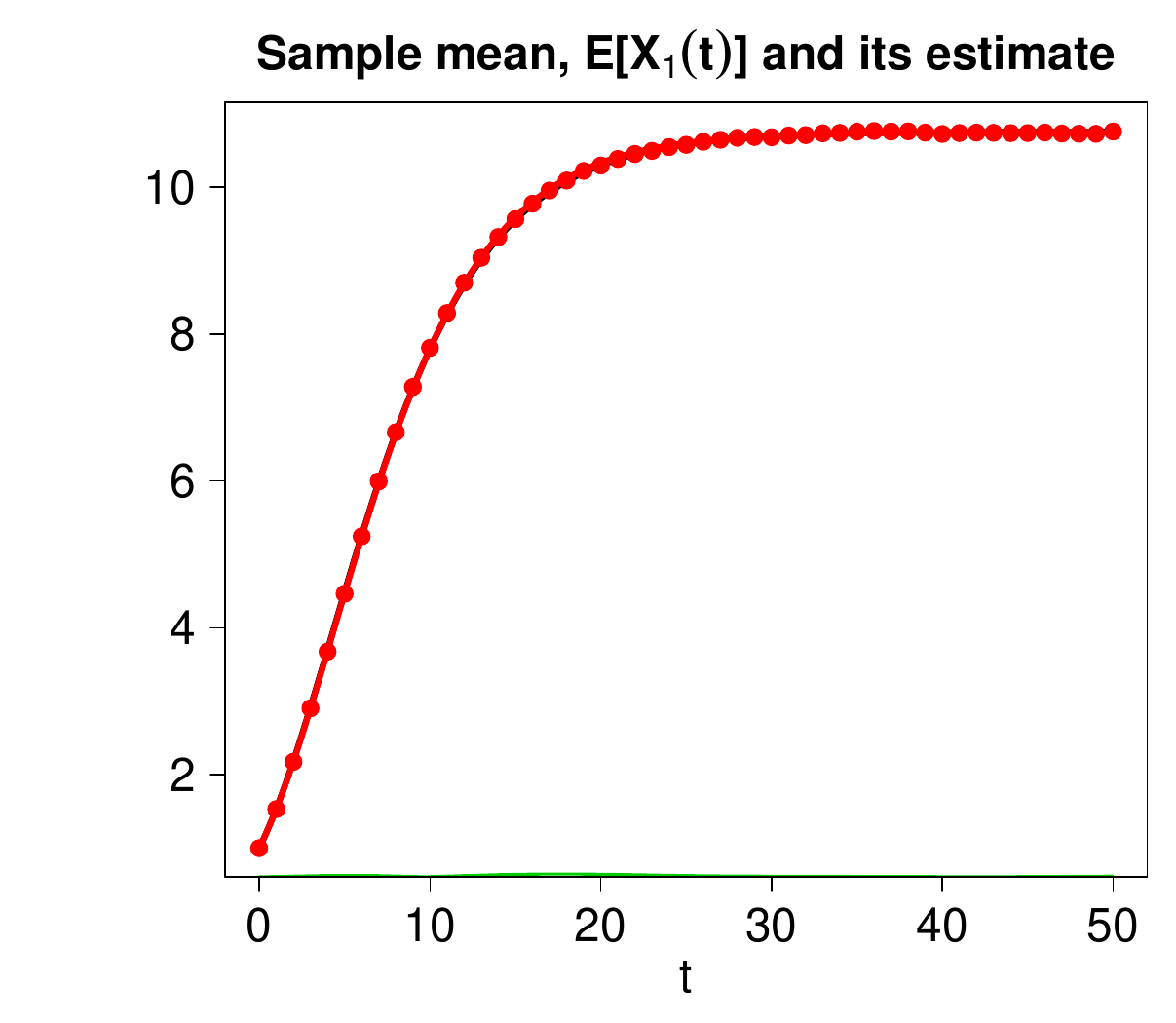} &
  \includegraphics[height=0.2\textheight]{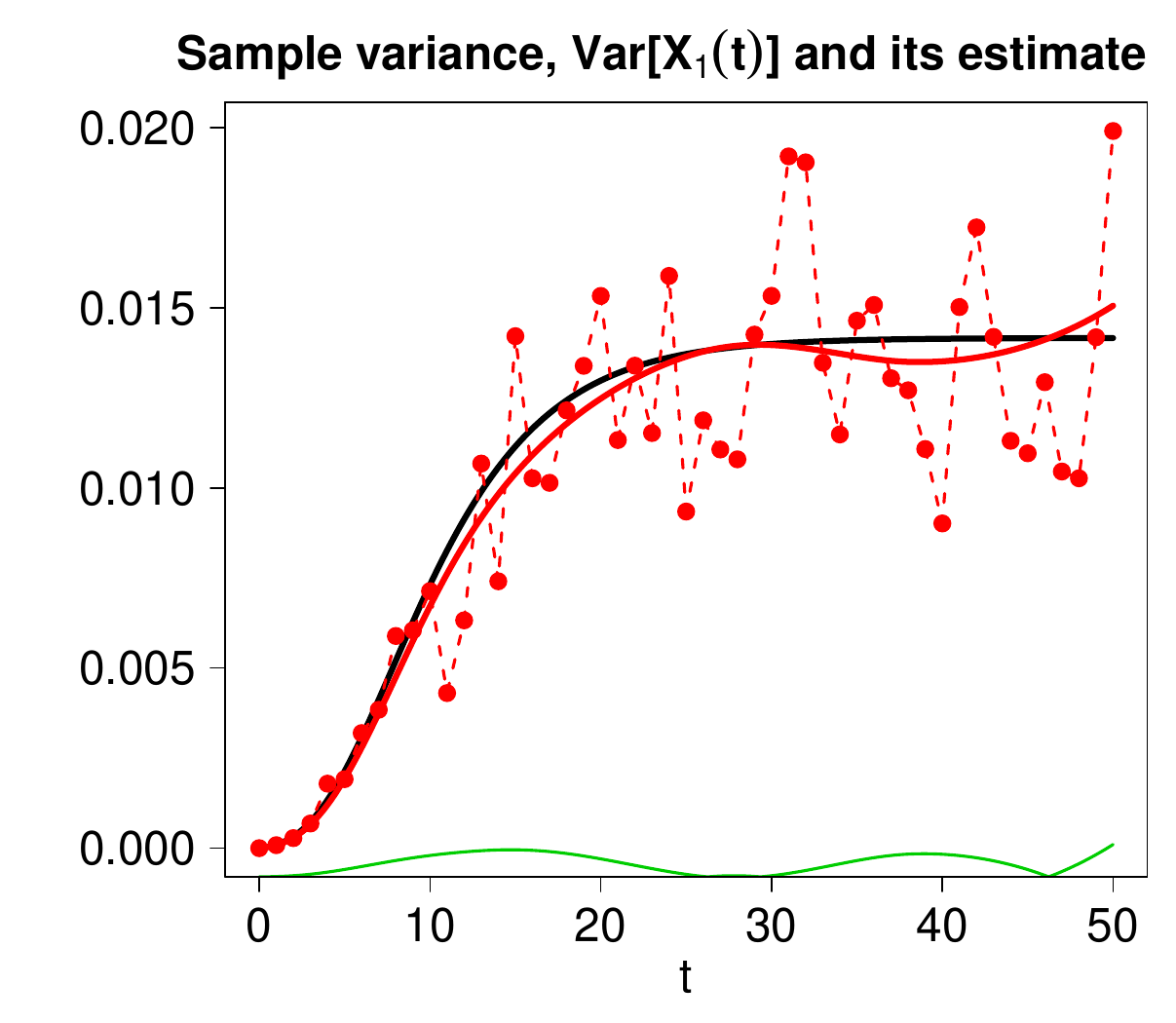}
  \end{tabular}
  \caption{Fit of simulated data of model (\ref{case_a_X1}) with $\alpha=0.5$, $\beta=0.2$, $\sigma=0.01$, $C(t)=0.025$ and $V_1(t)=0.49$.
  The absolute difference functions between the simulated and fitted function are represented in green.}
  \label{Fit_V1cte_Ccte}
\end{figure}

First we test the hypothesis $H_0: V_1(t) = v_1$ where $v_1 = 0.4761118$. The value of the
$D$-statistics is $D=1.0633344$ and the $p$-value is $0.996$, so there is no evidence to reject
that the effect of the anti-proliferative therapy on the infinitesimal variance of the process
$X_1(t)$ does not depend on time. \medskip

Then, under the assumption $V_1(t) = 0.4761118$, we test $H_0: C(t) =c$ with $c= 0.0248451$. The
bootstrap test provides $D=0.0234247$ and a $p$-value of $0.334$, so there is no evidence to
reject that the effect of the anti-proliferative therapy on the growth rate does not depend on
time. \medskip

Figure \ref{Fit_V1cte_Ccte_PostTest} shows the estimated mean and variance functions in the group
${\cal G}_1$ with $\widehat{\alpha} = 0.4972273$, $\widehat{\beta} = 0.1987757$, $\widehat{\sigma}
= 0.0100692$, $\widehat{C}(t)=0.0248451$ and $\widehat{V}_1(t)=0.4761118$, together with the
sample and the theoretical mean and variance functions of the process $X_1(t)$. It is clear that
now the estimated mean and variance functions reproduce more appropriately the theoretical ones.
\medskip

\begin{figure}[H]
  \centering
  \begin{tabular}{cc}
  \includegraphics[height=0.2\textheight]{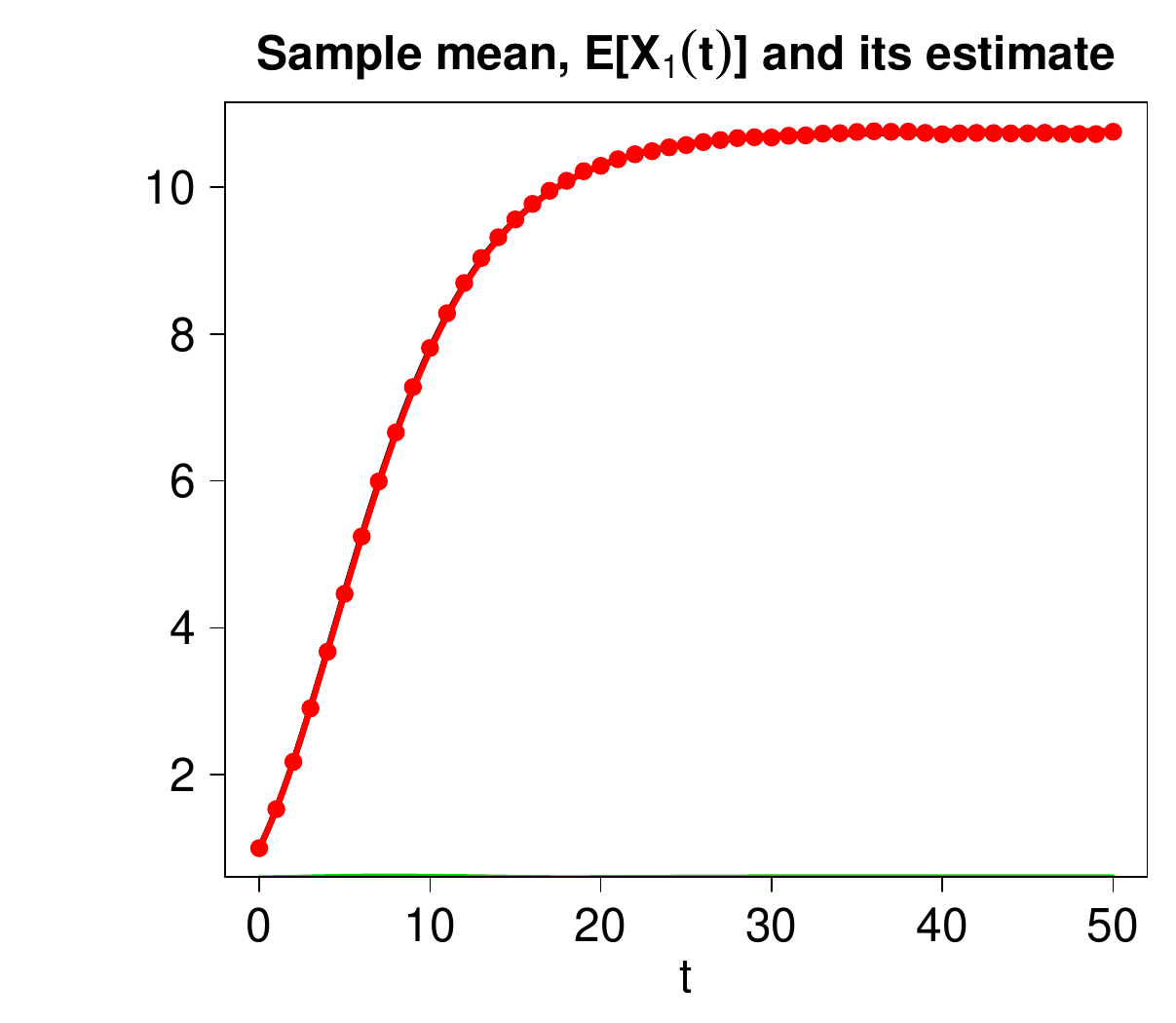} &
  \includegraphics[height=0.2\textheight]{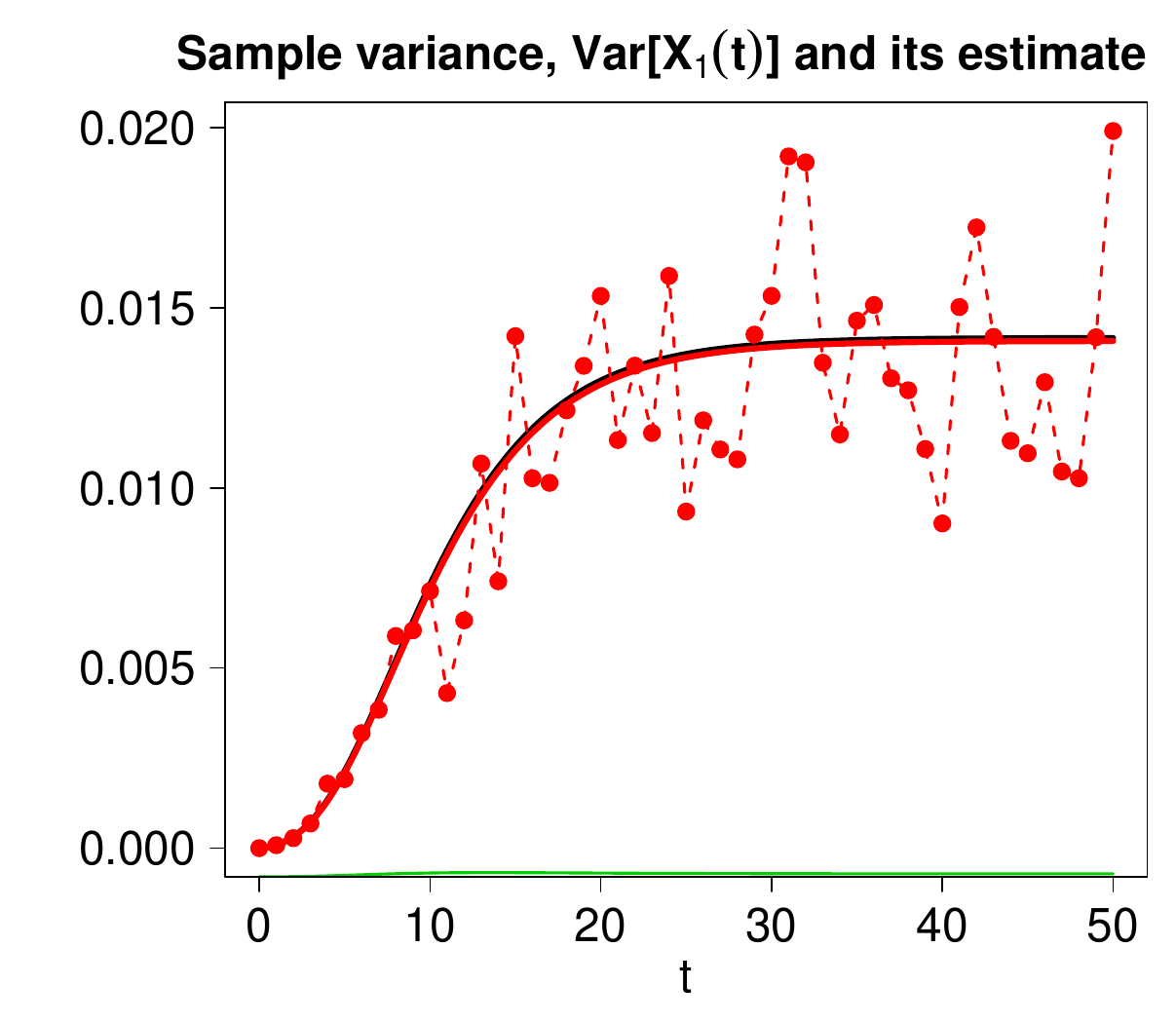}
  \end{tabular}
  \caption{Fit of simulated data of model (\ref{case_a_X1}) with $\alpha=0.5$, $\beta=0.2$, $\sigma=0.01$, $C(t)=0.025$ and $V_1(t)=0.49$,
  assuming that it is accepted first that $V_1(t)$ is constant, and then, that $C(t)$ is constant. The absolute difference functions between the simulated and fitted function are represented in green.} \label{Fit_V1cte_Ccte_PostTest}
\end{figure}

Figure \ref{Fit_V1cte_Ccte_V2cte_Dcte} shows the estimated functions $\widehat{D}(t)$ and
$\widehat{V}_2(t)$ in model (\ref{case_a_b_X2}) by using $\widehat{\alpha}$, $\widehat{\beta}$,
$\widehat {\sigma}$ and $\widehat{C}(t)=0.0248451$, as well as the estimated mean and variance
functions of process $X_2(t)$. The estimated values $\widehat{D}_i$ and $\widehat{V}_{2,i}$ in
this figure vary around values close to $-0.05$ and $0.49$, respectively, and it seems reasonable
to test whether the functions $D(t)$ and $V_2(t)$ are constant. \smallskip
\smallskip

\begin{figure}[h]
    \centering
    \begin{tabular}{cc}
    \includegraphics[height=0.2\textheight]{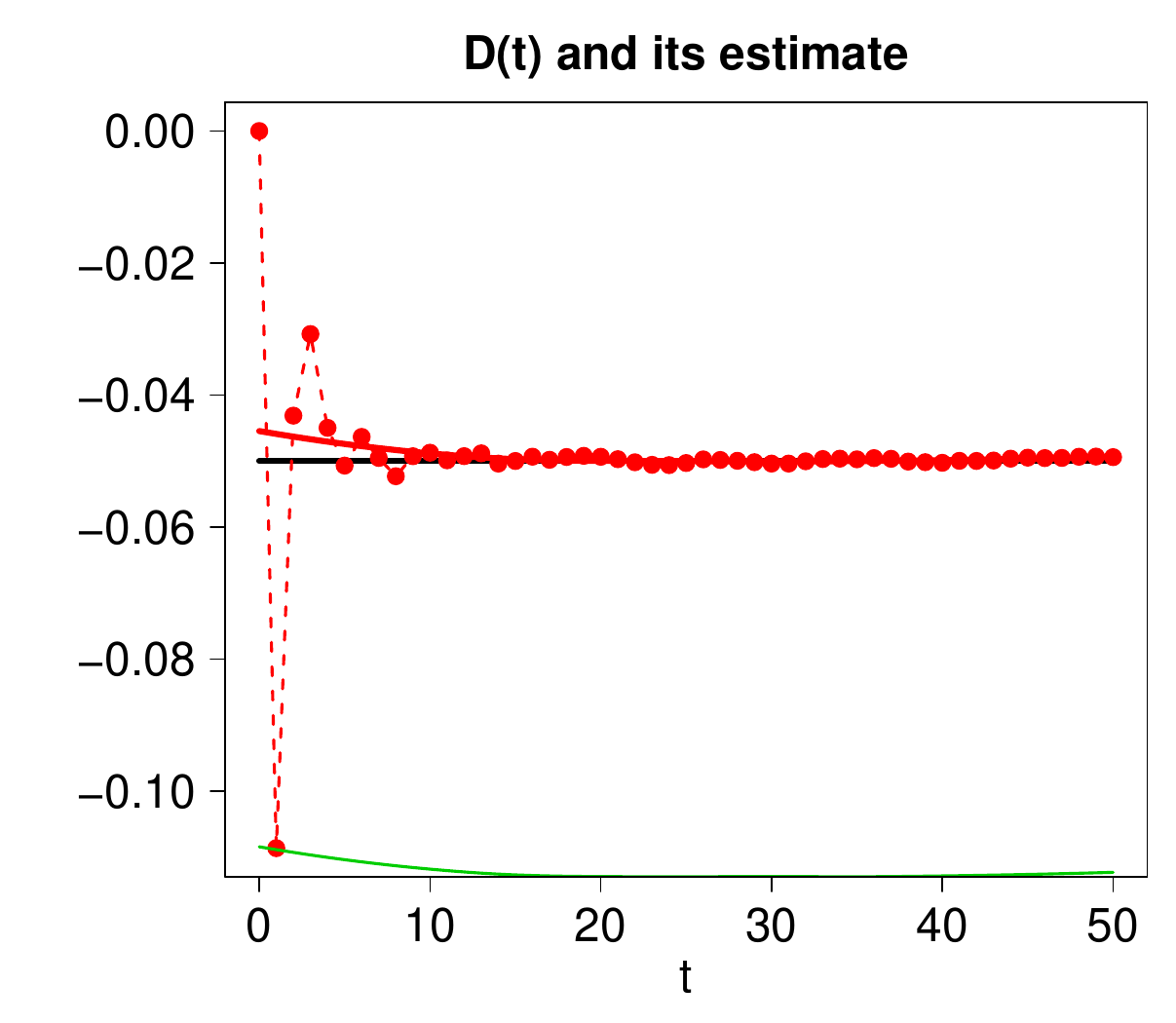} &
    \includegraphics[height=0.2\textheight]{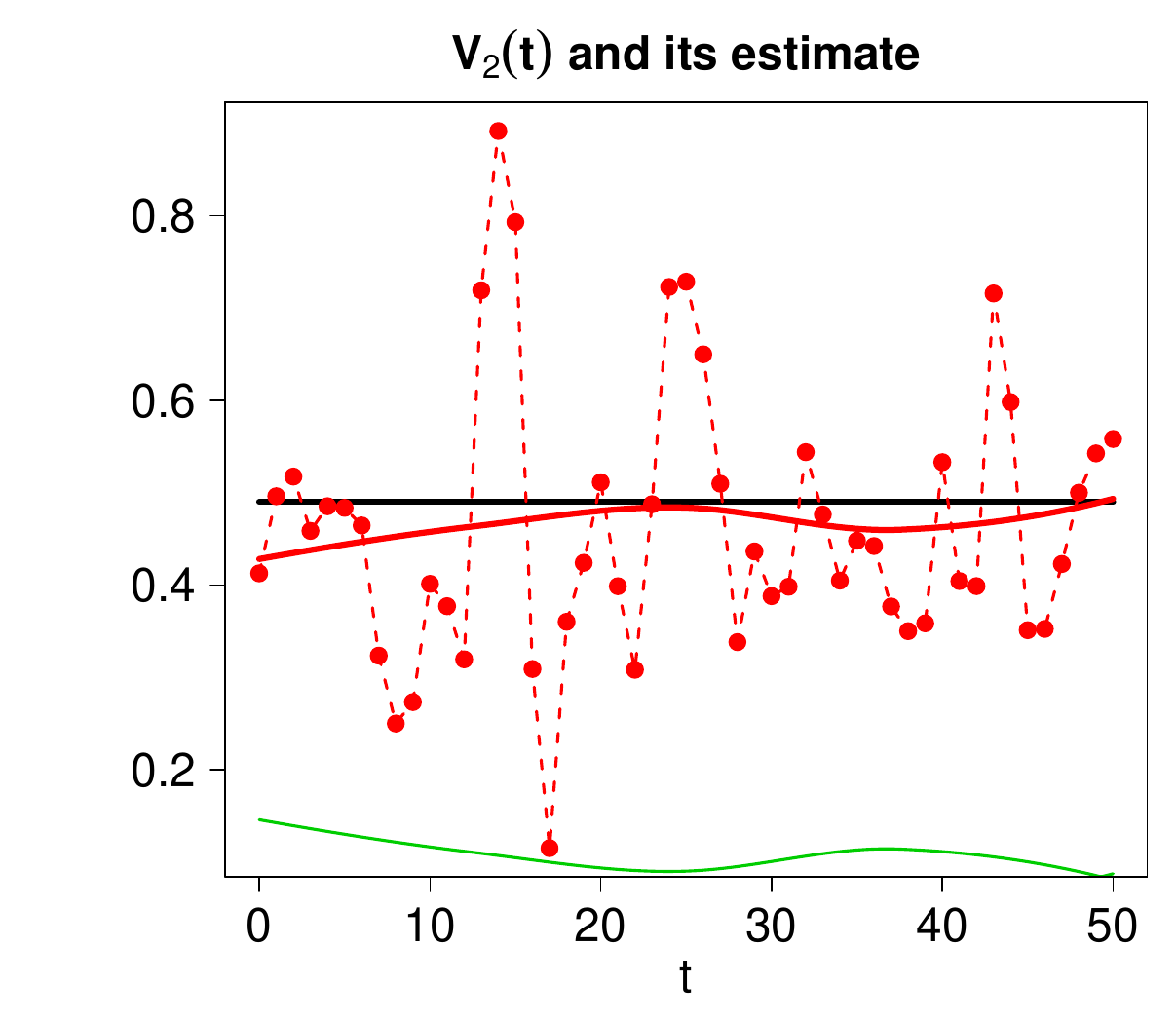} \\
    \includegraphics[height=0.2\textheight]{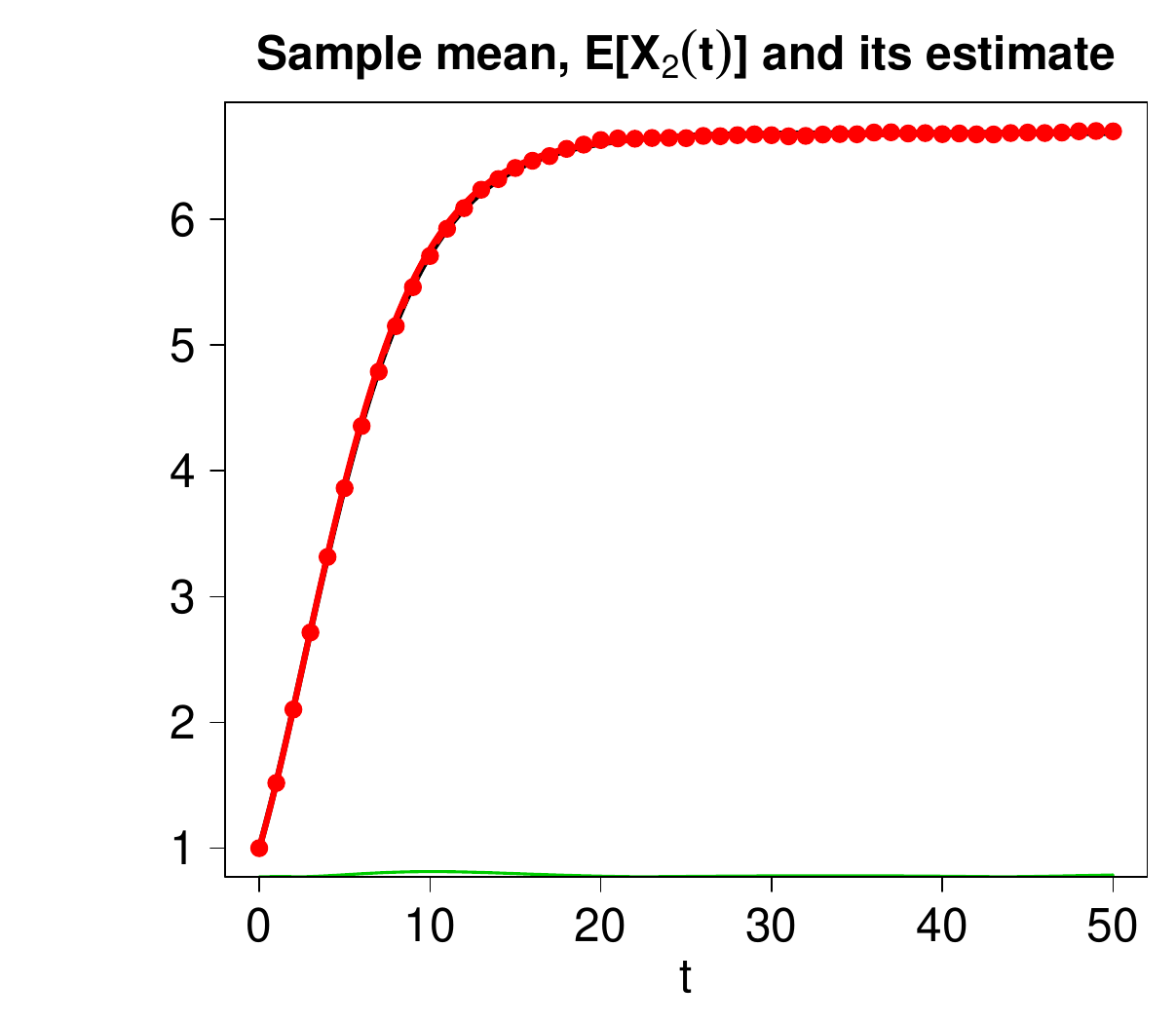} &
    \includegraphics[height=0.2\textheight]{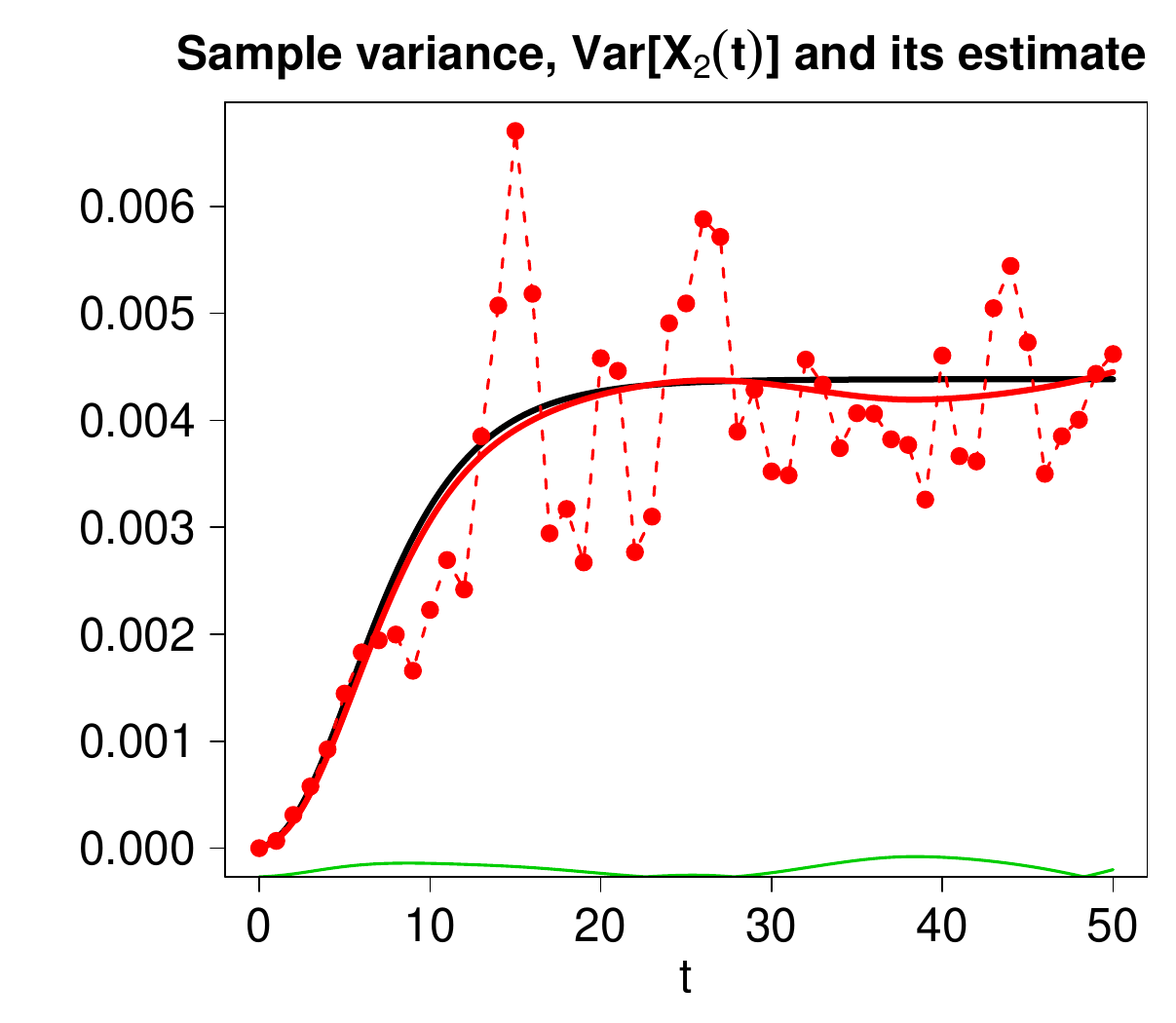}
    \end{tabular}
    \caption{Fit of simulated data of model (\ref{case_a_b_X2}) with $\alpha=0.5$, $\beta=0.2$, $\sigma=0.01$, $C(t)=0.025$, $D(t) = -0.05$ and $V_2(t)=0.49$.}
    \label{Fit_V1cte_Ccte_V2cte_Dcte}
\end{figure}

First we test $H_0: V_2(t) = v_2$ where $v_2 = 0.4842422$. The value of the
$D$-statistics is $D=0.9347132$ and the $p$-value is $0.981$, so that there is no evidence to
reject that the effect of the combination of therapies on the infinitesimal variance of the
process $X_2(t)$ is independent on time. \smallskip

Hence, under the assumption $V_2 (t) = 0.4842422$, we test $H_0: D(t) = d$ where the
proposed value for $d$ is $-0.0497542$. The bootstrap test results in $D=0.0359978$ and
$p$-value=$0.321$ and consequently there is no evidence to reject that the therapy that induces
the death of cancer cells does not depend on time. \smallskip

Figure \ref{Fit_V1cte_Ccte_V2cte_Dcte_PostTest} shows the estimated mean and variance functions in
group ${\cal G}_2$ with $\widehat{\alpha} = 0.4972273$, $\widehat{\beta} = 0.1987757$,
$\widehat{\sigma} = 0.0100692$, $\widehat{C}(t)=0.0248451$, $\widehat{V}_2(t)=0.4842422$ and
$\widehat{D}(t) = -0.0497542$, together with the theoretical and the sample mean and variance
functions of the process $X_2(t)$. Again, the estimated mean and variance functions better
reproduce the theoretical ones. \medskip

    \begin{figure}[h]
    \centering
    \begin{tabular}{cc}
    \includegraphics[height=0.2\textheight]{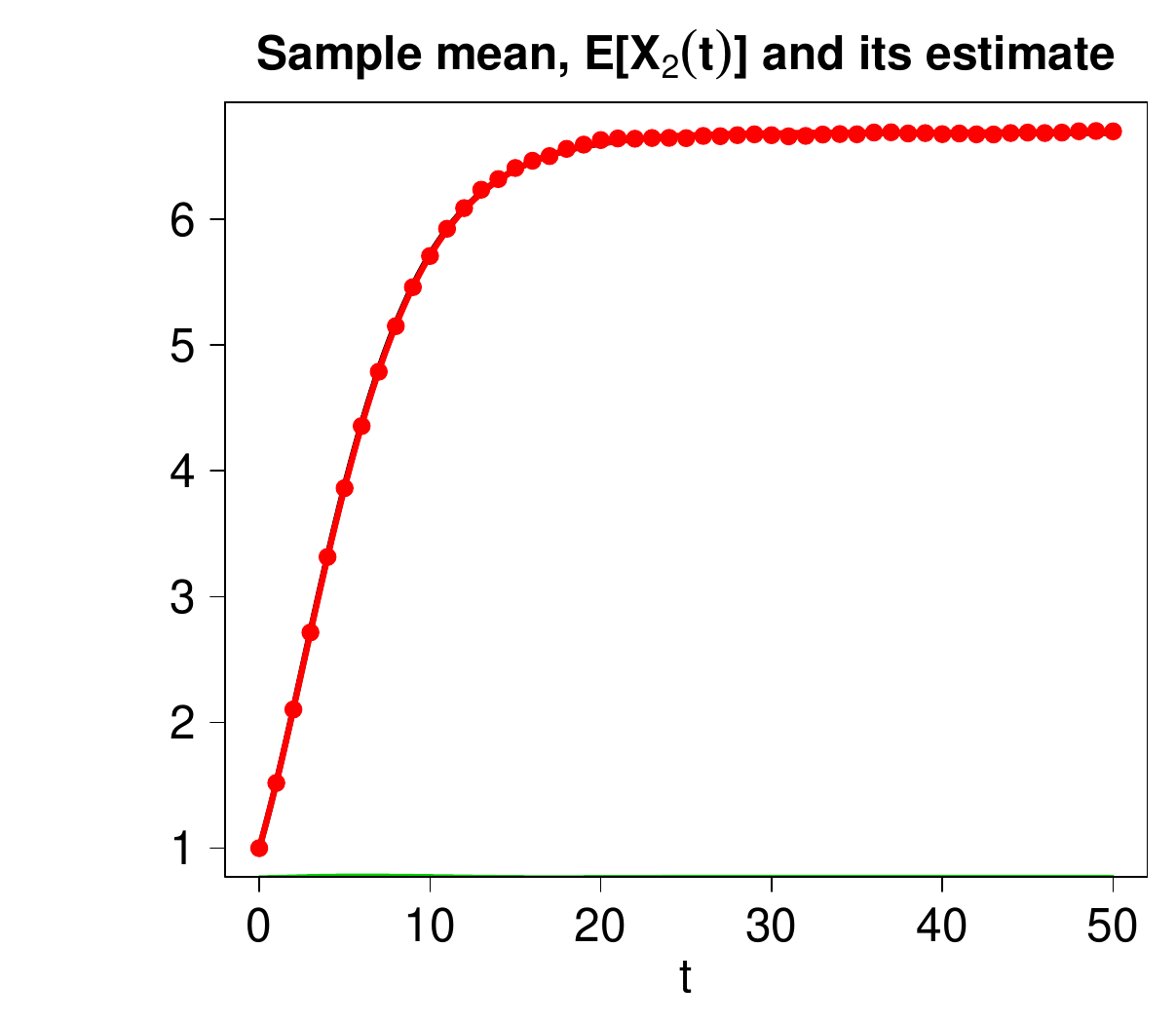} &
    \includegraphics[height=0.2\textheight]{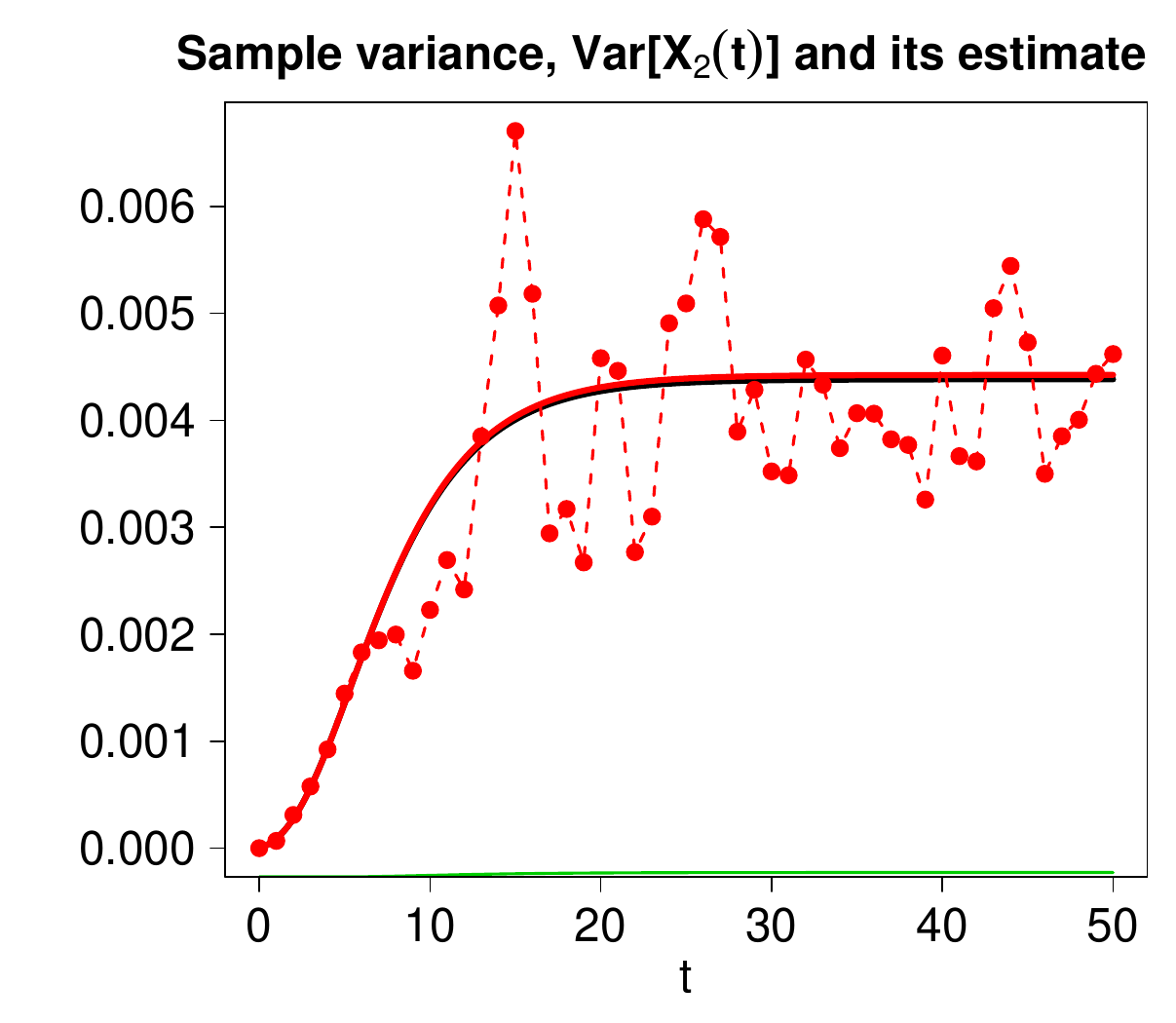}
    \end{tabular}
    \caption{Fit of simulated data of model (\ref{case_a_b_X2}) with $\alpha=0.5$, $\beta=0.2$, $\sigma=0.01$, $C(t)=0.025$, $D(t) = -0.05$ and $V_2(t)=0.49$
    knowing that $C(t)$ is constant and assuming that it is accepted first that $V_2(t)$ is constant, and then that $D(t)$ is constant.}
    \label{Fit_V1cte_Ccte_V2cte_Dcte_PostTest}
    \end{figure}

As in Figure \ref{kernelDensityCase1}, the Gaussian kernel density estimations of the
$D$-statistics for the tests just discussed are plotted in Figure \ref{kernelDensityCase2}.

 \begin{figure}[h]
    \centering
    \begin{tabular}{cc}
    \includegraphics[height=0.2\textheight]{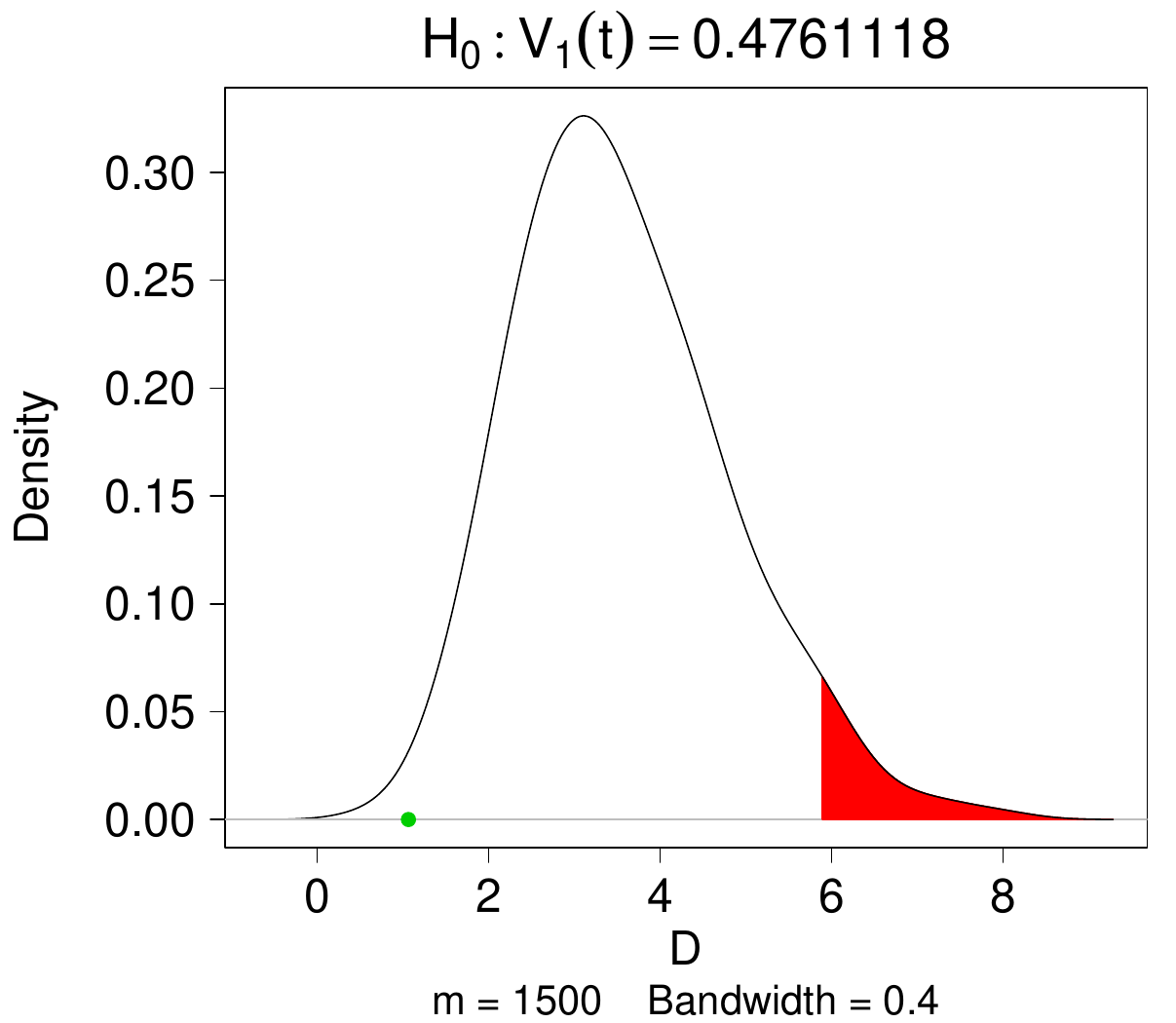} &
    \includegraphics[height=0.2\textheight]{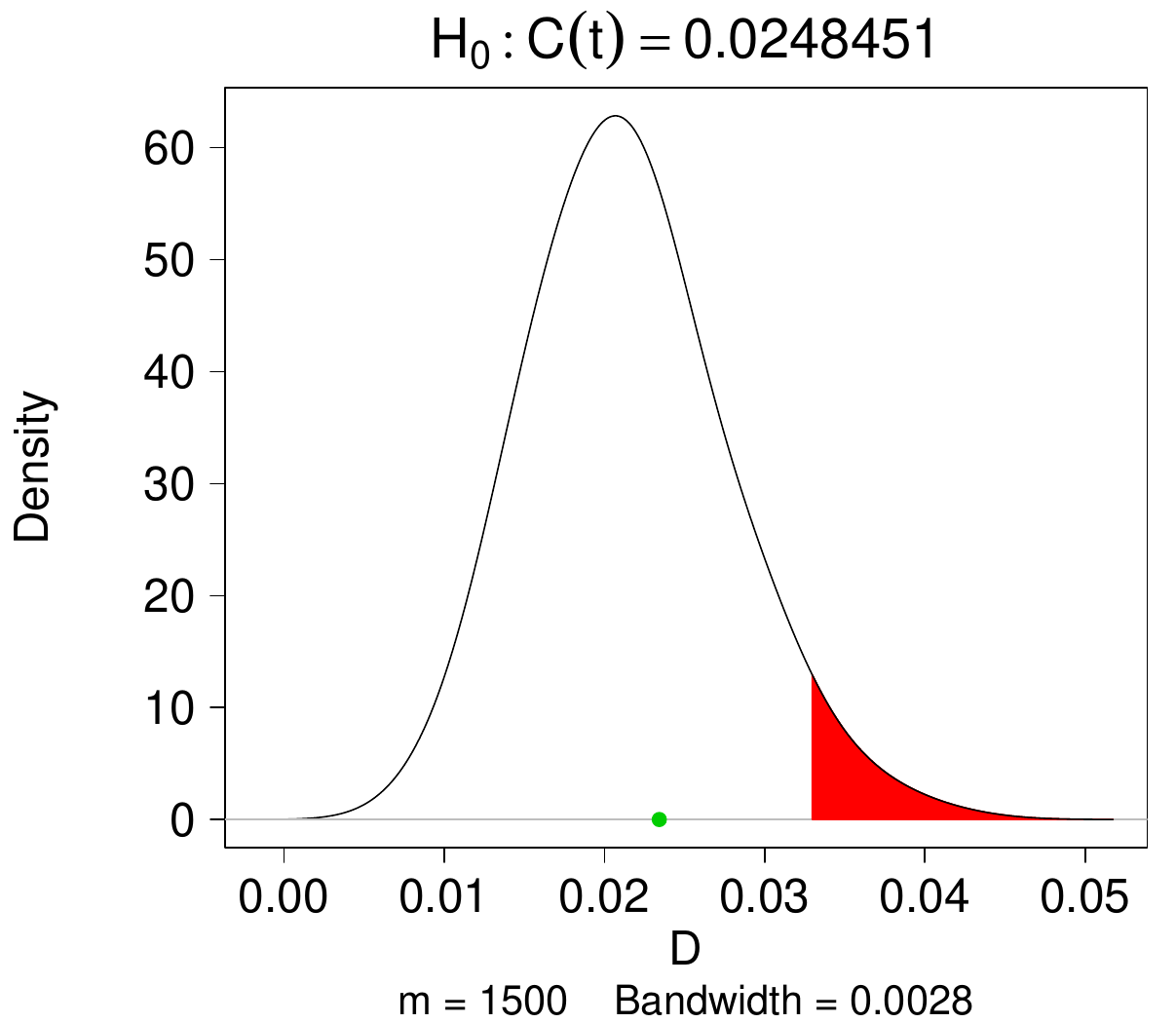}\\
\includegraphics[height=0.2\textheight]{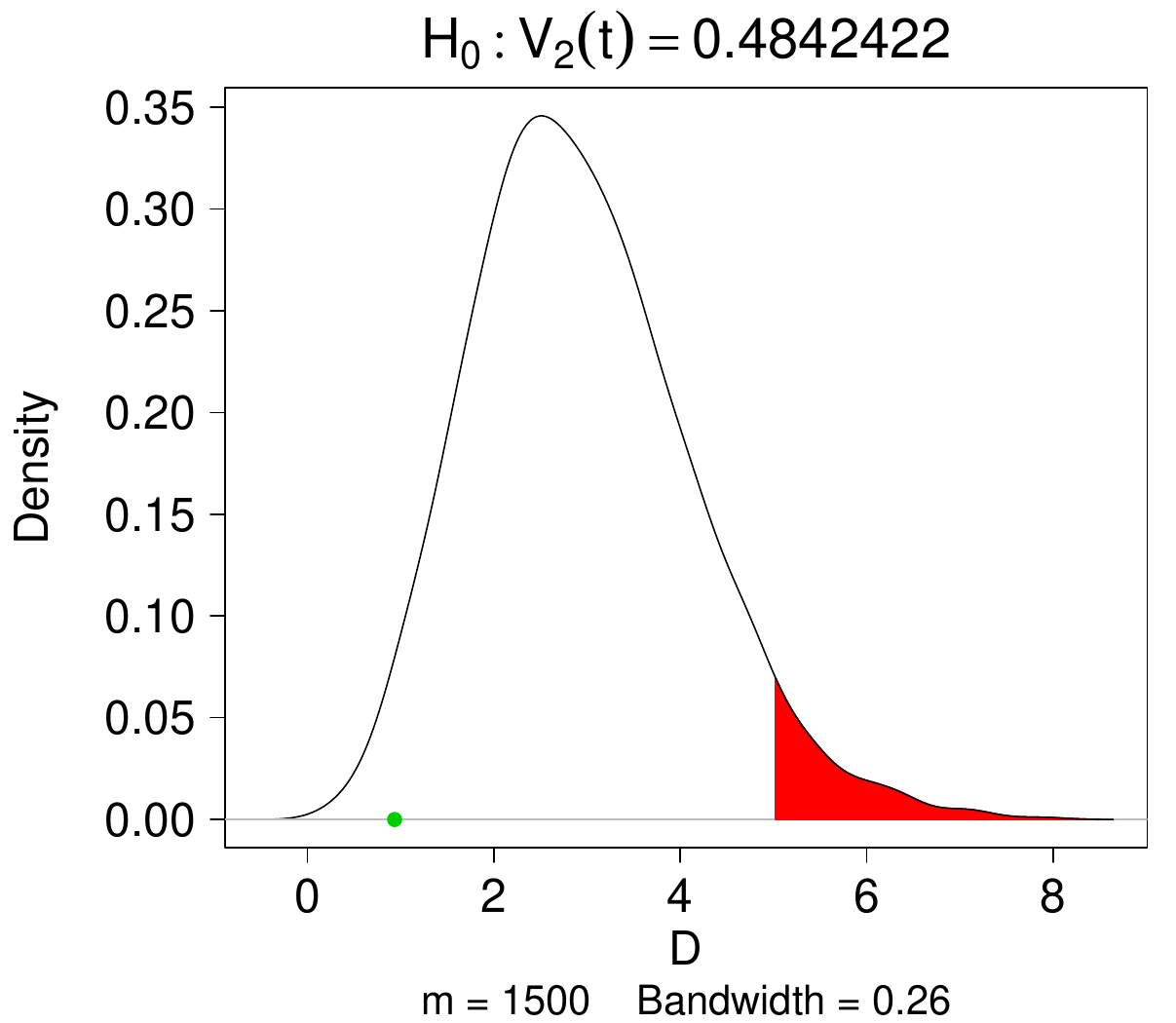} &
    \includegraphics[height=0.2\textheight]{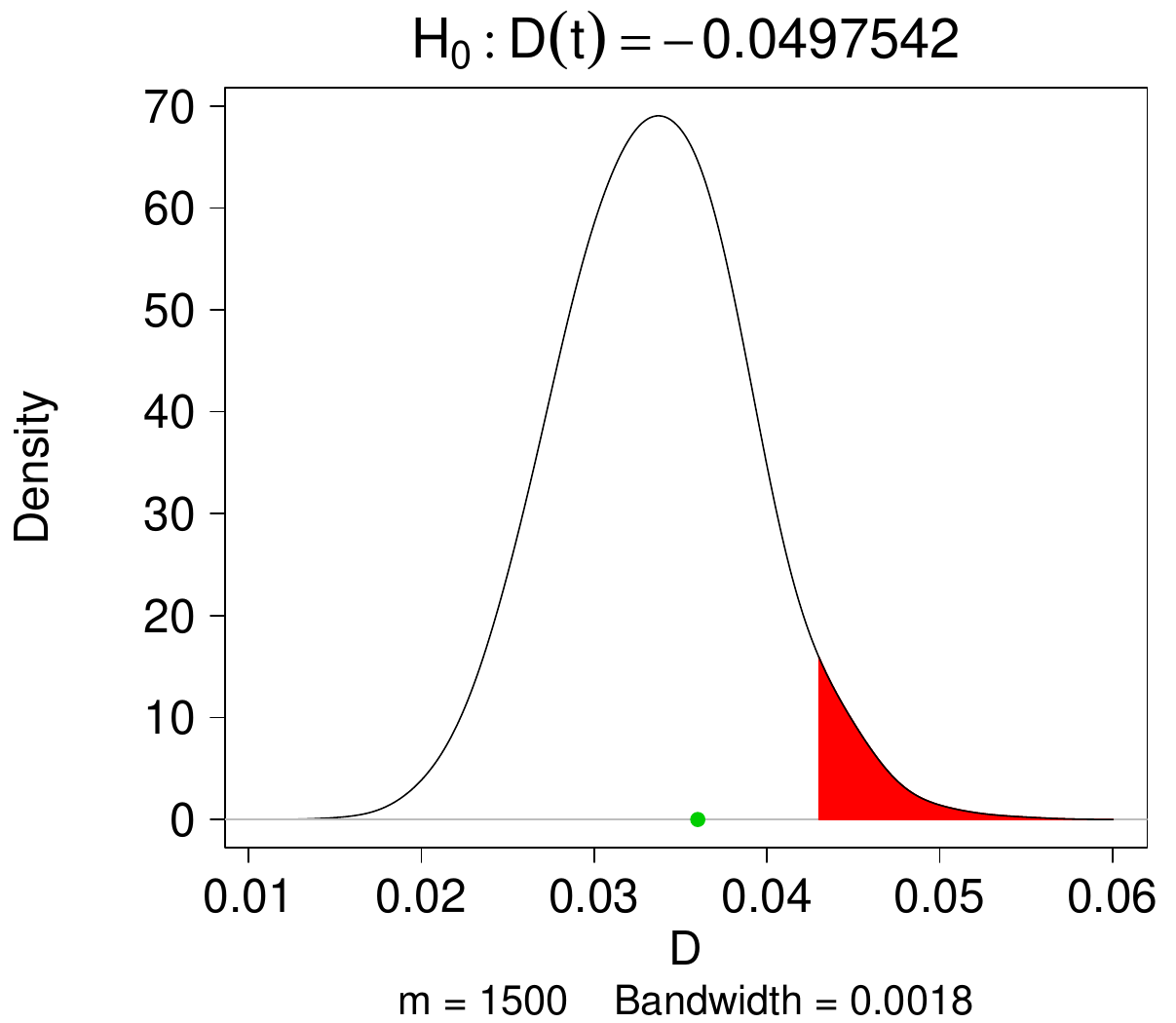}
    \end{tabular}
    \caption{Gaussian kernel density estimation of $D$-statistics for the tests associated to Case 2. Green points are the values of the D-statistics in our simulation experiment. In red the critical region with significance $0.05$ is shown.}
\label{kernelDensityCase2}
  \end{figure}

\end{itemize}

\section{Application to real data of tumor growth}

In this section we apply the stochastic process introduced in this paper to model experimental
data, obtained in mice, in order to study the effect of two treatments on ovarian cancer.
\smallskip

In particular, we analyze the effects of Carboplatin and Paclitaxel treatments on the growth of
OVA014HENp9 tumor from data of three experimental groups of 9, 8 and 8 mice. These data has been
provided by the Laboratory of Preclinical Investigation (LIP) that belongs to the Translational
Research Department of the Institute Curie, Paris. Carboplatin plus paclitaxel regimen remains the
standard chemotherapy for the initial treatment of ovarian cancer and it is less toxic and easier to
administrate compared to other drug combinations (Ozols et al. \cite{ClinicalOncology}).

The first group, ${\cal G}$, was a control (untreated); the second group, ${\cal G}_1$, was
treated with Carboplatin (66mg/kg/day the days 1 and 22); and the third group, ${\cal G}_2$,
received Carboplatin(idem)+Paclitaxel (12mg/kg/week over a period of six weeks). The relative
volume of tumor was measured at days 1, 4, 11, 16, 19, 31, 34, 38, 41, 53, 59 and 66. \medskip

\begin{figure}[h]
    \hfill{} \tabcolsep 2pt
    \begin{tabular}{ccc}
    \includegraphics[height=0.2\textheight]{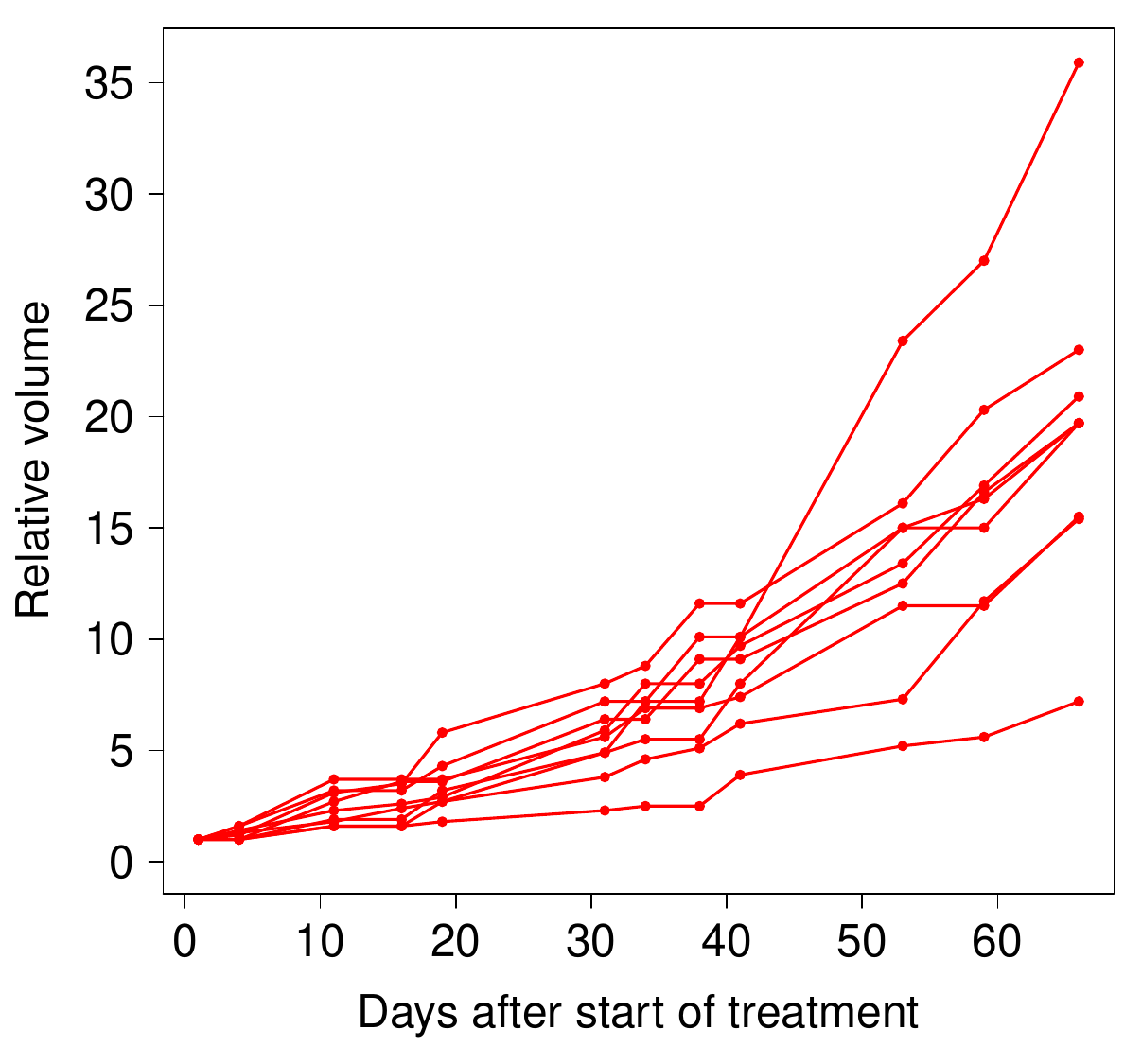} &
    \includegraphics[height=0.2\textheight]{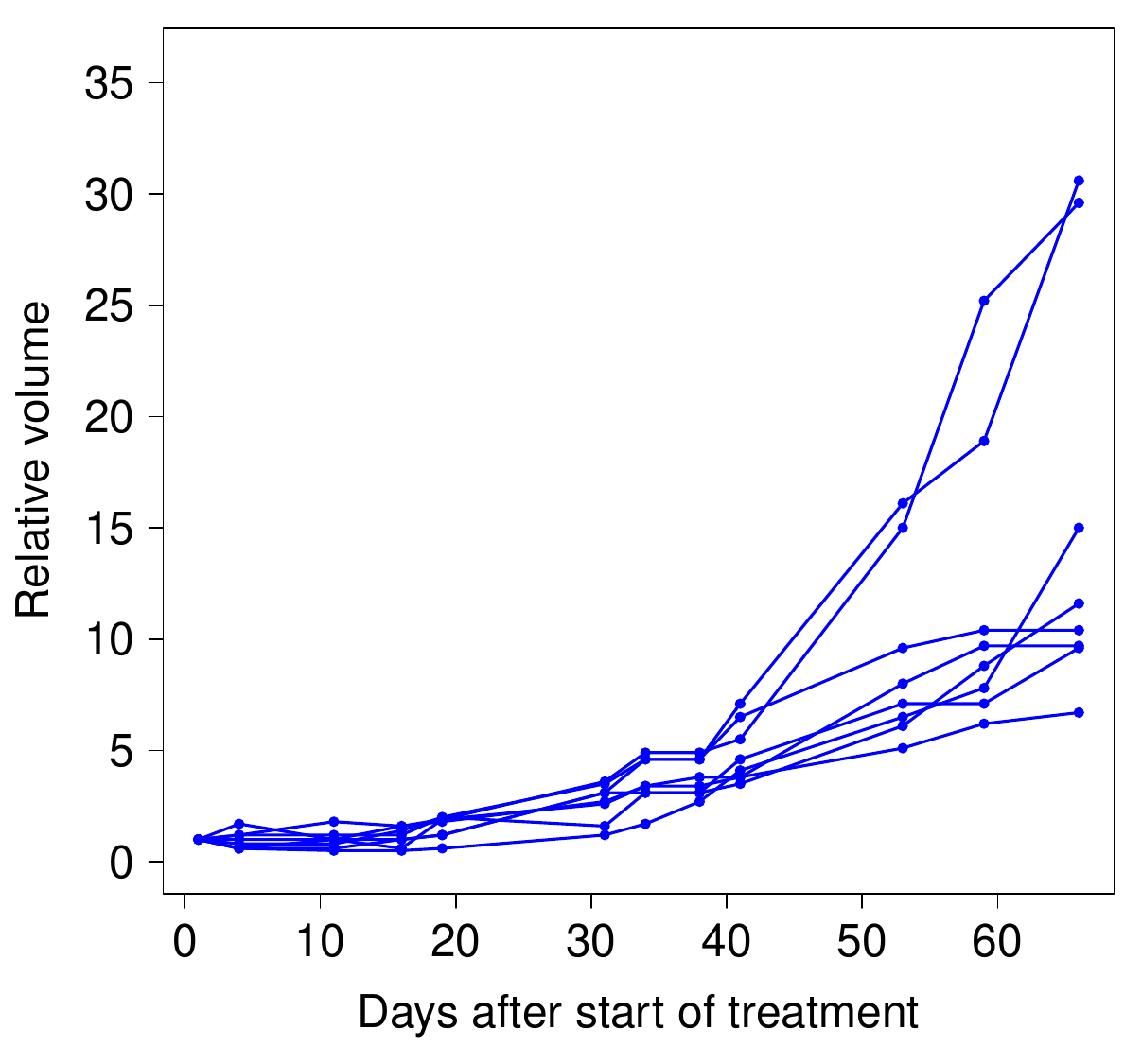} &
    \includegraphics[height=0.2\textheight]{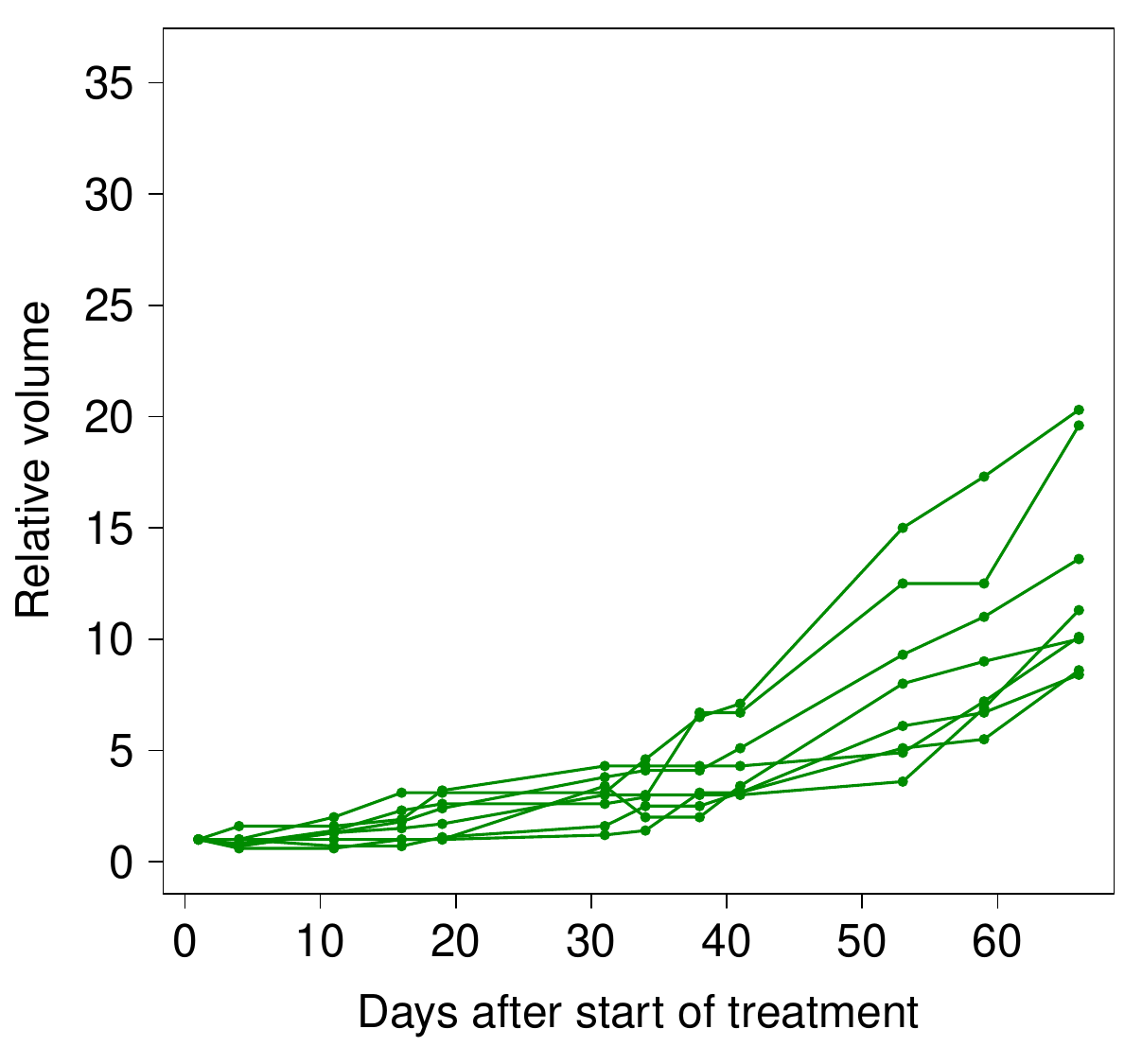} \\ {\scriptsize (a)} & \qquad \ {\scriptsize (b)} & \qquad \ {\scriptsize (c)} \\
    \end{tabular}
    \caption{Sample paths of relative volume of tumor in control group (a), and Carboplatin (b) and Carboplatin+Paclitaxel (c) treated groups.}
    \label{Paths-Exp930}
\end{figure}

\begin{figure}[h]
    \centering
    \begin{tabular}{cc}
    \includegraphics[height=0.2\textheight]{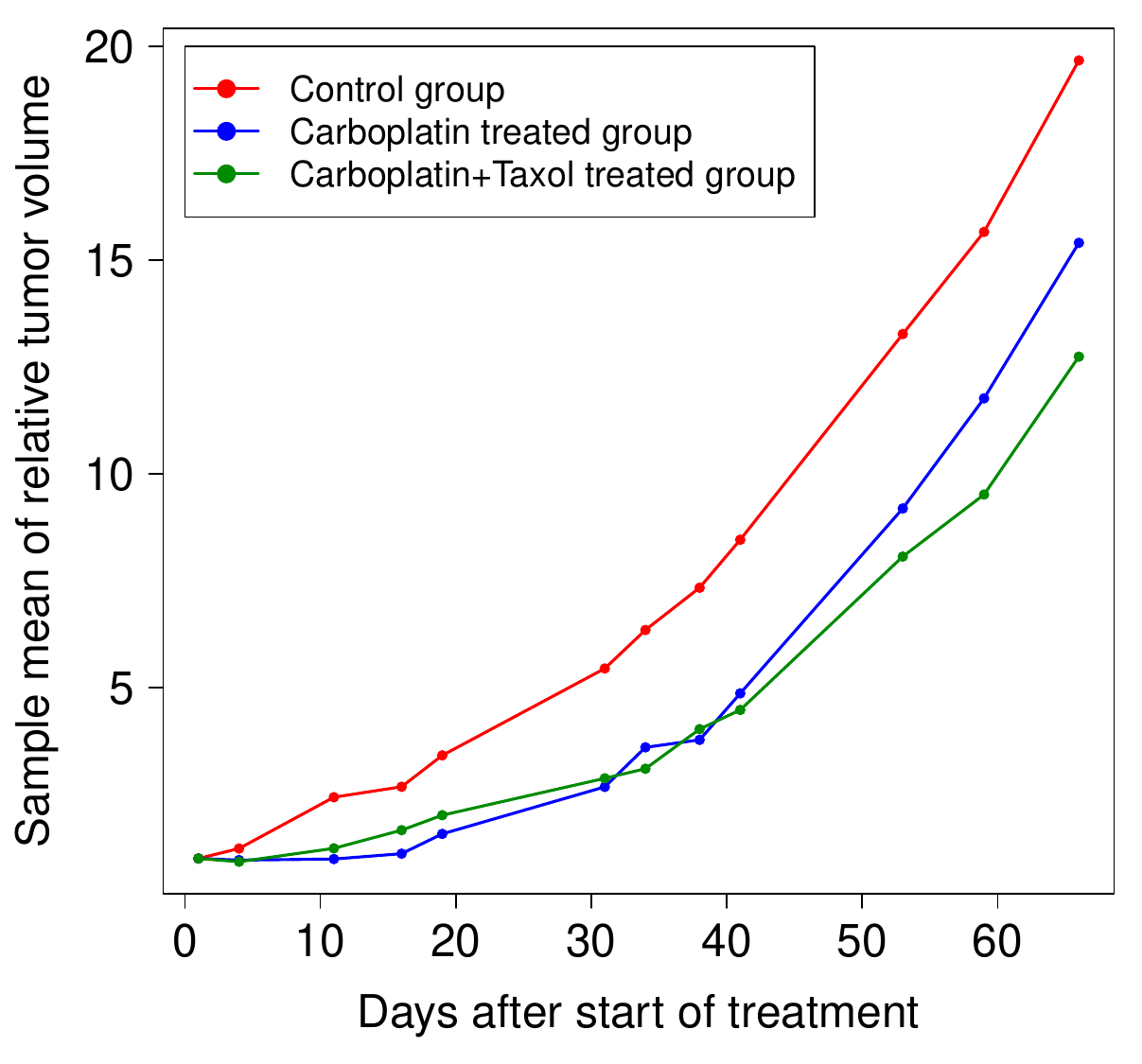} &
    \includegraphics[height=0.2\textheight]{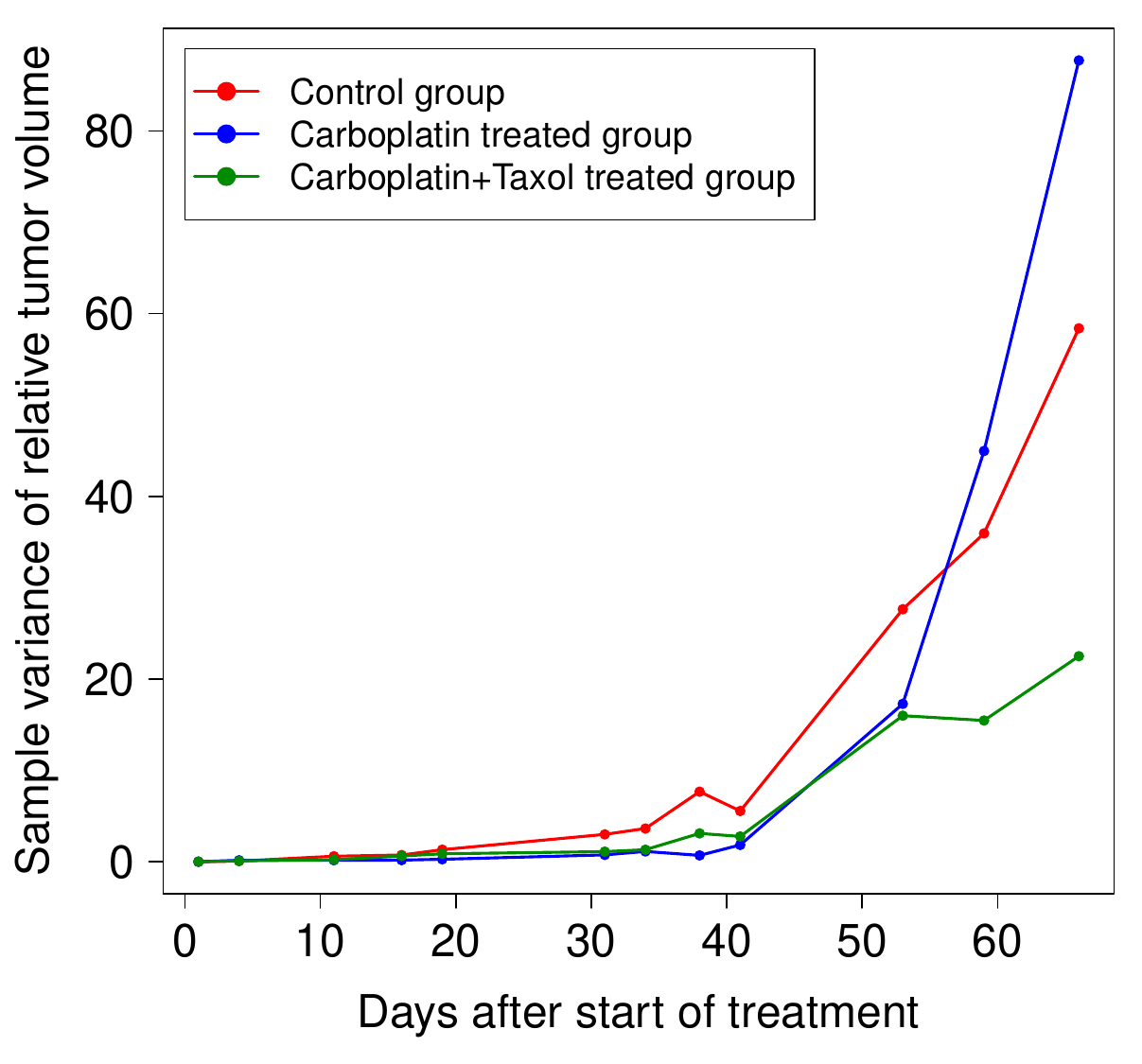}
    \end{tabular}
    \caption{Sample mean and variance of the relative tumor volume in control and treated groups.} \label{Mean-Var-Paths-Exp930}
\end{figure}

Figures \ref{Paths-Exp930} and \ref{Mean-Var-Paths-Exp930} show the sample paths and the sample
mean and variance, respectively, of the relative tumor volume for the three experimental groups as
a function of the days after starting the treatment. \medskip

The ML estimation of the parameters in control group provide $\widehat{\alpha} = 0.06964254$,
$\widehat{\beta} = 0.01238329$, $\widehat{\sigma} = 0.08964128$. \medskip

\begin{figure}[h]
    \centering
    \begin{tabular}{cc}
    \includegraphics[height=0.2\textheight]{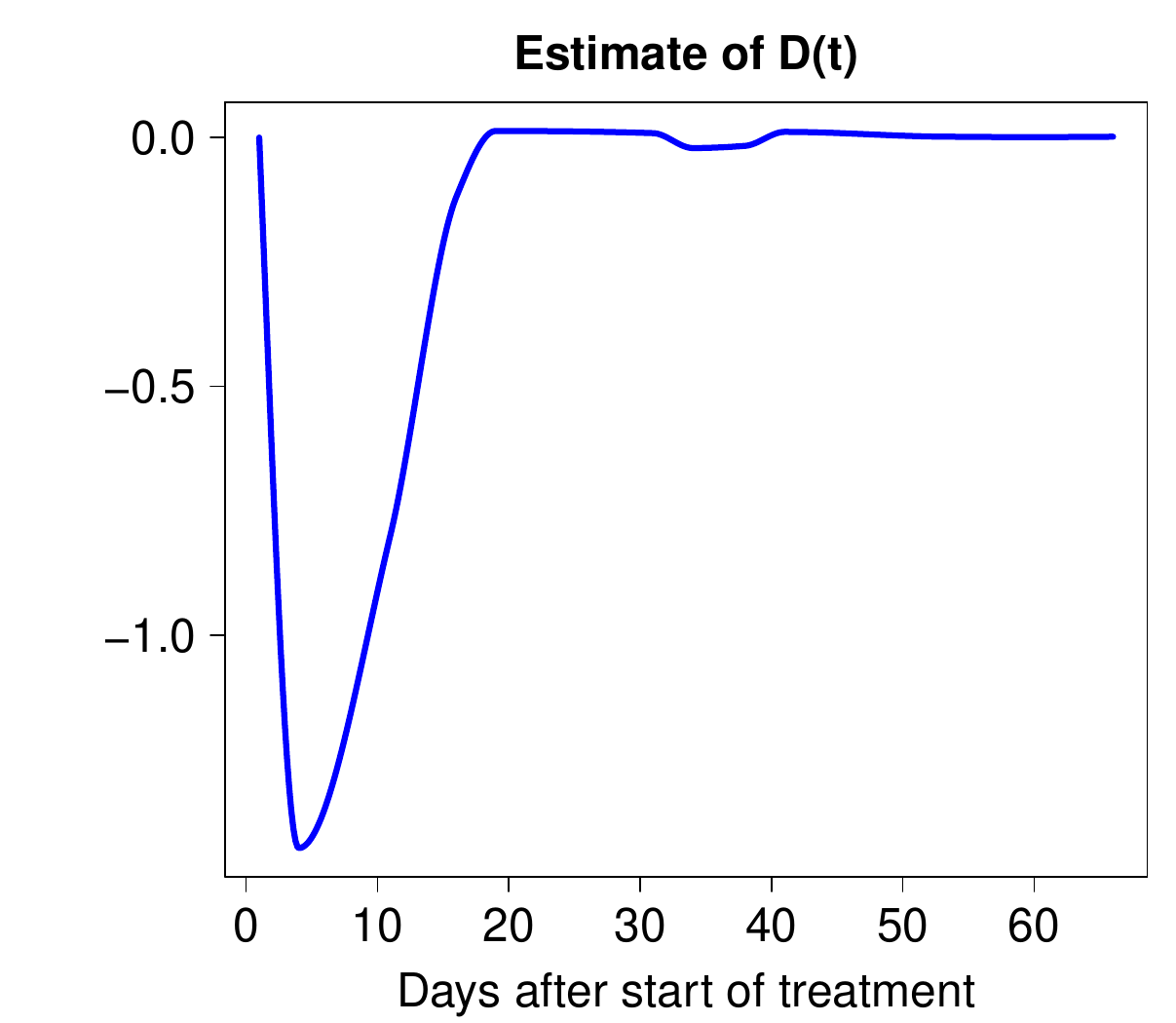} & \includegraphics[height=0.2\textheight]{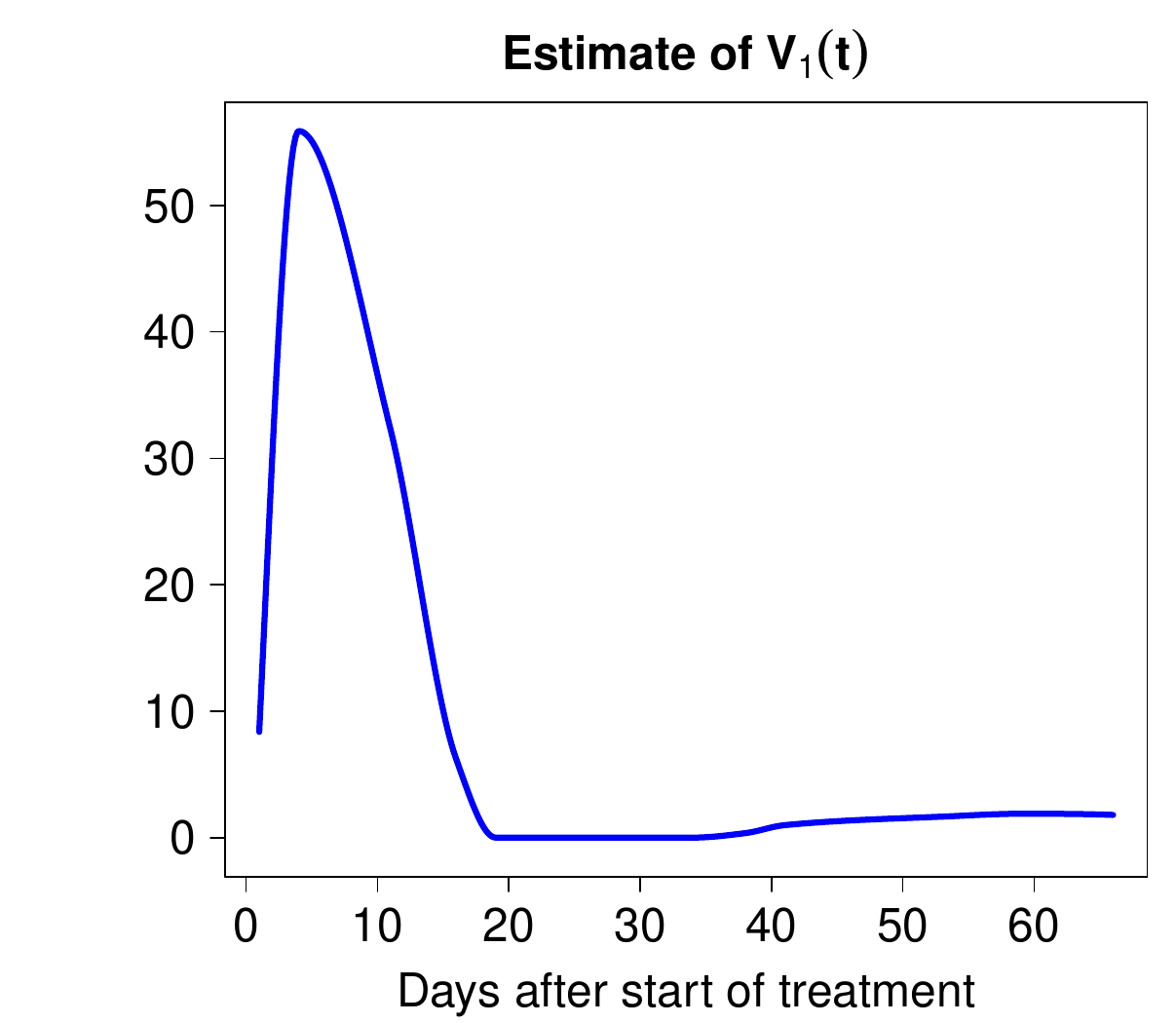} \\
    \includegraphics[height=0.2\textheight]{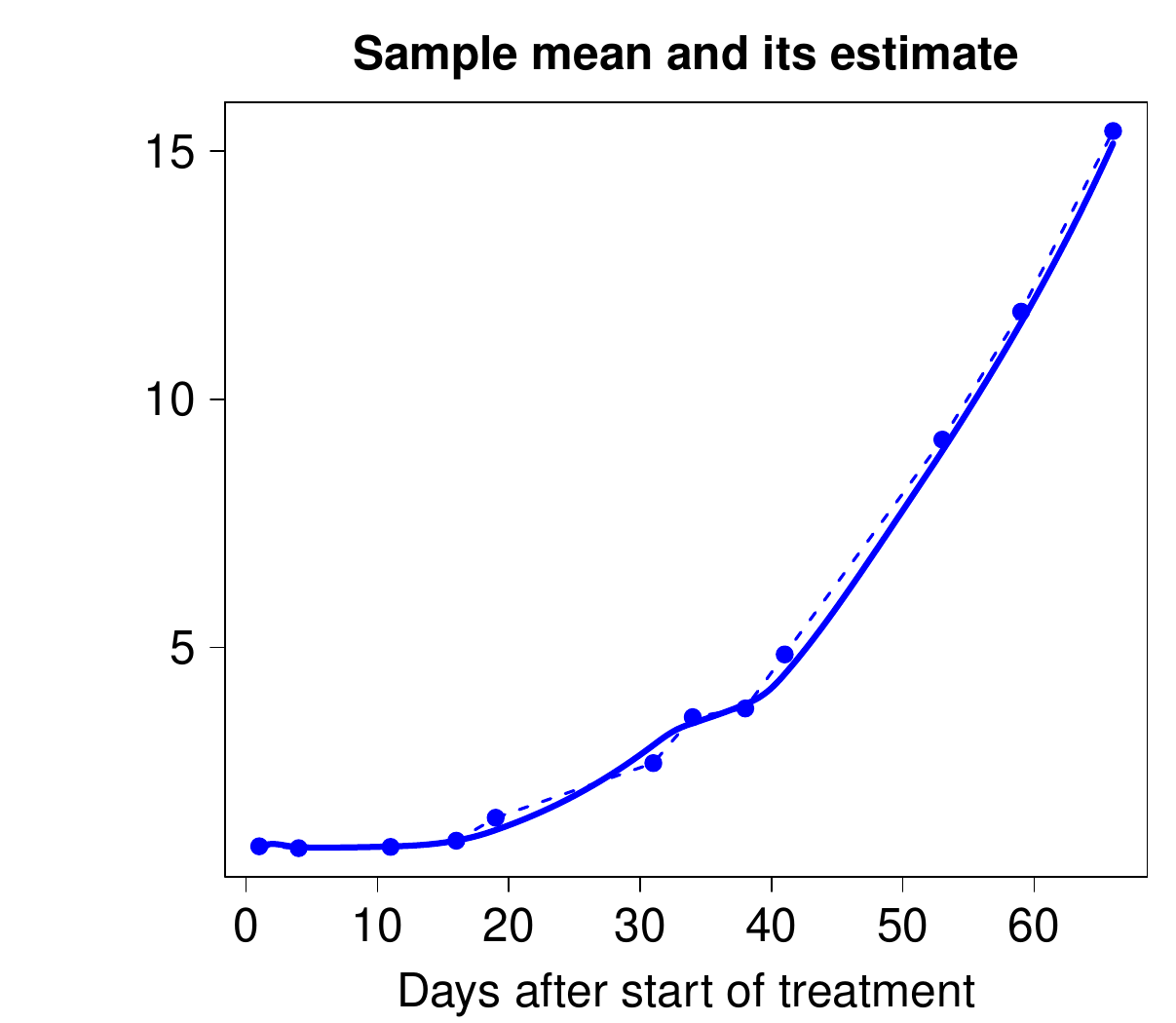} & \includegraphics[height=0.2\textheight]{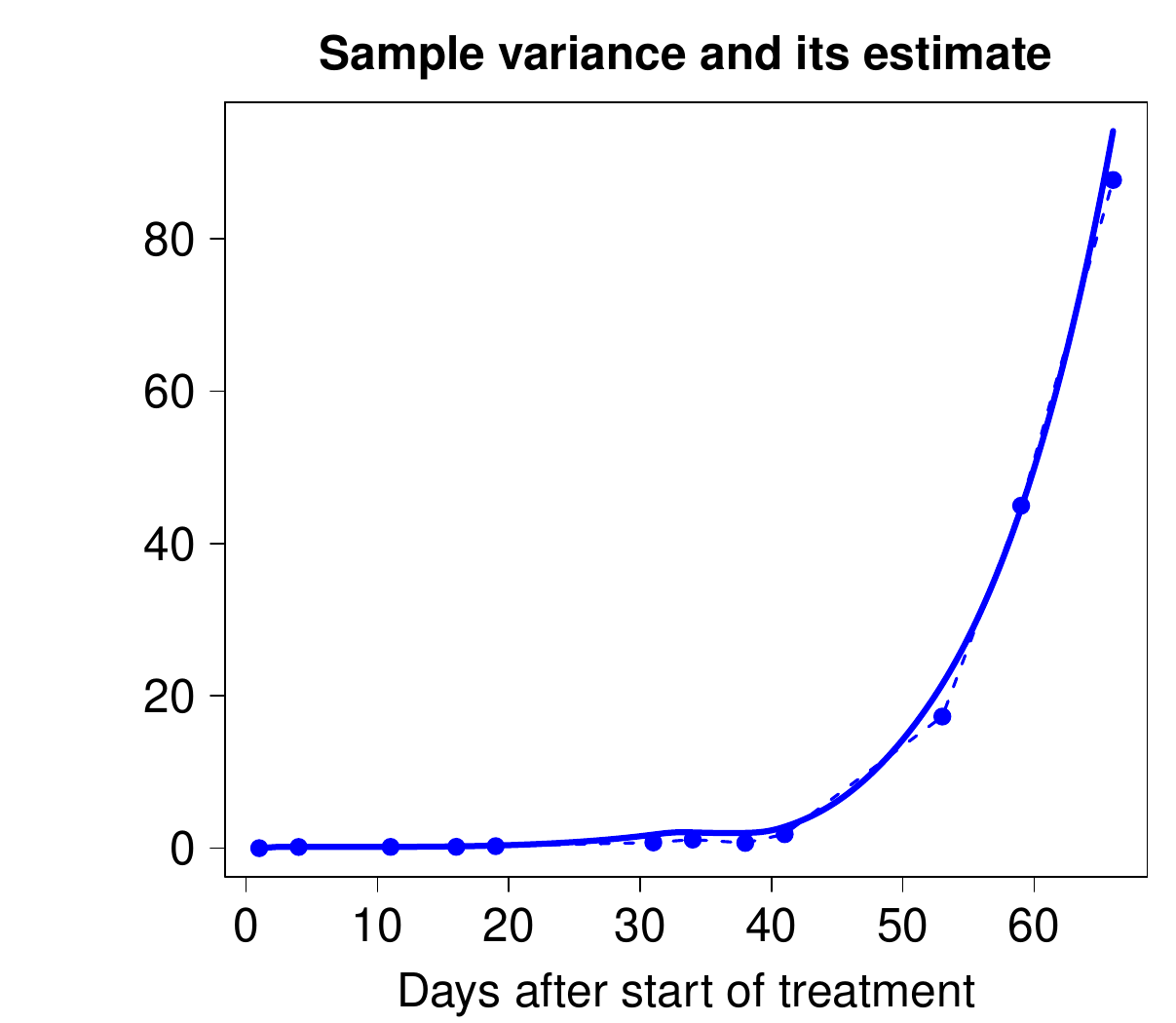}
    \end{tabular}
    \caption{Fit of model (\ref{case_b_X1}) in Carboplatin treated group (estimates in solid line).}
    \label{FitExp930-Carboplatin}
\end{figure}

Since the therapy with Carboplatin induces the death of cancer cells, we have adjusted the model
(\ref{case_b_X1}) to the data of treated group ${\cal G}_1$. Figure \ref{FitExp930-Carboplatin}
shows the estimates of the $D(t)$ and $V_1(t)$ functions as well as the fit of the sample means
and variances of data by using $E(\widehat{X}_1(t))$ and $Var(\widehat{X}_1(t))$, respectively.
\vspace{5pt}

\begin{figure}[h]
    \centering
    \begin{tabular}{cc}
      \includegraphics[height=0.2\textheight]{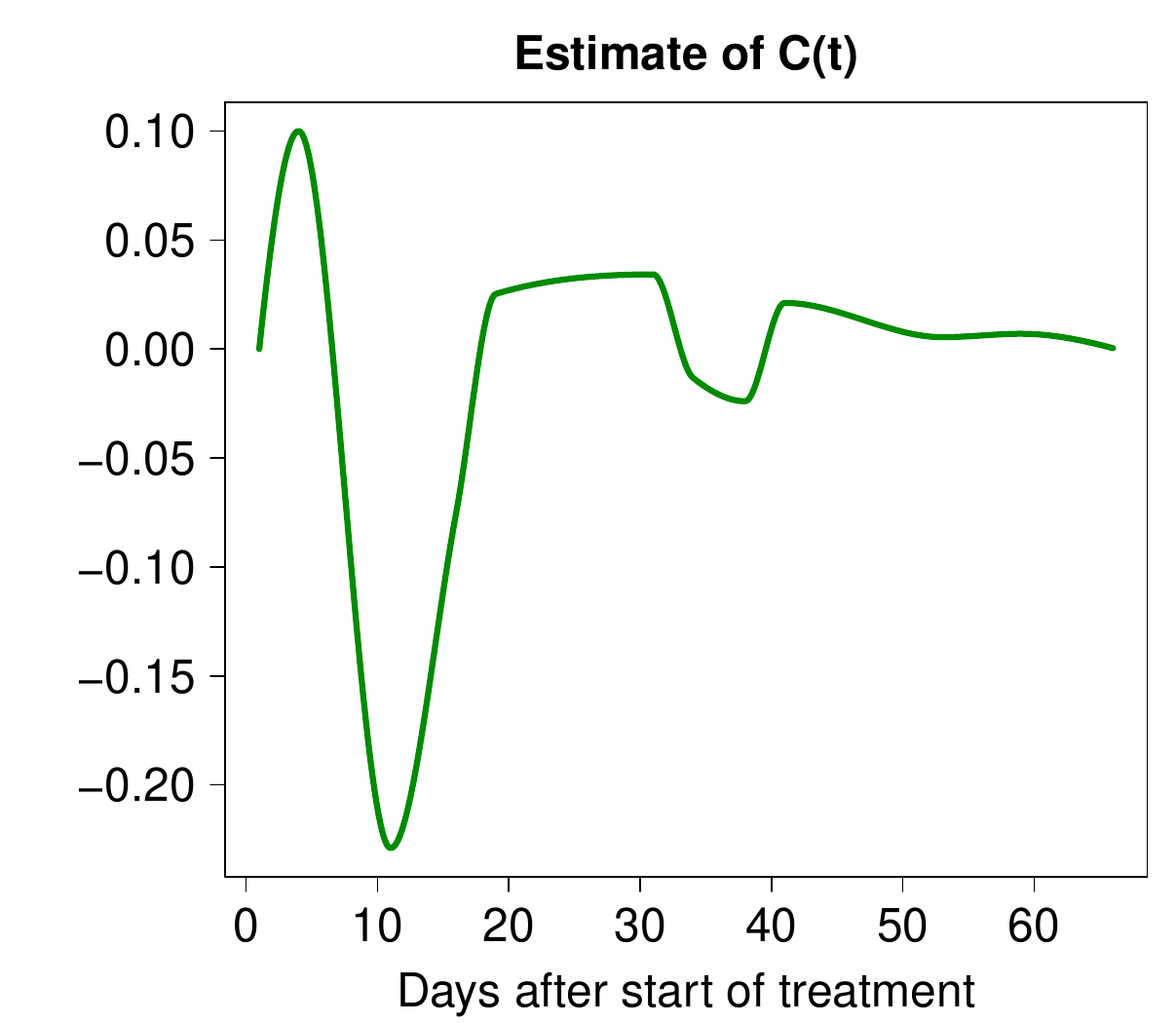} & \includegraphics[height=0.2\textheight]{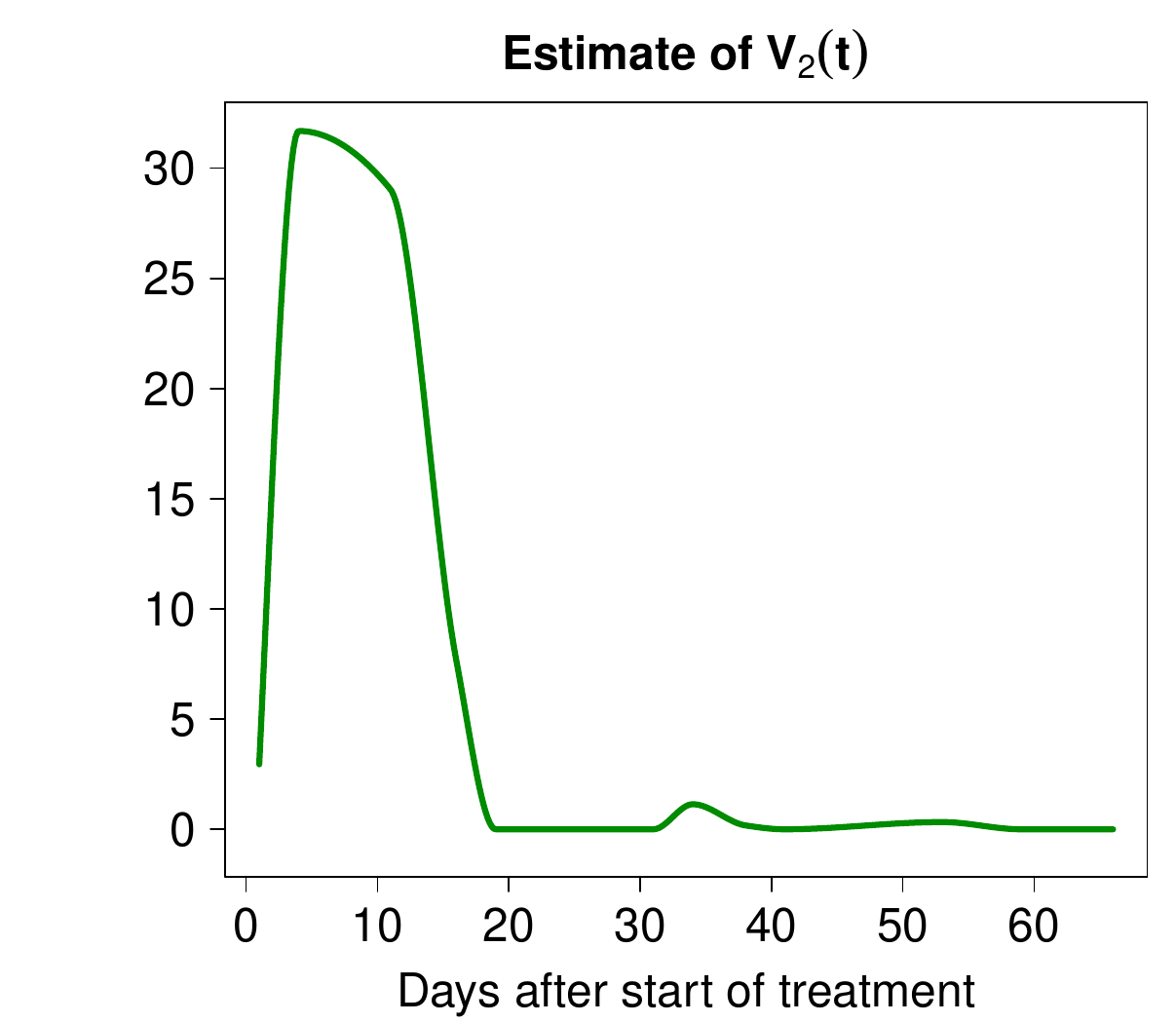} \\
      \includegraphics[height=0.2\textheight]{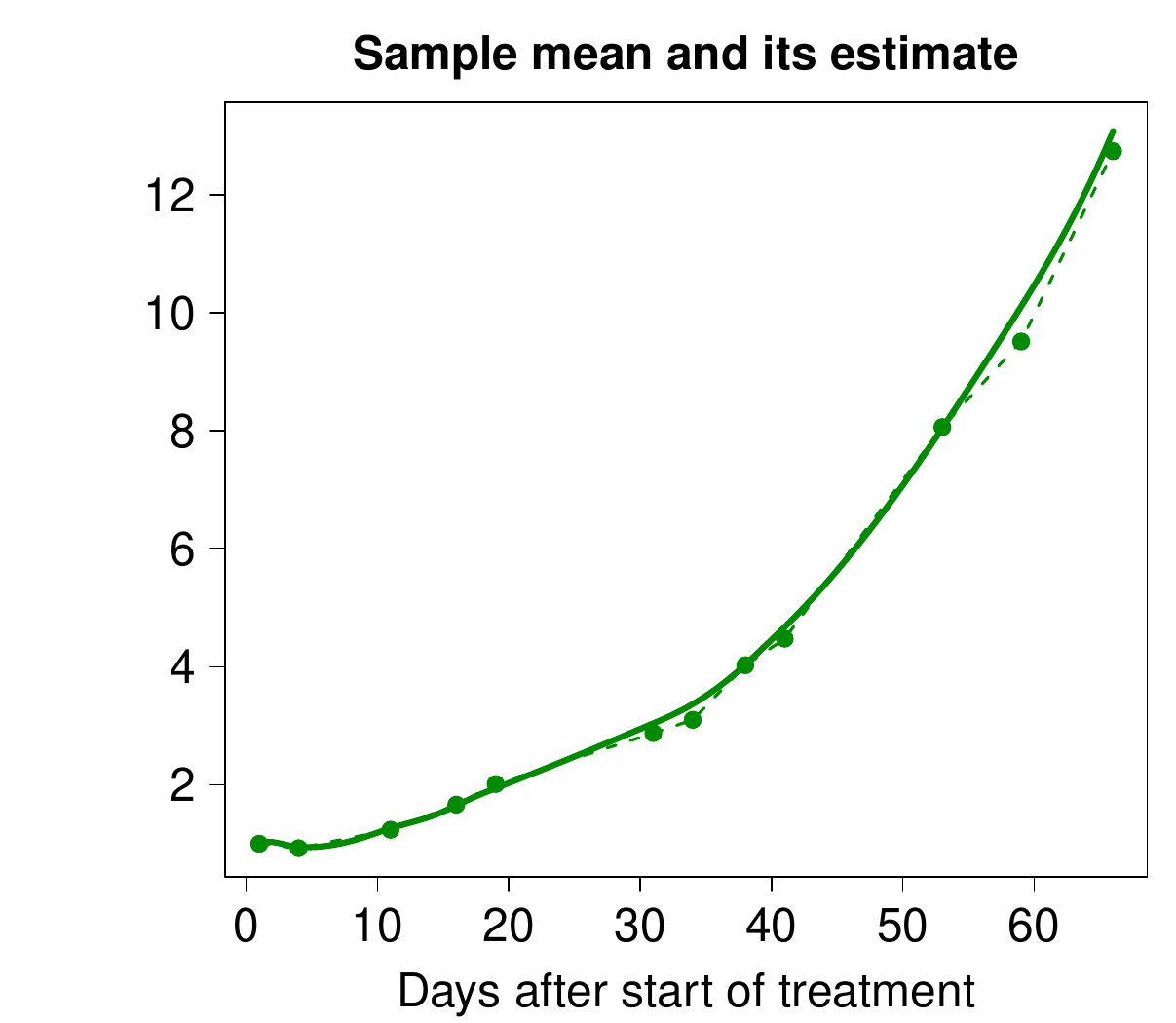} & \includegraphics[height=0.2\textheight]{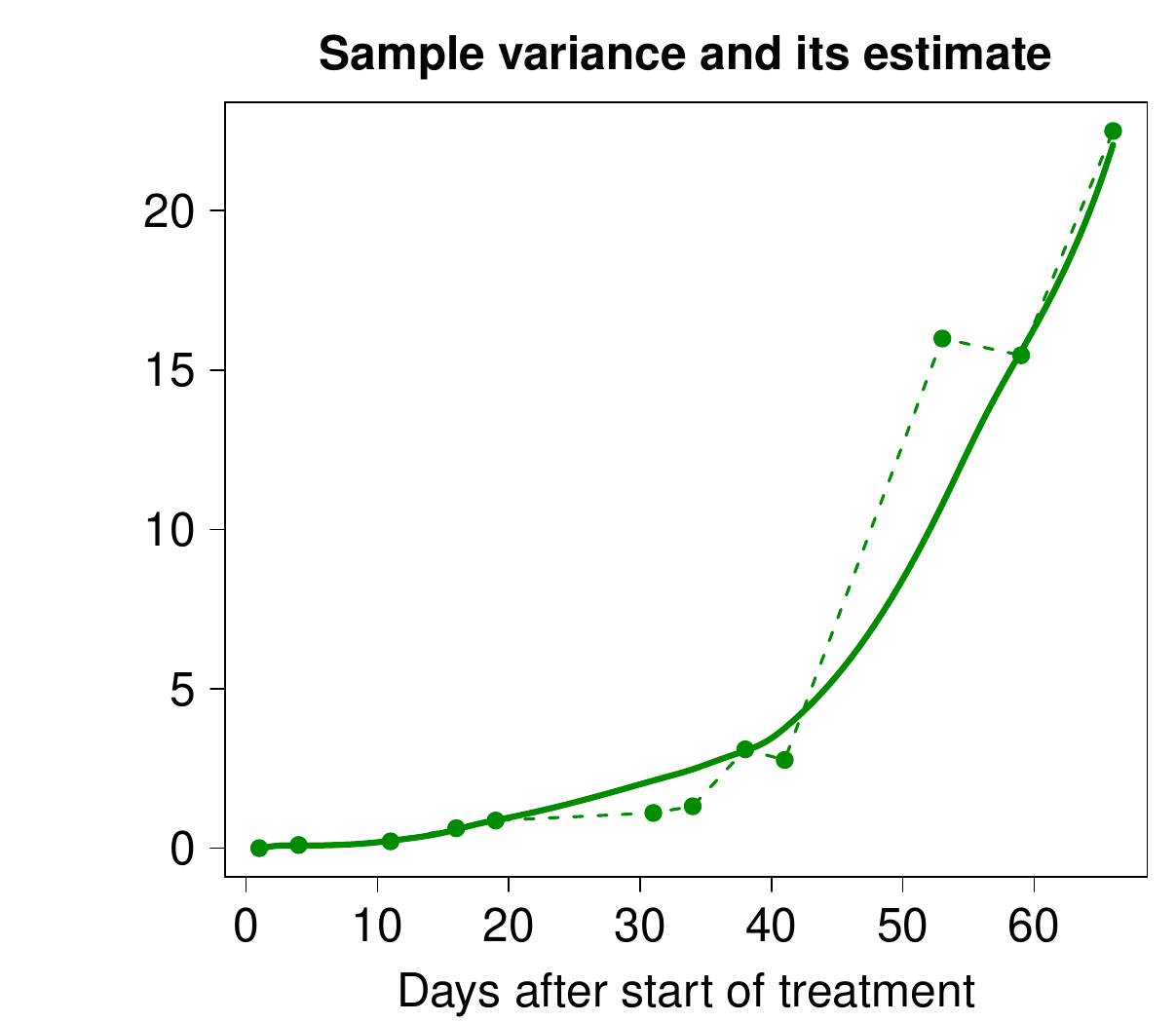}
    \end{tabular}
    \caption{Fit of model (\ref{case_a_b_X2}) in Carboplatin+Paclitaxel treated group (estimates in solid line).}
    \label{FitExp930-Carboplatin-Paclitaxel}
\end{figure}

In the same way, since the therapy with Paclitaxel is anti-proliferative, we have adjusted the
model (\ref{case_a_b_X2}) to the data of the treated group ${\cal G}_2$. Figure
\ref{FitExp930-Carboplatin-Paclitaxel} shows the estimates of the $C(t)$ and $V_2(t)$ functions as
well as the fit of the sample means and variances of data by using $E(\widehat{X}_2(t))$ and
$Var(\widehat{X}_2(t))$, respectively.

The results of the fitting function $D(t)$ (Figure \ref{FitExp930-Carboplatin}) show that the
Carboplatin treatment is effective in the first 15-20 days in which it present a negative peak,
then it becomes ineffective. The infinitesimal variance $V_1(t)$ seems to be greatly influenced by
the therapy when it is effective, after that the variability of the process restore to natural
constant values. Concerning the treated Carboplatin+Paclitaxel group ${\cal G}_2$ (Figure
\ref{FitExp930-Carboplatin-Paclitaxel}), we observe that the therapy is effective in the first
days of the treatment, then its effectivenes declines corresponding to a negative bump, followed
from values close to zero. The function $V_2(t)$ presents a similar behaviour with respect to
$V_1(t)$, although it shows a lower peak.

\vskip 5pt The estimated models in both treated groups provide a good fit of the sample means and
variances of the relative volume of tumor. Table \ref{MSE_Exp930} presents the mean squared errors between the sample mean and variance functions of the simulated process and the estimated ones, that is
$$
MeanMSE= \displaystyle\frac{1}{n}\sum_{j=1}^n (m_j - \hat{m}_j)^2, \qquad
VarMSE= \displaystyle\frac{1}{n}\sum_{j=1}^n (\sigma^2_j - \hat{\sigma}^2_j)^2
$$
where $(m_j,\sigma^2_j)$ are the values of the sample mean and variance functions at $t_j$, $j=1,\dots,n$ whereas $(\hat{m_j},\hat{\sigma}^2_j)$ are the estimated ones.
 \medskip

\begin{table}[h]
  \caption{Mean squared errors for the fits of the sample means and variances in the
  treated groups.} \label{MSE_Exp930} \smallskip
  \centering
  \begin{tabular}{lcc}
    \toprule
    MSEs & Carboplatin & Carboplatin+Paclitaxel \\ \midrule
    MeanMSEs & $0.0445$ & $0.0595$ \\
    VarMSEs  & $5.3507$ & $2.5464$ \\
    \bottomrule
  \end{tabular}
\end{table}

Taking into account the estimates of functions $D(t)$, $V_1(t)$, $C(t)$ and $V_2(t)$, it seems
reasonable to test only if $C(t)$ is constant. However, we have tested, in a concatenated form, if
each of the functions could be constant, as in the simulation study in Section 4.1. The
corresponding constants to be included in the null hypothesis are estimated by ML as described in
Section 4. \medskip

For group ${\cal{G}}_1$, $H_0: V_1(t)=9.612252$ was tested first. The associated $p$-value was
$0.04$ and therefore we reject that $V_1(t)$ be constant. Then, we have tested $H_0: D(t)
=d$, where $d=0.001296$ is determined by ML considering $V_1(t)
=\widehat{V}_1(t)$. The test produced a $p$-value of $0.03$ and we also reject that $D(t)$ be
constant. \medskip

Next, for group ${\cal{G}}_2$, $H_0: V_2(t)=8.097848$ was tested (making use of $D(t) =
\widehat{D}(t)$). This hypothesis is rejected with a $p$-value of $0.01$. Finally, we have tested
$H_0: C(t) = c$, where $c=0.023195$ is determined by ML considering $V_2(t)
=\widehat{V}_2(t)$. The resulting $p$-value was $0.03$ and we must also reject that $C(t)$ be
constant. \medskip

Thus, we can conclude that, in group ${\cal{G}}_1$, the effect of Carboplatin on the death of
cancer cells and the infinitesimal variability of relative volume of tumor is time-dependent. In a
similar way, in group ${\cal{G}}_2$, the same comment can be done about the combined effect of
Carboplatin and Paclitaxel on the growth and death of cancer cells, as well as on the
infinitesimal variability of relative volume of tumor. Therefore, based on the tests carried out, we can conclude that the therapies applied are time dependent, so the models \eqref{case_b_X1} and \eqref{case_a_b_X2} seem to be appropriated to describe the combined effect of the two therapies. \medskip

\section*{Appendix A. Maximum likelihood estimates of the parameters of the process}

The objective of this appendix is to provide the ML estimation of the parameters of the process in
model (\ref{SDE}) for known $C(t)$, $D(t)$ and $V(t)$ functions. In addition we provide the ML
estimation of each parameter for known values of the rest ones, which will be useful when
establishing the null hypotheses $H_0: H(t) = h$, with $H(t)$ any of the functions in model
(\ref{SDE}).

\vskip 8pt From \eqref{DistrX1_Xn} in Section 2.1, the transition pdf of the process can be
obtained, resulting in
\begin{equation}\label{tpdf}
X(t)|X(s)=y \sim \Lambda_1\left[\bar{k}(t|s)\ln\,y+\theta(t|s),\sigma^2\Omega(t|s)\right]
\end{equation}
where
$$
\begin{array}{l}
  \theta(t|\tau) =\displaystyle\int_{\tau}^{t} \left( \alpha-C(s)-\frac{\sigma^2}{2}V(s) \right) \bar{k}(t|s)\,ds, \smallskip \\
  \Omega(t|\tau) =\displaystyle\int_{\tau}^{t}\bar{k}^2(t|s)V(s)\,ds
\end{array}
$$

\noindent Let us consider a discrete sampling $\{x_{ij}, \, i=1,\ldots,d; \,j=0,\ldots,n_i-1\}$ of
the process based on $d$ sample paths at times $t_{ij}$, $(i = 1,\ldots, d, \ j=0, \ldots, n_i-1)$
with $t_{i0} = t_0$ and $x_{i0} = x_0$, $i = 1,\ldots,d$. Denote by
$\mathbf{X}=\left(\mathbf{X}_1^T|\cdots|\mathbf{X}_d^T\right)^T$, where
$\mathbf{X}_i=(X_{i0},\ldots,X_{i,n_i-1})^T$, $i=1,\ldots,d$, with $X_{ij}=X(t_{ij})$,
$j=0,\ldots,n_i-1$.

\vskip 10pt By taking $X(t_0)$ a degenerate random variable, i.e. $P[X(t_0) = x_0] = 1$, from
\eqref{tpdf}, the probability density function of $\mathbf{X}$ is
$$
f_\mathbf{X}(\mathbf{x}) = \prod_{i=1}^{d} \prod_{j=1}^{n_i-1} \displaystyle\frac{\exp \Bigg(
\negthinspace -\displaystyle\frac{ \big[ \delta_{\beta}^{ij} - \theta_{\bm\xi}^{ij}
\big]^2}{2\sigma^2\Omega_\beta^{ij}} \Bigg)} {x_{ij}\sigma\sqrt{2\pi\Omega_\beta^{ij}}}
$$
where $\delta_{\beta}^{ij} = \ln\,x_{ij}-\bar{k}_\beta^{ij}\ln\,x_{i,j-1}$, \,
$\theta_{\bm\xi}^{ij}=\theta(t_{ij}|t_{i,j-1})$ \, and \,
$\Omega_\beta^{ij}=\Omega(t_{ij}|t_{i,j-1})$, with $\bar{k}_\beta^{ij}=\bar{k}(t_{ij}|t_{i,j-1}) =
exp \left( -\int_{t_{i,j-1}}^{t_{ij}} (\beta - D(s)) ds \right)$, $i=1,2,\ldots,d$,
$j=1,\ldots,n_i-1$, and $\bm\xi=(\alpha,\beta,\sigma^2)^T$.

\vskip 10pt Then, for a fixed value $\mathbf{x}$ of the sample and known $C(t)$, $D(t)$ and $V(t)$
functions, the log-likelihood function is
\begin{equation}
\label{Vero2} L_{\mathbf{x}}(\bm\xi) = -\frac{n\ln(2\pi)}{2} - \frac{n\ln\sigma^2}{2} -
\frac{Z_\beta+\Phi_{\bm\xi}-2\\
\Gamma_{\bm\xi}}{2\sigma^2} - \dfrac{1}{2} \Upsilon_\beta -
\displaystyle \sum_{i=1}^d\sum_{j=1}^{n_i-1} ln \, x_{ij}
\end{equation}
where $n=\displaystyle\sum_{i=1}^d (n_i-1)$, $Z_\beta = \displaystyle
\sum_{i=1}^d\sum_{j=1}^{n_i-1} \dfrac{(\delta_{\beta}^{ij})^2}{\Omega_\beta^{ij}}$, \,
$\Phi_{\bm\xi} = \displaystyle \sum_{i=1}^d\sum_{j=1}^{n_i-1}
\dfrac{(\theta_{\xi}^{ij})^2}{\Omega_\beta^{ij}}$, \\
$\Gamma_{\bm\xi} = \displaystyle
\sum_{i=1}^d\sum_{j=1}^{n_i-1} \dfrac{\delta_{\beta}^{ij} \theta_{\xi}^{ij}}{\Omega_\beta^{ij}}$
and $\Upsilon_\beta = \displaystyle \sum_{i=1}^d\sum_{j=1}^{n_i-1} ln \, \Omega_\beta^{ij}$.

\vskip 10pt \noindent In order to obtain the ML estimate of $\alpha$, $\beta$ and $\sigma^2$ we
denote
$$
\Psi_{\beta,ij}^{l,m,p,q} = \displaystyle{\int_{t_{i,j-1}}^{t_{ij}}
(t_{ij}-s)^l\left(C(s)\right)^m\left(V(s)\right)^p\left(\bar{k}(t_{ij}|s)\right)^q\,ds,}
$$
from which we deduce:
$$
\begin{array}{l}
\dfrac{\partial\Psi_{\beta,ij}^{l,m,p,q}}{\partial\beta}=-q\Psi_{\beta,ij}^{l+1,m,p,q} \smallskip \\
\theta_{\bm\xi}^{ij} = \alpha \Psi_{\beta,ij}^{0,0,0,1} -
\Psi_{\beta,ij}^{0,1,0,1} - \dfrac{\sigma^2}{2} \Psi_{\beta,ij}^{0,0,1,1} \medskip \\
\Omega_\beta^{ij}=\Psi_{\beta,ij}^{0,0,1,2}
\end{array}
$$

\noindent The likelihood equations are:
$$
\begin{array}{l}
\dfrac{\partial L_{\mathbf{x}}}{\partial \alpha} = - \dfrac{1}{2\sigma^2} \left( \dfrac{\partial
\Phi_{\bm\xi}}{\partial \alpha} - 2\dfrac{\partial \Gamma_{\bm\xi}}{\partial \alpha} \right) = 0
\medskip \\ \dfrac{\partial L_{\mathbf{x}}}{\partial \beta} = - \dfrac{1}{2\sigma^2} \left(
\dfrac{\partial Z_\beta}{\partial \beta} + \dfrac{\partial \Phi_{\bm\xi}}{\partial \beta} -
2\dfrac{\partial \Gamma_{\bm\xi}}{\partial \beta} \right) + \dfrac{1}{2} \dfrac{\partial
\Upsilon_\beta}{\partial \beta}  = 0 \medskip \\ \dfrac{\partial L_{\mathbf{x}}}{\partial
\sigma^2} = - \dfrac{1}{2\sigma^2} \left( n - \dfrac{1}{\sigma^2} (Z_\beta + \Phi_{\bm\xi} -
2\Gamma_{\bm\xi}) + \dfrac{\partial \Phi_{\bm\xi}}{\partial \sigma^2}
  - 2\dfrac{\partial \Gamma_{\bm\xi}}{\partial \sigma^2} \right) = 0
\end{array}
$$
or equivalently, after calculus,
\begin{eqnarray}
2\alpha\,X_1^{\beta} - 2X_2^{\beta} - \sigma^2X_3^{\beta} - 2 X_4^{\beta} = 0 & \label{vero-alpha} \\
X_5^\beta - X_6^\beta - X_7^\beta - X_8^\beta - X_9^\beta + 2 X_{10}^\beta + X_{11}^\beta + \frac{\sigma^2}{2} X_{12}^{\beta} = 0 & \label{vero-beta} \\
\frac{\sigma^4}{4} X_{13}^\beta + n\sigma^2 - (Z_\beta + \alpha^2\,X_1^\beta - 2 \alpha\,X_2^\beta
- 2\alpha\,X_4^\beta + X_{14}^\beta + 2X_{15}^\beta) = 0 & \label{vero-sigma2}
\end{eqnarray}
where
$$
  \begin{array}{ll}
  X_1^\beta=\displaystyle \sum_{i=1}^d\sum_{j=1}^{n_i-1} \frac{\left(\Psi_{\beta,ij}^{0,0,0,1}\right)^2}{\Psi_{\beta,ij}^{0,0,1,2} }, &
  X_2^\beta=\displaystyle \sum_{i=1}^d\sum_{j=1}^{n_i-1} \frac{\Psi_{\beta,ij}^{0,1,0,1}\Psi_{\beta,ij}^{0,0,0,1}}{\Psi_{\beta,ij}^{0,0,1,2}
  }, \\ \\
  X_3^\beta=\displaystyle \sum_{i=1}^d\sum_{j=1}^{n_i-1} \frac{\Psi_{\beta,ij}^{0,0,1,1}\Psi_{\beta,ij}^{0,0,0,1}}{\Psi_{\beta,ij}^{0,0,1,2} }, &
  X_4^\beta=\displaystyle \sum_{i=1}^d\sum_{j=1}^{n_i-1} \frac{\delta_\beta^{ij}\Psi_{\beta,ij}^{0,0,0,1}}{\Psi_{\beta,ij}^{0,0,1,2}}, \\ \\
  X_5^\beta=\displaystyle \sum_{i=1}^d\sum_{j=1}^{n_i-1} \frac{\bar{k}_\beta^{ij}\Delta_{ij}\theta_{\bm\xi}^{ij}\ln\,x_{ij}}{\Psi_{\beta,ij}^{0,0,1,2}}, &
        X_6^\beta=\displaystyle \sum_{i=1}^d\sum_{j=1}^{n_i-1} \frac{\theta_{\bm\xi}^{ij} \varphi_{\bm\xi}^{ij}}{\Psi_{\beta,ij}^{0,0,1,2}},
        \\ \\
        X_7^\beta=\displaystyle \sum_{i=1}^d\sum_{j=1}^{n_i-1} \frac{\left(\theta_{\bm\xi}^{ij}\right)^2\Psi_{\beta,ij}^{1,0,1,2}}{\left(\Psi_{\beta,ij}^{0,0,1,2}\right)^{2}},
        & X_8^\beta=\displaystyle \sum_{i=1}^d\sum_{j=1}^{n_i-1} \frac{\left(\delta_\beta^{ij}\right)^2\Psi_{\beta,ij}^{1,0,1,2}}{\left(\Psi_{\beta,ij}^{0,0,1,2}\right)^2},
        \\ \\
        X_9^\beta=\displaystyle \sum_{i=1}^d\sum_{j=1}^{n_i-1} \frac{\delta_\beta^{ij}\bar{k}_\beta^{ij}\Delta_{ij}\ln\,x_{ij}}{\Psi_{\beta,ij}^{0,0,1,2}},
        & X_{10}^\beta=\displaystyle \sum_{i=1}^d\sum_{j=1}^{n_i-1} \frac{\delta_\beta^{ij}\Psi_{\beta,ij}^{1,0,1,2}\theta_{\bm\xi}^{ij}}{\left(\Psi_{\beta,ij}^{0,0,1,2}\right)^2},
        \\ \\ X_{11}^\beta=\displaystyle \sum_{i=1}^d\sum_{j=1}^{n_i-1} \frac{\delta_\beta^{ij} \varphi_{\bm\xi}^{ij}}{\Psi_{\beta,ij}^{0,0,1,2}}, & X_{12}^\beta = \displaystyle \sum_{i=1}^d\sum_{j=1}^{n_i-1} \frac{\Psi_{\beta,ij}^{1,0,1,2}}{\Psi_{\beta,ij}^{0,0,1,2}}, \\ \\
  X_{13}^\beta=\displaystyle \sum_{i=1}^d\sum_{j=1}^{n_i-1} \frac{\left(\Psi_{\beta,ij}^{0,0,1,1}\right)^2}{\Psi_{\beta,ij}^{0,0,1,2} }, &
  X_{14}^\beta=\displaystyle \sum_{i=1}^d\sum_{j=1}^{n_i-1} \frac{\left(\Psi_{\beta,ij}^{0,1,0,1}\right)^2}{\Psi_{\beta,ij}^{0,0,1,2} },
  \\ \\ X_{15}^\beta=\displaystyle \sum_{i=1}^d\sum_{j=1}^{n_i-1} \frac{\delta_\beta^{ij}\Psi_{\beta,ij}^{0,1,0,1}}{\Psi_{\beta,ij}^{0,0,1,2}},
  \end{array}
$$
being $\Delta_{ij}=t_{i,j+1}-t_{ij}$ and $\varphi_{\bm\xi}^{ij} = - \alpha
\Psi_{\beta,ij}^{1,0,0,1} + \Psi_{\beta,ij}^{1,1,0,1} + \frac{\sigma^2}{2}
\Psi_{\beta,ij}^{1,0,1,1}$.

\section{Conclusions}

Mathematical modeling of tumor growth can help investigators to improve the design of preclinical
or clinical trials and to better predict treatment outcome. Actually a comprehensive description
of tumor dynamics during therapy in preclinical setting allows to accurately compare different
schemes of drug administration. Ultimately, preclinical correlations between tumor growth dynamics
during treatment and the efficacy of drug(s) could help to tailor the schedule to be proposed to
patients in clinical trials.

In this paper a modified Gompertz diffusion process including exogenous factors in its
infinitesimal moments has been considered in order to model both the effect of anti-proliferative
and/or cell death-induced therapies. A procedure to estimate the parameters and time functions
included in the model has been proposed, and our simulation studies show how said procedure
adequately reproduces both the parameters and the form of the functions involved, as well as the
mean and variance of the simulated data. In addition, from the estimated model, we have provided
bootstrap tests about the form of the true functions in the model. The estimated functions $C(t)$,
$D(t)$ and $V(t)$ that finally result allow to understand how therapies affect tumor growth.
Thus, our model could constitute a valuable tool to adjust the drug administration scheme in the
preclinical setting, in order to improve the efficacy of treatment, and to optimize the schedule
to be proposed to patients in clinical trials.

\section*{Acknowledgments}
The authors are very grateful to Dr. Didier Decaudin (Institute Curie, Paris) for providing the
data used in this research. They acknowledge the constructive criticism of anonymous reviewers on
the earlier version of this paper.\par\noindent This work was supported in part by the Ministerio
de Econom\'ia, Industria y Competitividad, Spain, under Grant MTM2017-85568-P and by the
Consejer\'ia de Econom\'ia y Conocimiento de la Junta de Andaluc\'ia, Spain under Grant
A-FQM-456-UGR18 and by MIUR - PRIN 2017, Project ``Stochastic Models for Complex Systems'' no.
2017JFFHSH. The authors Albano and Giorno are members of the research group GNCS-INdAM.


\bibliography{biblio}

\end{document}